\documentclass{aa}
\usepackage{txfonts} 
\usepackage{graphicx}
\usepackage{natbib} 
\usepackage{color}
\usepackage{upgreek}

 \makeatletter \newcommand \listoftodos{\section*{Automatic list of \textcolor{red}{Todo} items in text} \@starttoc{tdo}}
  \newcommand\l@todo[2]
    {\par\noindent \textit{#2}, \parbox{10cm}{#1}\par} \makeatother

 \makeatletter \newcommand \listoftocites{\section*{Automatic list of \textcolor{blue}{Tocite} items in text} \@starttoc{tct}}
  \newcommand\l@tocite[2]
    {\par\noindent \textit{#2}, \parbox{10cm}{#1}\par} \makeatother

\bibpunct{(}{)}{;}{a}{}{,} 
\title{Identifying gaps in flaring Herbig Ae/Be disks using spatially resolved mid-infrared imaging}
\subtitle{Are all group I disks transitional?}

\author{K. M. Maaskant\inst{1,}\inst{2} \and M. Honda\inst{3} \and L.B.F.M. Waters\inst{4,}\inst{2} \and A.G.G.M. Tielens\inst{1} \and C. Dominik\inst{2,}\inst{5} \and M. Min\inst{2} \and A. Verhoeff\inst{2} \and G. Meeus\inst{6} \and M. E. van den Ancker \inst{7}}
\institute{	Leiden Observatory, Leiden University, P.O. Box 9513, 2300 RA Leiden, The Netherlands \and 
		Anton Pannekoek Astronomical Institute, University of Amsterdam, P.O. Box 94249, 1090 GE Amsterdam, The Netherlands \and
		Department of Mathematics and Physics, Faculty of Science, Kanagawa University, 2946 Tsuchiya, Hiratsuka, Kanagawa, 259-1293, Japan \and
		SRON Netherlands Institute for Space Research, Sorbonnelaan 2, 3584 CA Utrecht, The Netherlands \and
		Department of Astrophysics/IMAPP, Radboud University Nijmegen, PO Box 9010 6500 GL Nijmegen, The Netherlands \and
		Universidad Autonoma de Madrid, Dpt. Fisica Teorica, Campus Cantoblanco, Spain \and
		European Southern Observatory, Karl-Schwarzschild-Str. 2, D-85748 Garching b. M\"unchen, Germany}

\abstract
{The evolution of young massive protoplanetary disks toward planetary
systems is expected to correspond to structural changes in 
observational appearance, which includes  the formation of gaps and
the depletion of dust and gas.}
{A special group of disks around Herbig Ae/Be stars do not show
prominent silicate emission features, although they still bear signs of
flaring disks, the presence of gas, and small grains.
We focus our attention on four key Herbig Ae/Be stars
to understand the structural properties responsible for the absence
of silicate feature emission.}
{We investigate Q- and N-band images taken with Subaru/COMICS,
Gemini South/T-ReCS and VLT/VISIR.  We perform radiative transfer modeling
to examine the radial distribution of dust and PAHs.  Our solutions
require a separation of inner- and outer- disks by a large gap.
From this we characterize the radial density structure of dust
and PAHs in the disk.}
{The inner edge of the outer disk has a high surface brightness and a
typical temperature between $\sim$100--150 K and therefore dominates the
emission in the Q-band. All four disks are characterized by large
gaps.  We derive radii of the inner edge of the outer disk of 34$^{+4}_{-4}$, 23$^{+3}_{-5}$, 30$^{+5}_{-3} $  and 63$^{+4}_{-4} $ AU for HD\,97048, HD\,169142, HD\,135344\,B and Oph IRS 48 respectively.  
For HD 97048 this is the first detection of a disk gap. 
The large gaps deplete the entire population of silicate particles with
temperatures suitable for prominent mid-infrared feature emission, while small
carbonaceous grains and PAHs can still show prominent emission at mid-infrared wavelengths. The continuum
emission in the N-band is not due to emission in the wings of
PAHs. This continuum emission can be due to VSGs or to thermal
emission from the inner disk.  We find that PAH emission is not always dominated by PAHs on the surface of the outer disk.}
{The absence of silicate emission features is due to the presence of
large gaps in the critical temperature regime.  Many, if not all Herbig disks
with Spectral Energy Distribution (SED) classification `group I' are disks with large gaps and can
be characterized as (pre-) transitional.  An evolutionary path from the
observed group I to the observed group II sources seems no longer
likely. Instead, both might derive from a common ancestor.}

\keywords{circumstellar matter --- stars: pre-main sequence ---protoplanetary disks---stars: individual (HD\,97048, HD\,169142, HD\,135344B, Oph IRS 48)--planet-disk interactions--stars: variables: T-Tauri, Herbig Ae/Be}
\date{\today} 
 \date{Received \dots / Accepted \dots}

\begin{document}

\maketitle


\section{Introduction}

Planetary systems form in the dusty disks around pre-main-sequence stars. The structure and evolution of these disks depend upon a wide range of parameters, such as the initial mass, size, and chemical characteristics of the disk and the properties of the central star. In addition, evolutionary processes such as grain growth, grain settling, photo-evaporation and planet formation can take place in disks. Their interplay will eventually transform gas rich disks into debris disks with possibly planetary systems, such as in Fomalhaut \citep{2012Boley}, $\upbeta$ Pictoris \citep{2010Lagrange} or HR8799 \citep{2010Marois}. Characterization of pathways toward mature planetary systems are key in understanding protoplanetary disks evolution. 

The evolution of intermediate-mass pre-main-sequence stars, known as Herbig Ae/Be stars, is different from that of lower- and higher-mass stars due to the differences in stellar and circumstellar physics \citep{2007Natta}. \citet{2001Meeus} classified the Herbig Ae/Be stars into two groups: group I sources, which require blackbody components up to far-infrared wavelengths to fit the SED, and group II sources, which can be well fitted with only a single power law at mid- to far-infrared wavelengths. 

Several explanations have been proposed to understand the evolutionary link between group I and group II objects. \citet{2004bDullemond, 2005Dullemond} showed that spectral energy distributions (SEDs) of group I sources can be interpreted by hydrostatic disks with a flaring geometry while group II sources are thought to be the evolved version of group I sources: as dust grains coagulate and settle to the mid-plane, the disk becomes flatter producing the steeper, bluer mid- to far-infrared SEDs. Additionally, a decrease of mid-to far-infrared flux in the SED can be explained by a puffed-up inner disk that shadows large parts of the outer disk \citep{2004aDullemond}. 

Gaps have been found in an increasing number of Group I sources such as in AB Aur \citep{2010Honda}, HD\,142527 \citep{2006Fukagawa, 2006Fujiwara}, HD\,135344 \citep{2009Brown}, HD\,36112 \citep{2010Isella}, HD\,169142 \citep{2012Honda}, Oph IRS 48 \citep{2007aGeers} and HD\,100546 \citep{2003Bouwman, 2010Benisty}. Recently, \citet{2012Honda} proposed that the classification as a group I source could mean a disk geometry with a strongly depleted inner disk (i.e. a transitional disk). Due to gaps and inner holes, huge vertical walls arise at the edge of the outer disk and significantly contribute to the mid-to far-infrared flux. In view of the possibility that all group I disks are transitional disks, the relation between group I and group II sources may have to be revisited.

To get a better understanding of the geometry of group I disks and the use of PAHs to trace the geometry of disks, we examine a representative set of four group I objects: HD\,97048, HD\,169142, HD\,135344\,B and IRS48 for which we have obtained a homogeneous set of N- and Q-band data. These targets have been observed and resolved before in the N- and Q-band by \citet{2011Marinas} and \citet{2007aGeers}. Our study adds a careful examination of the disk sizes in the continuum and in the PAH filter at 11.2 $\upmu$m by comparing to radiative transfer models. In Section \ref{sec:sample}, we introduce the sources in our sample. Thereafter we describe the observations, which are used as an input for the analysis in Section \ref{sec:observations}. We start in Section \ref{sec:modelQ} by fitting the Q-band size and demonstrate that the flux at 18.8 and 24.5 $\upmu$m is dominated by the inner edge of the outer disk and thus gives a solid constraint on the location where the outer disk starts. In Section \ref{sec:modelN}, we show that the disk size in the N-band at 10.5 $\upmu$m is given by emission from the inner disk, outer disk, and very small grains and is therefore less suitable to derive the gap radius. We show in Section \ref{sec:modelPAH} that the FWHM of the 8.6 and 11.2 $\upmu$m PAH features in our sample do not correlate with the disk geometry which implies that imaging in the PAH bands cannot always be used as a reliable tracer of the disk structure. In Section \ref{sec:discussion}, we discuss the implications of our findings. Section \ref{sec:conclusions} summarizes our conclusions.

\begin{table*}[htdp]
\tiny
\caption{  \label{tab:sample} Star \& disk parameters used in this study }
\begin{center}
\begin{tabular}{lcccccccccc}
\hline
\hline
Object		& R.A. (J2000)	& Decl. (J2000)			&Teff [K]		& L$_{*}$				& Av		& d 			& M$_*$ 				&	i 						\\
			& h     m      s	& $^\circ$     '     ''  		&	[K]		&	[L$_\odot$]		& 		& [pc]		&  [M$_{\odot}$]		&	[$^{\circ}$]				\\
\hline
HD\,97048		&11 08 03.32	& -77 39 17.48			&10 010		&40.70$\pm$5.9		&1.21	&158$^{+16}_{-14}$ & 1.84				&43$^{\circ}$		 \\
HD\,169142	& 18 24 29.79	&-29 46 49.22			&8200		&15.33$\pm$2.17		& 0.46	& 145$^{+15}_{-15}$	& 2.28 				&13$^{\circ}$	\\
HD\,135344\,B 	&15 15 48.40	& -37 08 55.86			&6590		& 8.26$\pm$1.17		& 0.32 	&140$^{+14}_{-14}$	& 3.30				&11$^{\circ}$					\\
Oph IRS 48	&  16 27 37.19	& -24 30 34.80			&9000		&14.3$\pm$5			& 11.50	& 120$^{+5}_{-5}$	& 2.0					&48$^{\circ}$		\\

\end{tabular}
\end{center}
\end{table*}%

\begin{figure*}[htbp]
   \centering
   \includegraphics[width=0.4\textwidth]{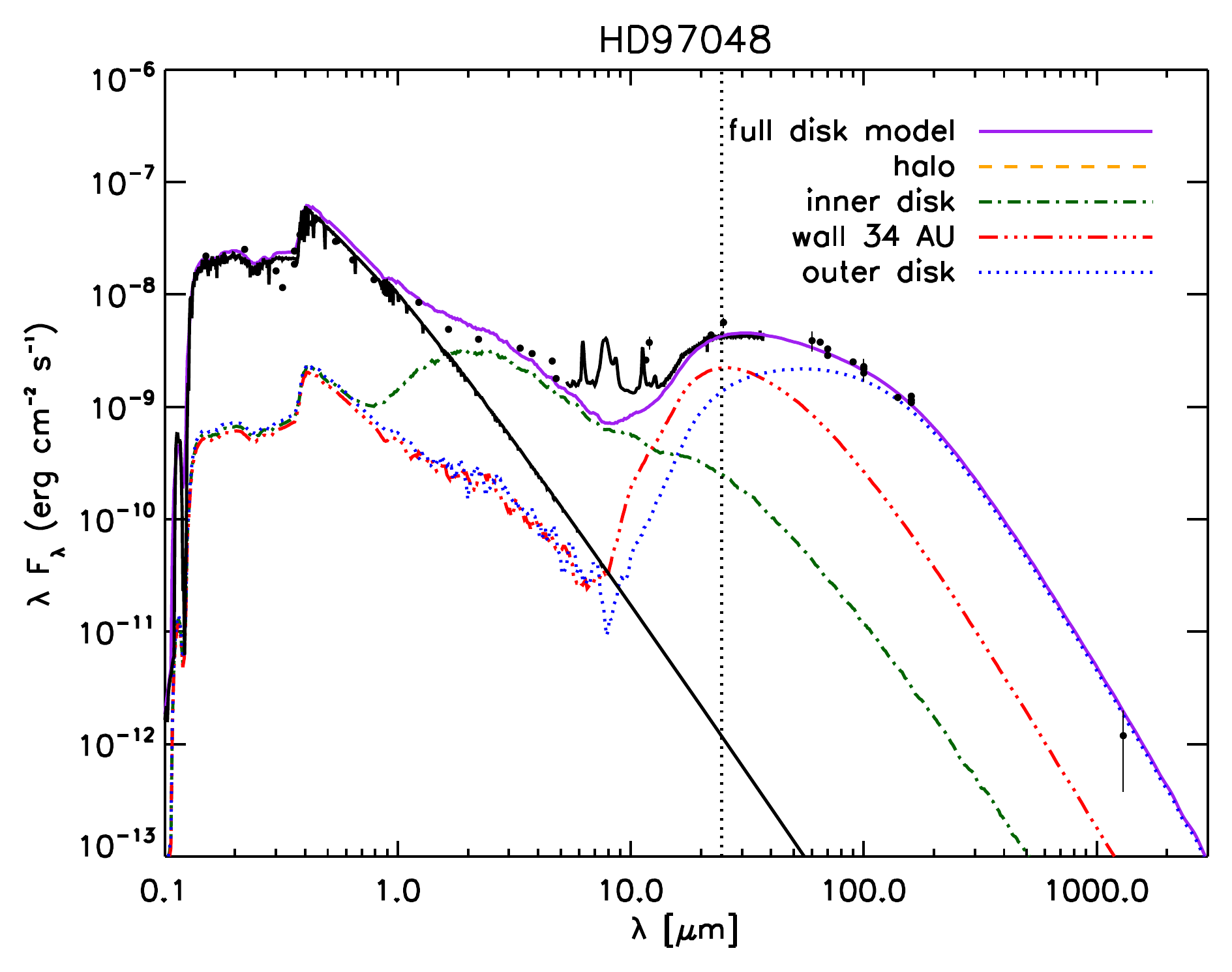} 
      \includegraphics[width=0.4\textwidth]{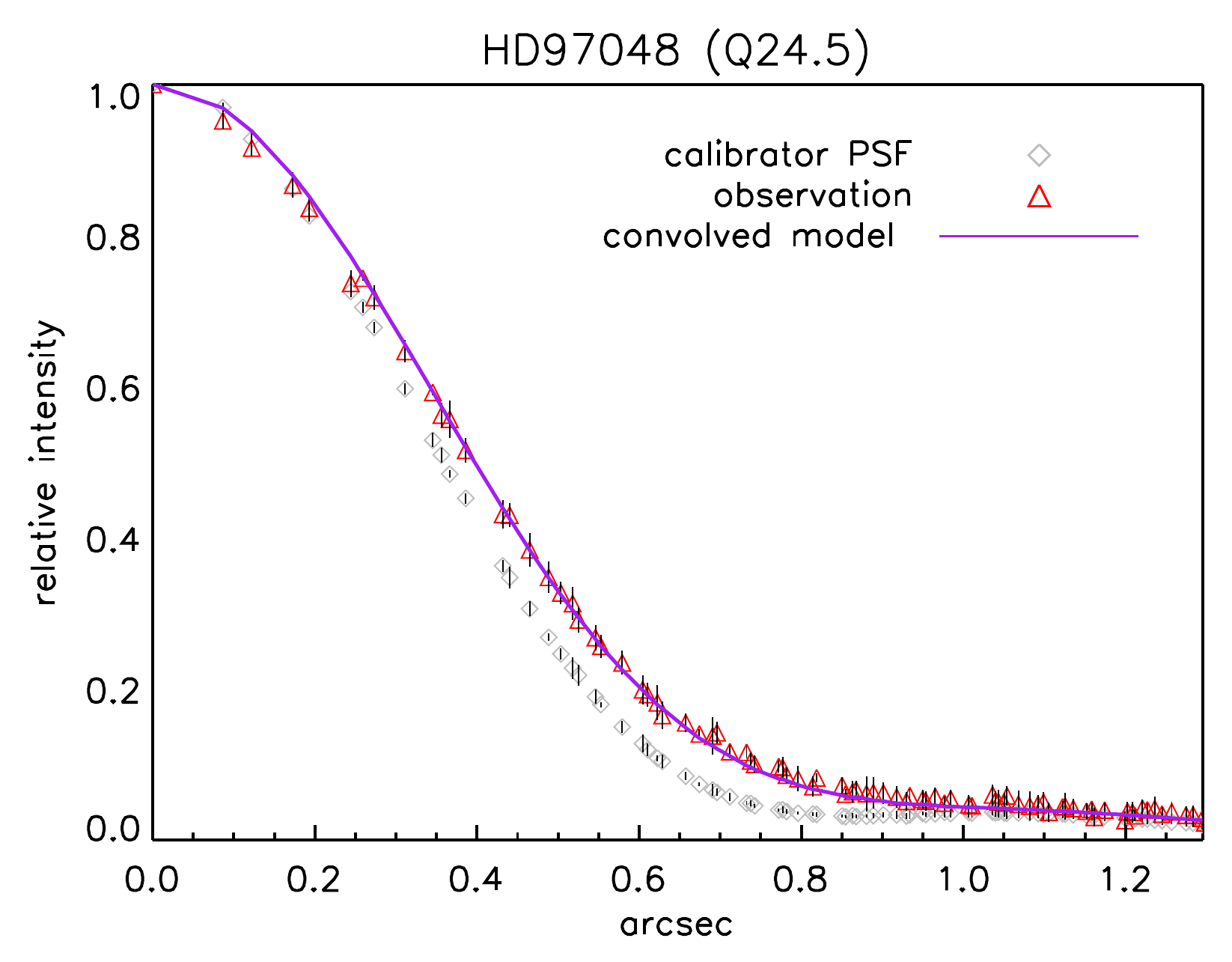} 

   \includegraphics[width=0.4\textwidth]{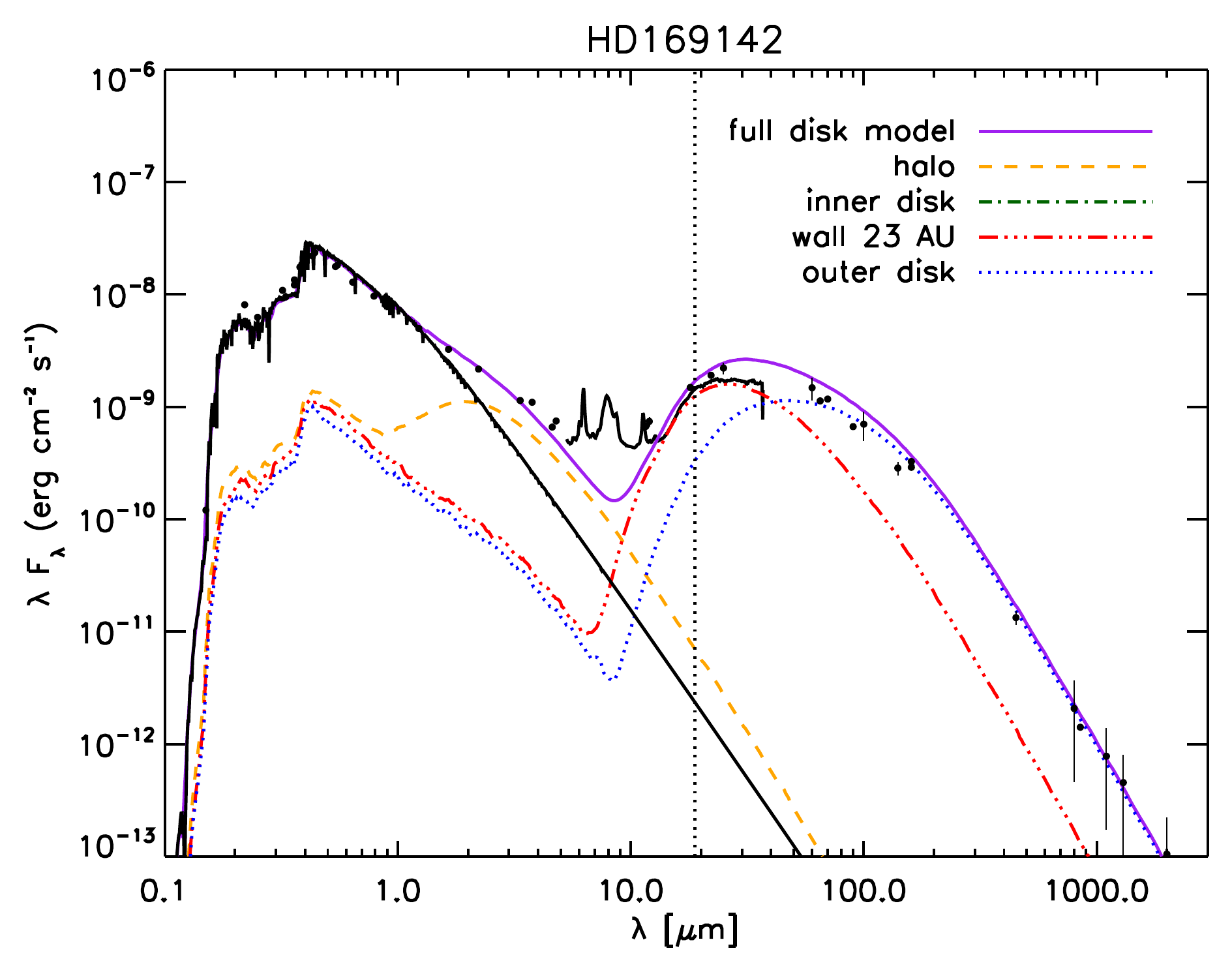} 
      \includegraphics[width=0.4\textwidth]{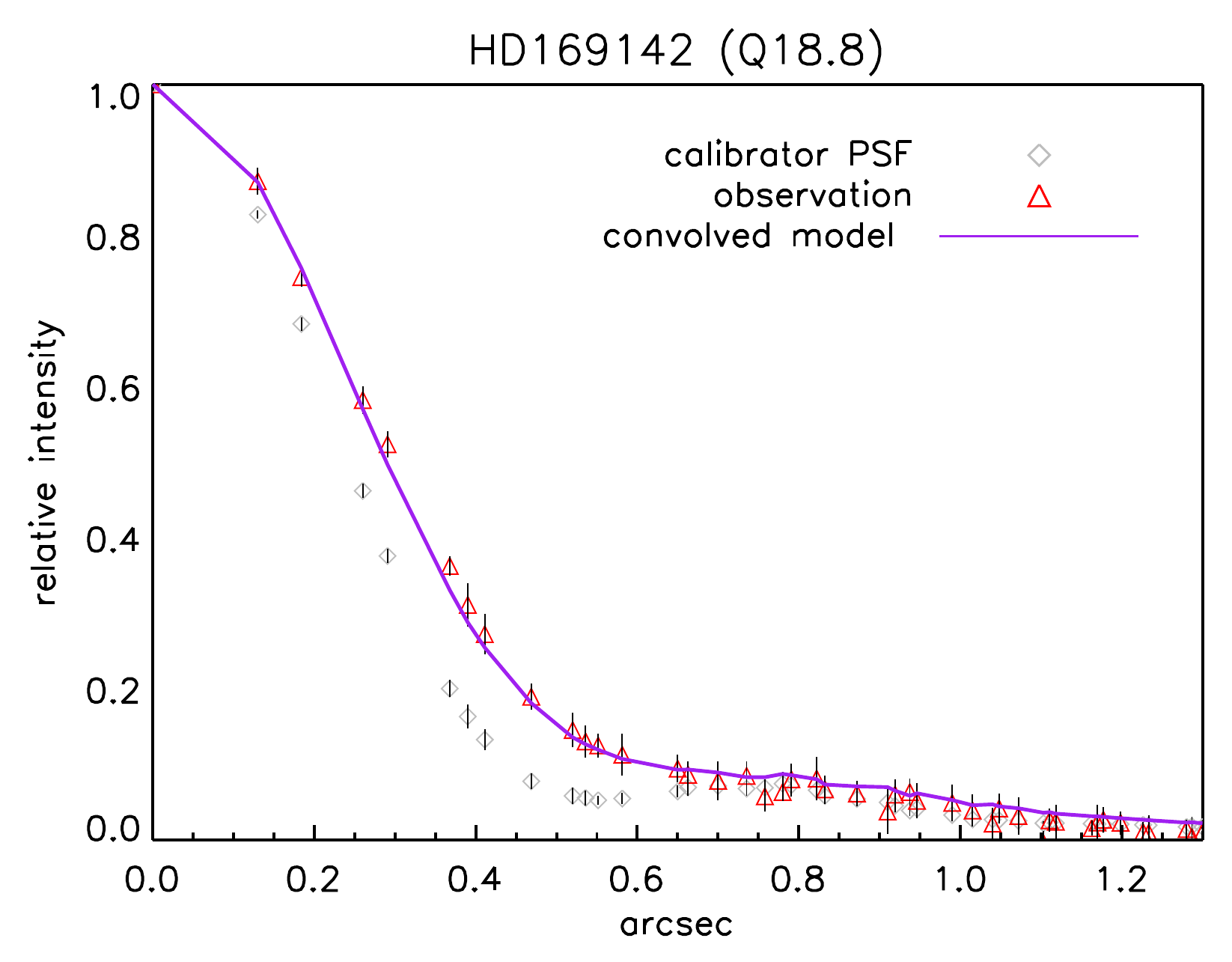} 

   \includegraphics[width=0.4\textwidth]{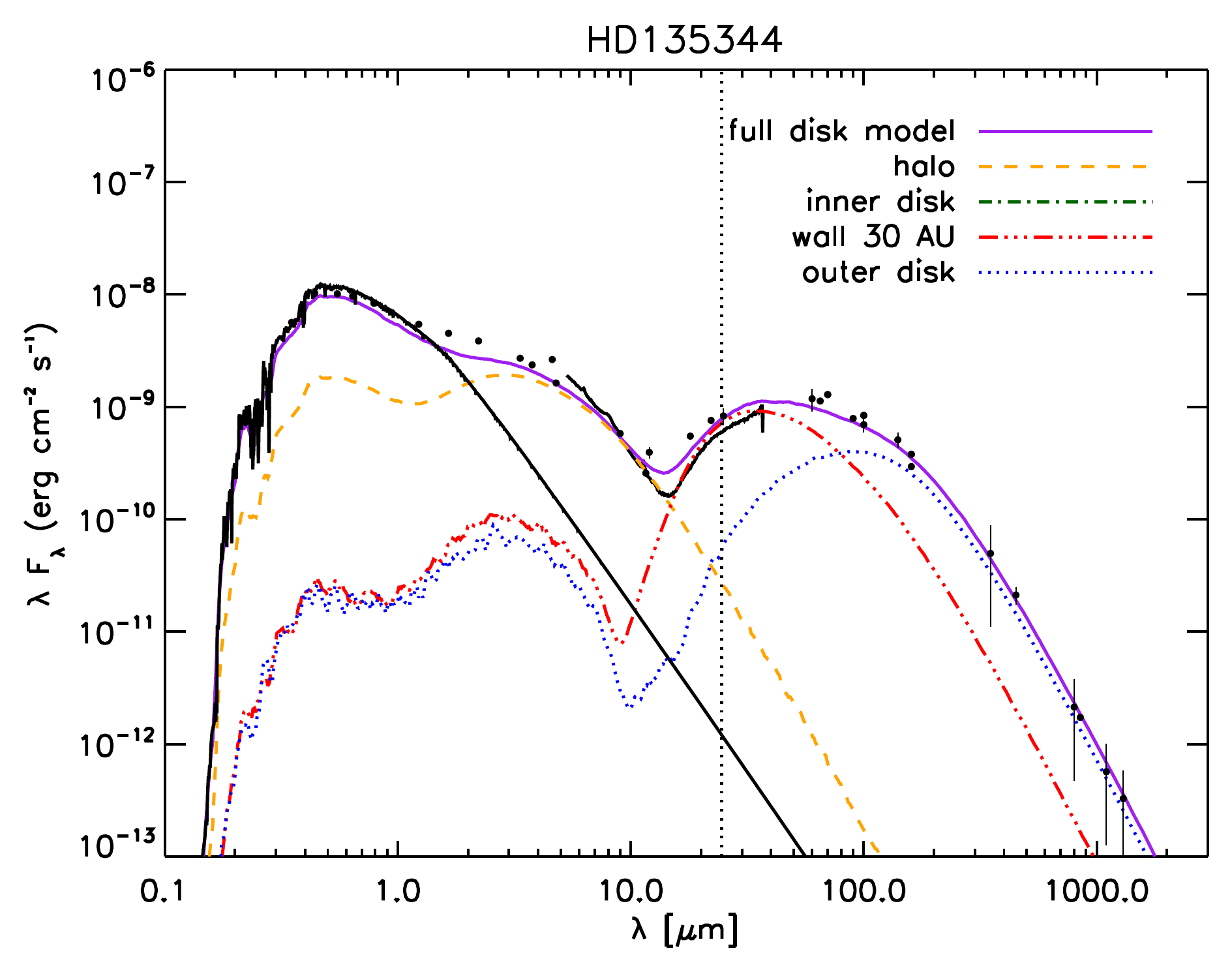} 
      \includegraphics[width=0.4\textwidth]{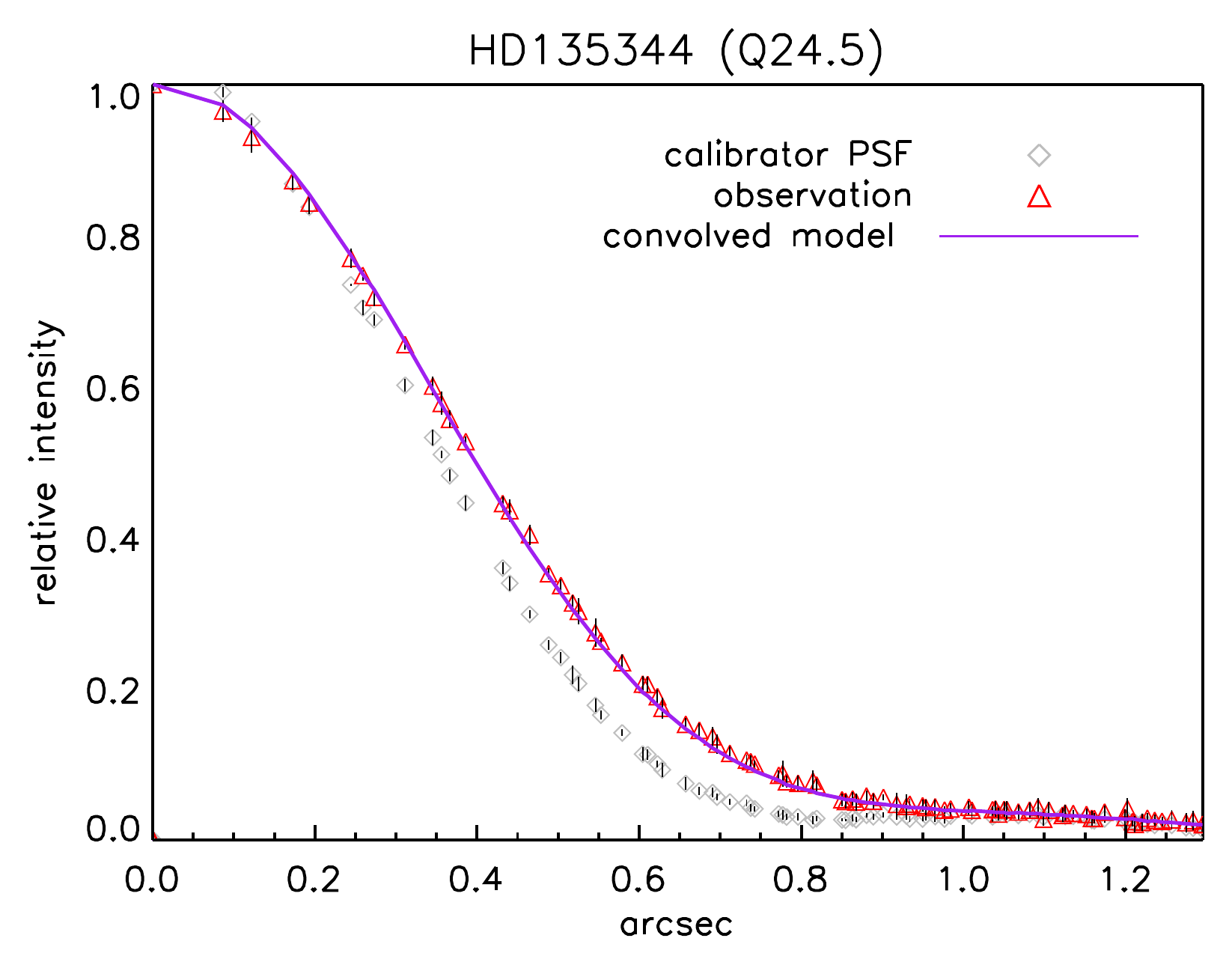} 

   \includegraphics[width=0.4\textwidth]{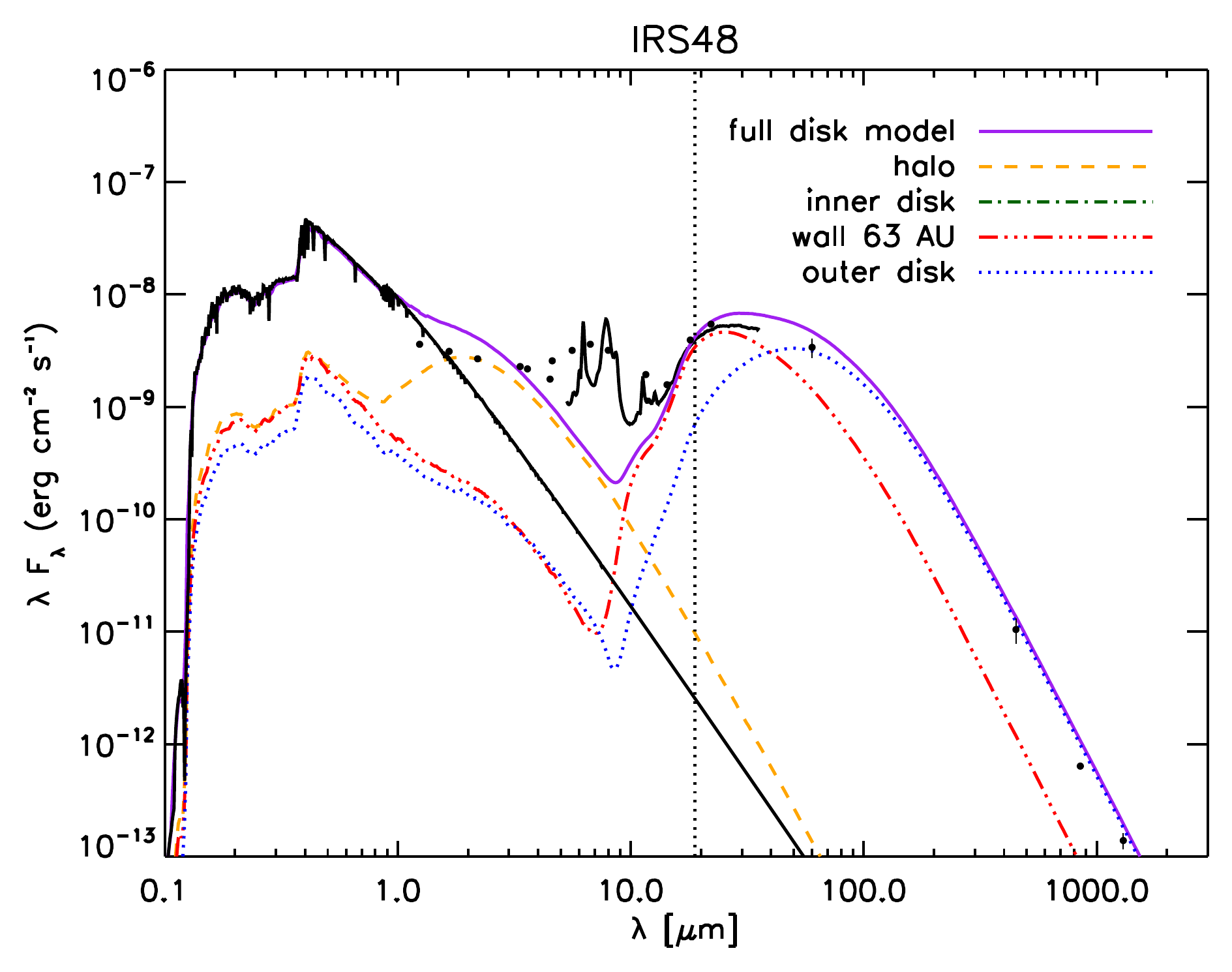} 
      \includegraphics[width=0.4\textwidth]{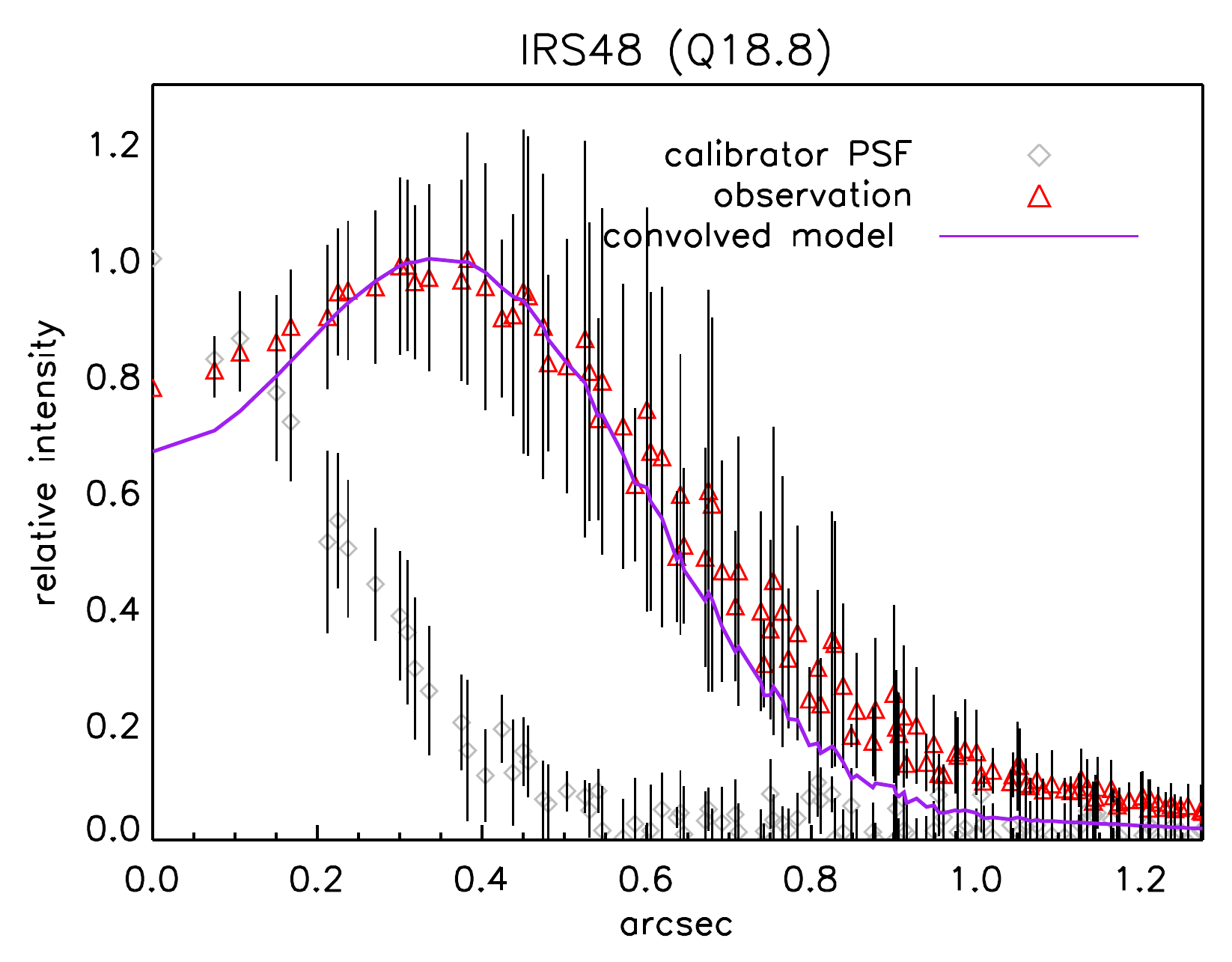} 
   
   \caption{Left: decomposed SEDs. The black solid lines show the Kurucz models and the Spitzer IRS spectra. The black dots represent the observed photometry corrected for extinction (See also Appendix \ref{sec:photometry}.). The solid purple line is the sum of the flux of all the components. The dashed yellow and dashed-dotted green lines are the halo and inner disk components, respectively. The triple dotted dashed red line indicates the emission from the wall (from the wall radius to 3 AU outwards). The dotted blue line shows the emission from the rest of the outer disk. Note that the IRAS photometry and Spitzer spectrum do not match very well for HD\,97048, this may be caused by the difference in beam sizes. Right: azimuthally averaged radial brightness profiles of the Q-band relative to the maximum flux. The central wavelengths of these images are 18.8 $\upmu$m for HD\,169142 and Oph IRS 48  and 24.5 $\upmu$m for HD\,97048 and HD\,135344\,B. The 24.5 $\upmu$m fit to HD\,169142 can be found in \citet{2012Honda}. The grey diamonds indicate the PSF of the calibration star. The red triangles show the observation of the science targets.  The error bars indicate the one sigma error on the azimuthally averaged radial brightness profiles. The solid purple line shows our best fit model. }
   \label{fig:SEDQ}
\end{figure*}

     \begin{figure*}[htbp]
   \centering
     \includegraphics[width=0.4\textwidth]{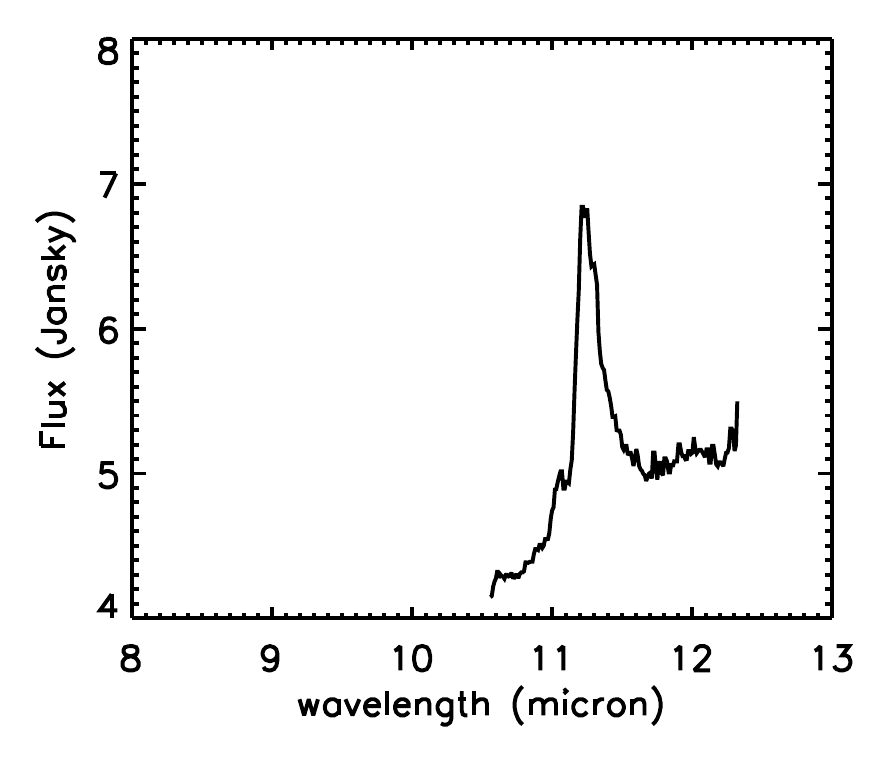} 
     \includegraphics[width=0.4\textwidth]{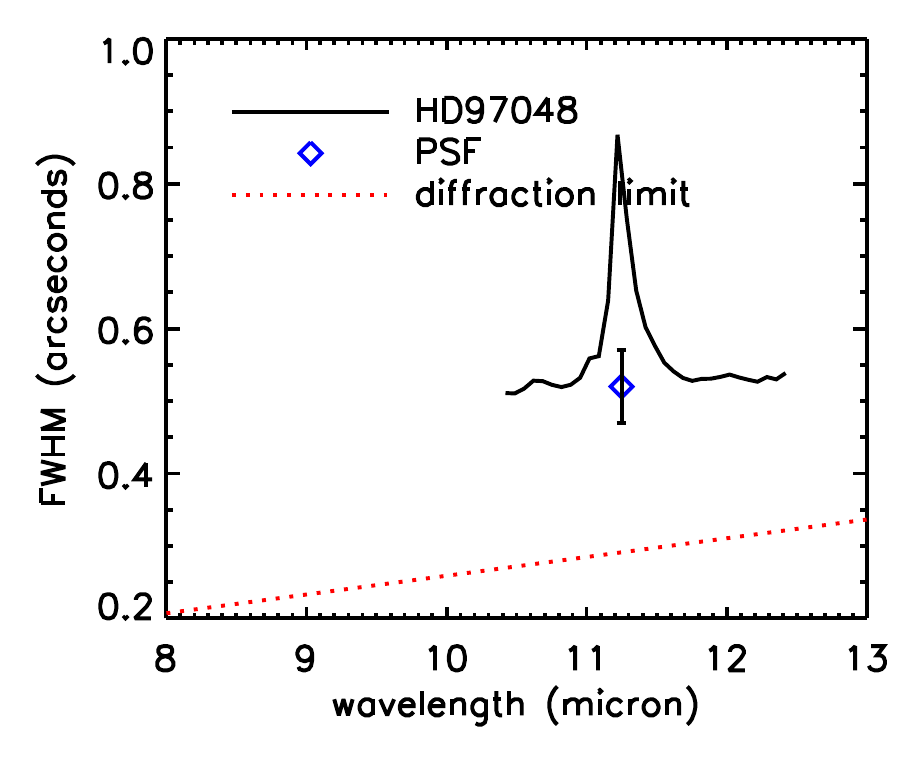} 
     \includegraphics[width=0.4\textwidth]{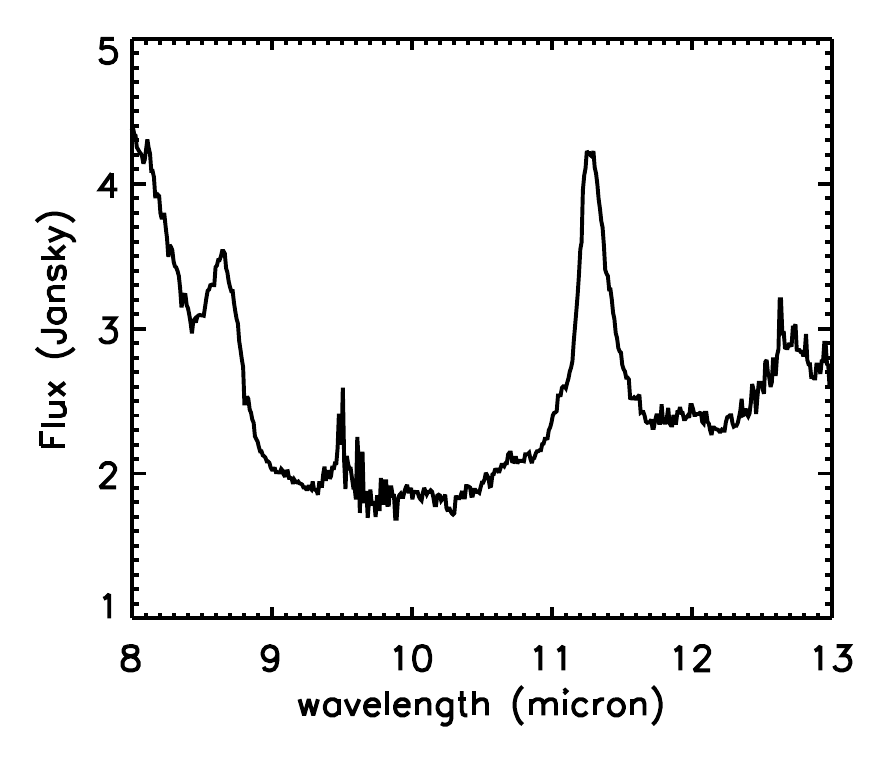} 
     \includegraphics[width=0.4\textwidth]{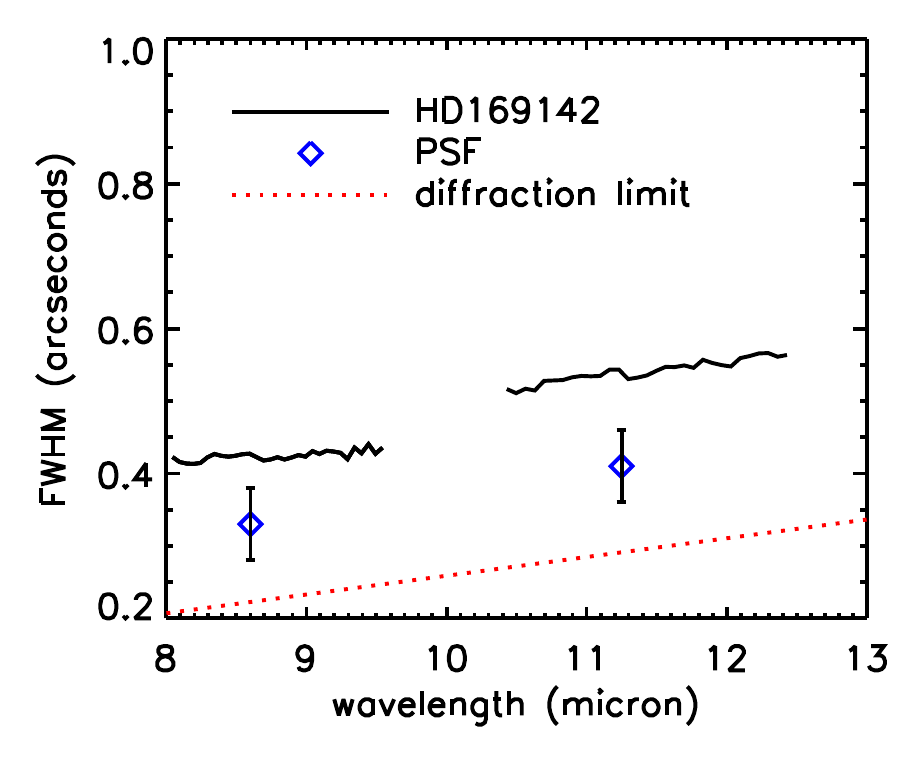} 
     \includegraphics[width=0.4\textwidth]{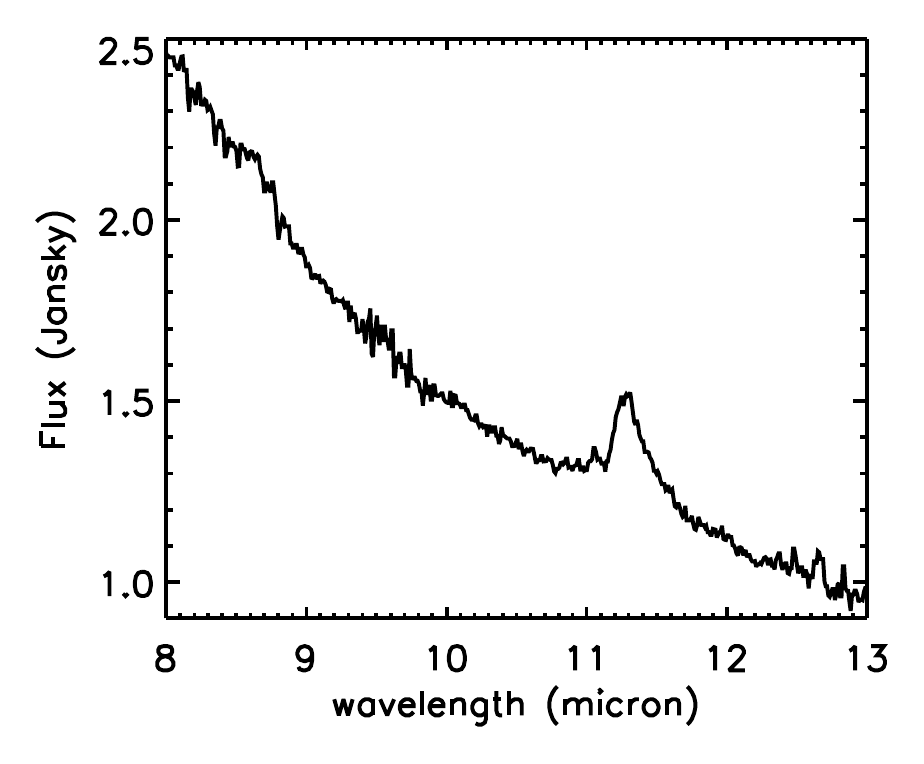} 
     \includegraphics[width=0.4\textwidth]{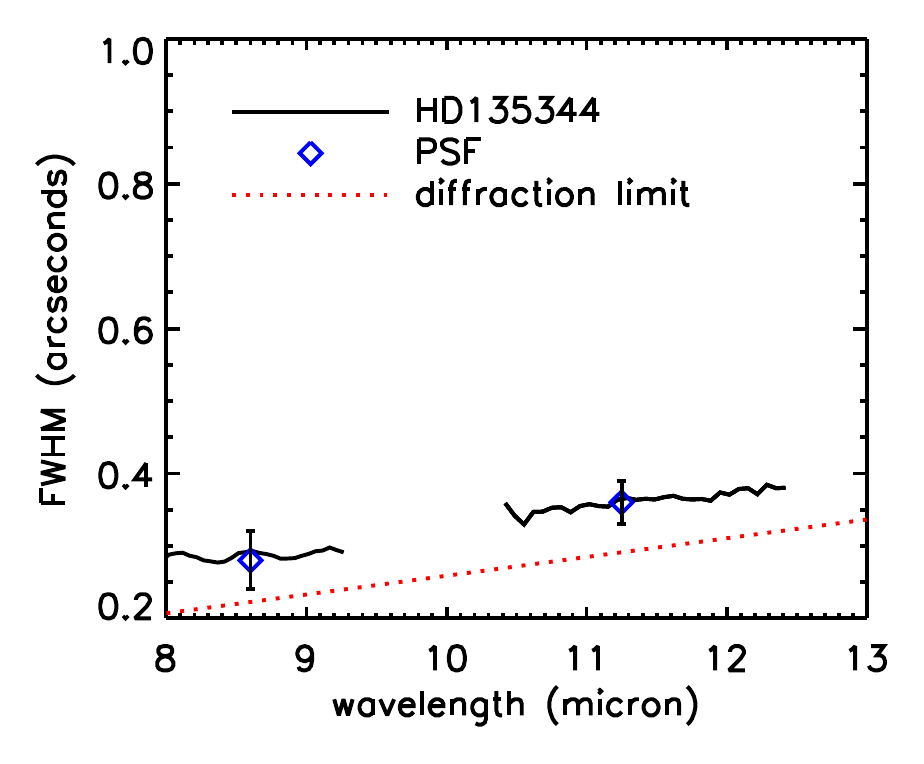} 
     \includegraphics[width=0.4\textwidth]{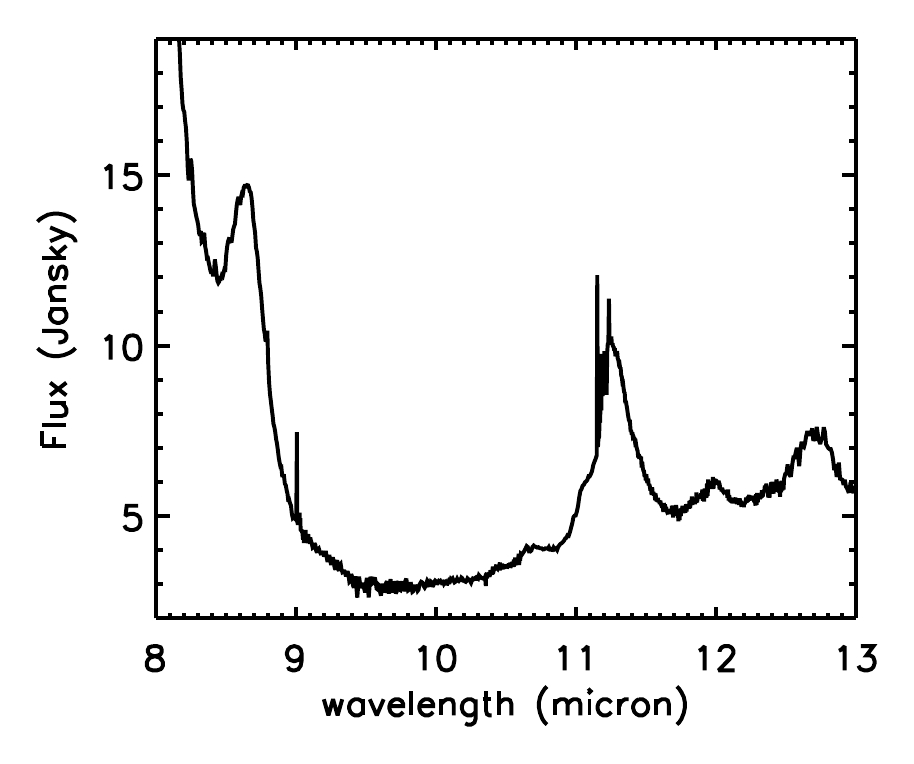} 
     \includegraphics[width=0.4\textwidth]{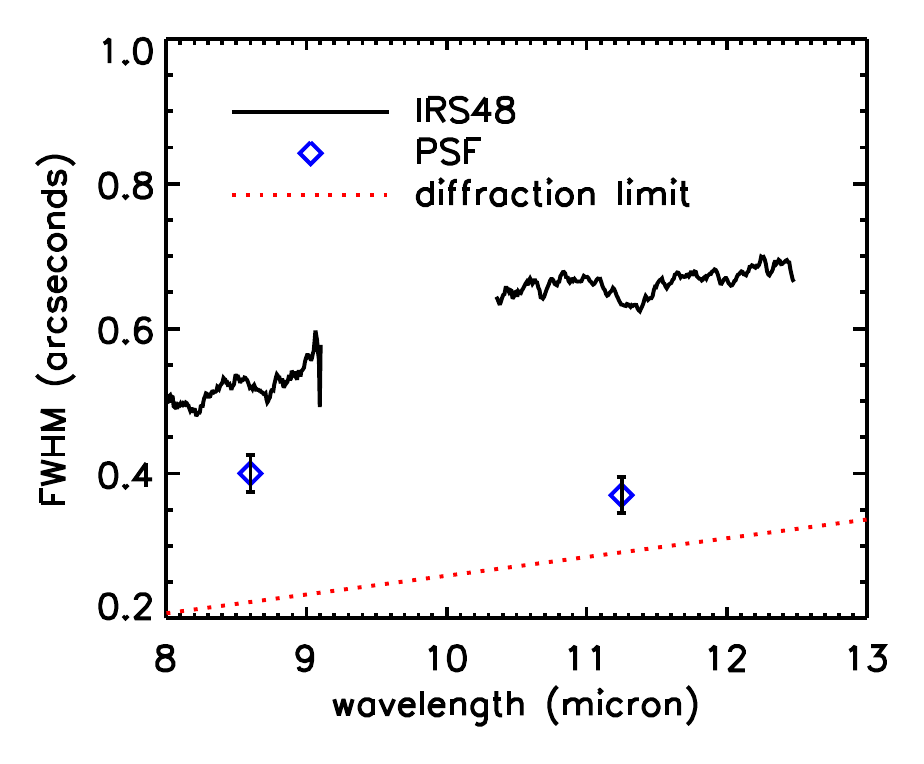} 
     
   \caption{ \label{fig:FWHMprofiles} VISIR observations in the N-band. Left: the observed spectra, right: the corresponding FWHM (in arcseconds) as a function of wavelength. From top to bottom, the VISIR observations of HD\,97048, HD\,169142, HD\,135344\,B, and Oph IRS 48 are shown. The red line shows the diffraction limit for a 8.2 m telescope; the blue diamonds show the PSF values of the statistical calibrator. The 11.2$\upmu$m PAH feature in HD97048 is more extended than the continuum (which is unresolved) and as large as the continuum for HD\,169142, but smaller than the continuum in Oph IRS 48. For HD\,135344\,B, the continuum and PAH features are both unresolved.}
\end{figure*}


\section{The Sample}
\label{sec:sample}
We selected a sample of four Herbig Ae/Be objects for study: HD\,97048, HD\,169142, HD\,135344\,B, and Oph IRS 48. The SEDs are shown in Figure \ref{fig:SEDQ}. We collected a comprehensive data set for these objects, which consists of Spitzer IRS spectra and photometry. See Appendix \ref{sec:photometry} for an overview of available photometric UV, optical, infrared, and millimeter data and their references. Spitzer IRS spectra are taken from \citet{2010Juhasz}. The Spitzer IRS spectrum of Oph IRS 48 has been reduced with the latest calibration version (S18.18.0, \citealt{2013Sturm}). These four sources are representative for the sample of flaring Herbig Ae/Be objects, which lack the silicate features in their spectra. We have chosen these stars, since we could acquire a homogeneous set of MIR data for study. The four selected stars are listed in Table \ref{tab:sample} with the adopted spectral types, temperatures, luminosities, visual extinctions, distances, stellar masses, and disk inclinations. The SEDs are bright at mid- to far-infrared and sub-mm wavelengths and all show a similar `bump' of excess emission at $\sim$20 micron. Detailed radiative transfer modeling suggests that this kind of SEDs are consistent with disks with partially depleted inner region (i.e., a cavity or a gap), while a Òwall-likeÓ structure at the outer edge of this region can be responsible for the abrupt rise of the SED at MIR \citep{2003Bouwman, 2005Calvet, 2007Espaillat}. In the remainder of this paper, we present and discuss the targets ordered by their sub-mm emission above the photosphere, which is strongest for HD\,97048, followed by HD\,169142, HD\,135344\,B and Oph IRS 48. With all other properties being equal, the sub-mm excess can be used as a crude estimate of the total disk mass. Thought we are aware that the sub-mm excess is sensitive for some other properties, such as the differences in stellar temperature, luminosity and the temperatures and opacities in the outer disk, we use the sub-mm excess as a rough indication of its evolutionary position between a massive young flaring disk towards an evolved debris disk.

\subsection{HD\,97048}

TIMMI2 observations by \citet{2004Boekel} resolved the disk at 8-13 $\upmu$m. They suggested that the dominating source of continuum emission is from very small carbonaceous grains since thermal emission from grains in the outermost region of the disk are too cold to dominate the SED in the continuum. Resolved VISIR images in the N-band \citep{2006Lagage, 2007Doucet} showed a detection of the circumstellar disk with an inclination of i $\sim$43$^{\circ}$ from pole-on. It was shown that the extended PAH emission at 8.6 $\upmu$m could be explained by surface emission from a hydrostatic disk model with a flaring geometry. 
HST/ACS V-band images \citep{2007Doering} revealed scattered light out to a radial distance of $\sim$4'' in almost all directions. Recent polarimetric differential imaging detected polarized scattered light emission between $\sim$0.1''--1.0'' ($\sim$16--160 AU) \citep{2012Quanz}. No detection of dust depleted inner regions have been yet reported. 

Several gas emission lines ([OI], CO, and H2) have been detected and were, at least partly, spectrally and spatially resolved \citep{2006AckeAncker, 2009Plas, 2007Martin-Zaidi, 2011Carmona}. The spatial distribution of gas is peculiar: analysis of the CO emission profile shows that the CO gas sets in at greater than 11 AU, while 80\% of the [OI] emission originates between 0.8 $<$ R $<$ 20 AU. This was taken as a possible sign of photo-dissociation of CO in the inner regions and heating by high-energy UV and X-ray photons.

\subsection{HD\,169142}
The disk around HD\,169142 has been resolved with Subaru/COMICS in the Q-band \citep{2012Honda}. Fitting the Q-band images with radiative transfer models, they derive an inner radius of 23$^{+3}_{-5}$ AU for the outer disk, which is in agreement with SED modeling by \citet{2010Meeus}. H-band VLT/NACO polarized light images \citep{2013Quanz} confirmed this result where it is shown that the inner disk ($\lesssim20$ AU) appears to be depleted of scattered dust grains and an unresolved disk rim is imaged at $\sim$25 AU. In addition \citet{2012Quanz} found an extra gap extending from $\sim$40 -- 70 AU. Modeling of the 1.3 mm continuum emission and CO line emission shows that the disk has an outer radius of $\sim$235 AU and is viewed close to pole-on with an inclination of i $\sim$13$^{\circ}$ \citep{2006Raman, 2008Panic}. This is in agreement with scattered light detected out to $\sim$200 AU with differential polarimetry \citep{2001Kuhn}, HST/ASC observations \citep{2007Grady}, and polarized light images \citep{2013Quanz}. The PAH emission feature at 3.3 $\upmu$m is extended with a FWHM of 0.3'' ($\sim$50 AU) \citep{2006Habart}, though the underlaying continuum is not. 

Near-infrared (NIR) excess emission indicates that the inner region ($<$23 AU) is not empty of dust and gas. Modeling by \citet{2012Honda} favored the idea that an optically thin, but geometrically thick component close to the star is the best dust-solution to model the NIR excess. Detections of [OI] \citet{2005Acke}, H$\alpha$ \citep{1997Dunkin} and Br$\gamma$ \citep{2006GarciaLopez} emission indicate the presence of gas in the inner disk of HD\,169142. Higher resolution interferometric observations are needed for a better understanding of the inner disk. 

\subsection{HD\,135344\,B }

The disk of HD\,135344\,B (SAO 206462) has recently been imaged in the H-band with Subaru/HiCIAO \citep{2012Muto}. Scattered light components with two small-scale spiral structures are observed close to 0.2'' ($\sim$ 28 AU) from the central star. This result was surprising since SMA sub-mm imaging \citep{2009Brown} showed a larger inner hole of $\sim$ 39 AU. These different radii of the beginning of the outer disk may indicate different radial distributions of small and large grains \citep{2012Dong, 2012Pinilla}. 
Observations of CO line profiles \citep{2005Dent, 2008Pontoppidan, 2011Lyo} indicate an almost pole-on geometry (i $\sim$11$^{\circ}$). From 20.5$\upmu$m imaging, \citet{2006Doucet} derived an outer radius of 200 AU for the dust disk.

Compared to the other sources in our sample, the MIR spectrum of HD\,135344\,B shows relatively weak PAH features. This has been attributed to its lower stellar temperature and thus lower UV field (e.g. \citealt{2010Acke}). For our study, HD\,135344\,B serves as a perfect comparison object of a protoplanetary disk in which emission from the stochastically heated PAHs and very small grains is significantly weaker.    

For the inner disk structure, interferometric N-band observations \citep{2008Fedele} suggested a much higher inclination (53$^{\circ}$--61$^{\circ}$) with estimated inner and outer radii of 0.05 and 1.8 AU respectively. \citet{2008Pontoppidan} observed molecular CO gas ($\lambda$ = 4.7 $\upmu$m) at 0.3--15 AU. Though variabilities in the NIR \citep{2012Sitko} may indicate planet-disk interactions, the physical origin of the inner disk remains unclear. 

\subsection{Oph IRS 48} 

The disk around the Herbig Ae star Oph IRS 48 (WLY 2-48) in the $\rho$ Ophiuchi star formation region was imaged and resolved using VLT/VISIR \citep{2007aGeers}.  A distance of 120 pc has been derived for the $\rho$ Ophiuchi star formation region \citep{2008Loinard}. The 18.8 $\upmu$m image reveals a ring structure with an $\sim$1'' diameter (120 AU). From the east-west orientation of the continuum image an inclination of 48$^\circ$ was derived. The brightness distribution of the image is not symmetric, the flux on the east side of the star is $\sim$30\% higher than on the west side. The spectral type of A0$\pm$3 has been derived from optical spectroscopy by \citet{2010McClure}. 

The size of the gap is unclear as observations at 18.8 $\upmu$m dust continuum, PAH emission and 4.6 $\upmu$m CO all indicate very different gap sizes. \citealt{2007bGeers} found that the PAH emission at 8.6, 9.0 and 11.2 $\upmu$m peaks closer towards the central star than the dust continuum ring at 18.8 $\upmu$m. A study by \citet{2012aBrown} using high spatial and spectral resolution VLT/CRIRES data found vibrationally excited CO gas in a ringlike structure at 30 AU around the star. No CO was found closer than 30 AU from the star. SMA observations by \citet{2012bBrown} show a 12.9 AU inner hole radius at 880 $\upmu$m. Observations of higher spatial resolution are needed to differentiate between potential scenarios to explain the different radial distributions of CO gas and large and small dust grains.


\section{Observations}
\label{sec:observations}

\begin{table*}[htdp]
\tiny
\caption{\label{tab:FWHM}FWHM sizes }
\begin{center}
\begin{tabular}{ l l l l l l}

\hline
\hline
Object			&			&FWHM 10.5$\upmu$m			& FWHM 11.2$\upmu$m 				& FWHM 18.8$\upmu$m 			&FWHM 24.5$\upmu$m \\		 
				&			& ['']							&				 ['']				& 				 ['']			&			 ['']		 \\
\hline

HD\,97048		&PSF		& 0.270$\pm$0.015 $^a$			& 0.55$\pm$0.05$^b$				&\dots						& 0.720$\pm$0.006	 $^e$	\\
				&obs			&0.330$\pm$0.015  $^a$			&0.83$\pm$0.05$^b$ 				&\dots						&0.789$\pm$0.017		 $^e$	\\
\vspace{0.1cm}\\
HD\,169142		&PSF		& 0.40 $\pm$0.05$^b$			&  0.33$\pm$0.05$^b$				 &0.493$\pm$0.006 $^d$			& 0.628$\pm$0.007  $^d$	\\
				&obs			&0.50 $\pm$0.05$^b$			& 0.50$\pm$0.05 $^b$				& 0.604$\pm$0.017  $^d$			& 0.680$\pm$0.034  $^d$	\\
\vspace{0.1cm}\\
HD\,135344\,B 		&PSF		&0.30 $\pm$0.03 $^b$			&0.30$\pm$0.03 $^b$				& \dots						& 0.719$\pm$0.004 $^e$	\\
				&obs			& 0.30 $\pm$0.03 $^b$			&0.30$\pm$0.03 $^b$				& \dots						&0.802$\pm$0.007	 $^e$\\
\vspace{0.1cm}\\	
IRS48			&PSF		&  0.41$\pm$0.025$^c$			&0.41$\pm$0.025 $^c$				&0.5$\pm$0.1 $^c$				&\dots \\
				&obs			&  0.66$\pm$0.025$^c$			&  0.63$\pm$0.025$^c$				& 0.7 $\pm$0.1$^c$				&\dots \\
		\vspace{0.1cm}\\
		
\end{tabular}
\end{center}
\textbf{References: a)} \citet{2007Doucet}  \textbf{b)} \citet{2009Verhoeff}  \textbf{c)} \citet{2007bGeers} \textbf{d)} \citet{2012Honda} \textbf{e)} This study

\end{table*}%

\begin{figure*}[htbp]
   \centering
   \includegraphics[width=1\textwidth]{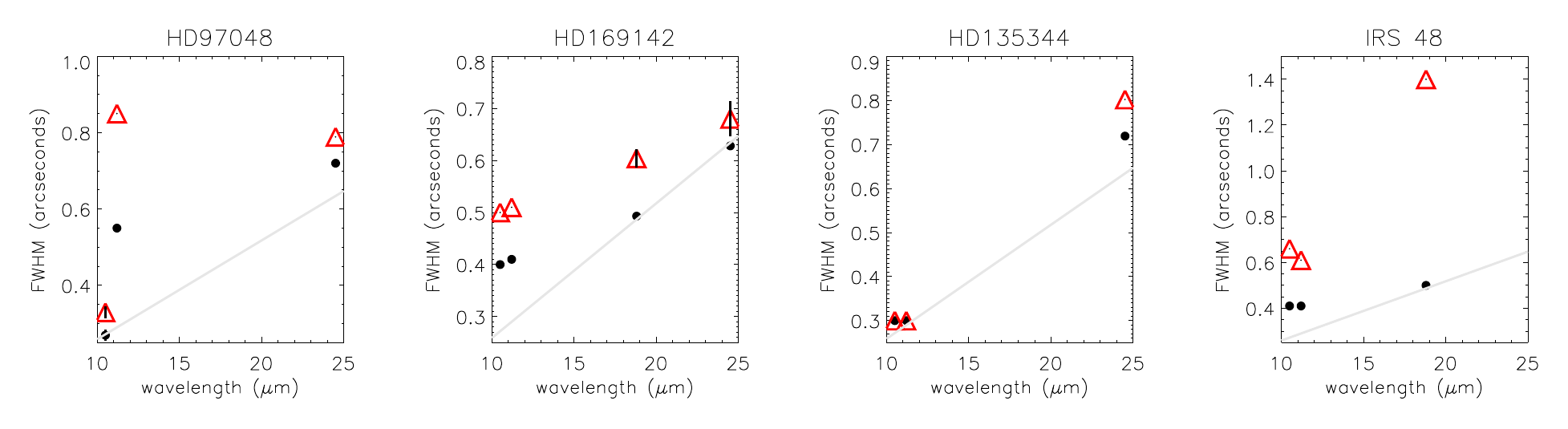} 
   
   \caption{The FWHM of the four targets as function of wavelengths are given by the red triangles. Their corresponding PSF is given by the black dots. The FWHM sizes are taken from different observations (References are given in Table \ref{tab:FWHM}.). The solid grey line is the FWHM size of the diffraction limit of a 8.2 m telescope. }    \label{fig:sizes}
\end{figure*}

\subsection{Q-band imaging}
All objects in our sample are resolved in the Q-band. An overview of the observed sizes is included in Table \ref{tab:FWHM} and Figure \ref{fig:sizes}. Images at 24.5$\upmu$m for HD\,97048 and HD\,135344\,B were obtained with the T-ReCS on the Gemini South telescope on UT 2011 June 28 and 14, and the total integration times were 637 s and 1361 s, respectively. The calibration stars used for PSF measurements were Gamma Cru for HD\,97048 and Alpha Cen for HD\,135344\,B. The filter used for the T-ReCS observation is Qb centered at 24.56 $\upmu$m ($\Delta \lambda$ = 1.92$\upmu$m). For T-ReCS imaging data, the chop and nod technique was used. The reduction of the T-ReCS imaging data has been done using the pipeline in the Gemini IRAF MIDIR package provided by Gemini. Each observation is accompanied with the PSF (Point Spread Function) references obtained just before or after the main object. In all the PSF data taken with Gemini/T-ReCS in 24.56 $\upmu$m during the observing runs, the FWHM size is relatively stable (0.72'' $\pm$ 0.03''). Images at 18.8 $\upmu$m and 24.5 $\upmu$m for HD\,169142 were taken from \citep{2012Honda}, which were observed using COMICS on the 8.2 m Subaru Telescope on Mauna Kea, Hawaii. An 18.8$\upmu$m image of Oph IRS 48 is published in \citet{2007aGeers}, which was taken with VISIR on VLT. 
 
\subsection{N-band VLT/VISIR}
\label{sec:VISIR}

During the nights of March 26, 2005 and June 16, 2005, and July 9, 2007, HD\,169142, HD\,97048, HD\,135344\,B  have been observed respectively with the VLT Imager and Spectrometer for the mid-IR (VISIR, \citealt{2004Lagage}). These observations were part of the VISIR GTO program on circumstellar disks (see e.g. \citealt{2009Verhoeff}). Long-slit low-resolution (R$\approx$300) spectroscopy was performed using the 8.8 and 11.4\,$\upmu$m setting with a 0.75$\arcsec$ slit in the North-South direction. For HD\,135344\,B  the 8.5\,$\upmu$m setting was used. Standard parallel chopping and nodding was employed with a chopper throw of 8$\arcsec$ to correct for the atmospheric background. The observing conditions were good for HD\,169142 and HD\,135344 in which the airmass was in between 1.0-1.1 and the optical seeing $\sim$0.6$\arcsec$ and$\sim$0.8$\arcsec$ respectively, and poor for HD\,97048 in which the airmass was in between 1.6-1.7 and the optical seeing $\sim$1.3$\arcsec$.  The spatial FWHM of the spectra and their PSF was extracted with the method described in \cite{2012Verhoeff}. VLT/VISIR data of Oph IRS 48 is taken from \citet{2007bGeers}. In this study, we do not fit the absolute sizes of the N-band VISIR observations because of the large uncertainty of the statistical calibrator. However, relative size differences in the observations can be trusted.  The spectra and FWHM are shown on Figure \ref{fig:FWHMprofiles}.


\section{Description of the model}
\label{sec:modeldiscription}
For modeling the disks, we use the radiative transfer tool MCMax \citep{2009Min}. MCMax uses an axially symmetric density setup and performs 3D Monte Carlo radiative transfer using the scheme outlined by \citet{2001BjorkmanWood} to solve the temperature structure and scattering of light in the disk. MCMax has been successfully applied to model observations of protoplanetary disks over a large range of wavelengths and observational techniques (e.g. \citealt{2011Verhoeff, 2012Honda}). It is successfully benchmarked against a large number of radiative transfer codes \citep{2009Pinte}. The mass accretion rates of the disks that we study is sufficiently low to ignore viscous heating through accretion. Thereby, our only energy source is the central star, which is described by a Kurucz model for the stellar photosphere using the stellar parameters listed in Table 1.

For the radial density profile we use the so-called similarity solution \citep{2008Hughes, 2011Andrews},
\begin{equation}
\Sigma(r) \propto r^{-p}\exp \left\{- \left( \frac{r}{R_0} \right) ^{2-p}\right\},
\end{equation}
for $r < R_\mathrm{out}$. Here $R_0$ is the turnover point from where an exponential decay of the surface density sets in and $p$ sets the power law in the inner region. We fix this to $p=1$, which is a commonly used value \citep[see e.g.][]{2006Dullemond}. We allow for a gap in the radial surface density profile.
We solve for the vertical structure of the disk assuming vertical hydrostatic equilibrium. We assume turbulence is sufficiently high such that the gas and dust are well mixed and there is no size dependent dust settling.

The hydrostatic equilibrium causes the inner edge of a disk to be puffed up, because it is directly irradiated by the central star. This is not only true for the innermost disk edge but also for the inner edge of the outer disk. This allows such a wall to capture and reprocess significant amounts of radiation at a single temperature.

To compute the optical properties of the dust grains in the disk, we use the distribution of hollow spheres \citep[DHS; see][]{2005Min} with an `irregularity parameter' $f_\mathrm{max}=0.8$. This method simulates the properties of irregularly shaped particles by breaking the symmetry of a homogeneous sphere. For the grain composition we take a simple dust mixture of 80\% silicates \citep{1995Dorschner, 1996HenningStognienko, 1998Mutschke} and 20\% amorphous carbon \citep{1993Preibisch}. Scattered light contribution is included in the total flux, and for simplicity, we have assumed isotropic scattering. Opacities are calculated from the optical constants using a size distribution from a$_{min}$ to $a_{max}$ proportional to $f(a) \propto a^{-p}$. For each object, we choose a maximum grain size of 1 mm and fit the minimum grain size (between 0.1 and 1.0 $\upmu$m) to the SED. Dust larger than millimeter grains do not leave signatures on the observed SED and are therefore of little relevance. The size distribution of the grains is fitted to the observed spectral slope from the far-IR to the submm using power-laws of $p = -3.5$ and $p = -4.0$. The power-law index is close to the size distributions ($p = -3.5$) of interstellar grains \citep{1977Mathis}.

The PAHs and VSGs optical properties are taken from \citet{2007DraineLi}. The spectral feature profiles of the PAH and VSG models are based on observations that attempt to mimic the spectra for the central regions of galaxies in the SINGS galaxy sample \citep{2007Smith} at wavelengths $\uplambda > 5.5 \upmu$m. We used a PAH and VSG size of 0.001 $\upmu$m  ($N_{\mathrm{C}} \sim$ 460) and 0.01$\upmu$m  ($N_{\mathrm{C}}  \sim 4.6 \times 10^4$) respectively. We have chosen single sized PAHs and VSGs to be able to distinguish their effect to the spectrum. \citet{2007DraineLi} note that the broad feature seen in the opacity profile of VSG near 30 $\upmu$m may not apply to realistic carbonaceous grains since the measured absorption in amorphous carbon grains does not appear to show a peak near 30 $\upmu$m  \citep{1983Tanabe, 1999Mennella}. However, studies of well-developed PDRs, such as the Orion Bar and the reflection nebula \mbox{NGC 2023} have indicated that $\sim$ 20 -- 30 $\AA$ VSG ($N_{\mathrm{C}} \sim 10^3 -10^4$) may be responsible for a 25 $\upmu$m cirrus \citep{2008Tielens}. There is yet no clear evidence for this broad feature in protoplanetary disks. 

In the radiative transfer code, the PAHs and VSGs are treated as transiently heated particles in the temperature distribution approximation. This means that the particles are heated using single photon events. We use the cooling timescale of the particles to determine the average temperature distribution before the next photon hits the particles. In this way, we get a temperature distribution function for these transiently heated grains that can be used to compute their emissivity. This emissivity is used again in the Monte Carlo radiative transfer to include these grains fully self-consistently. This procedure included multiphoton events (i.e., a photon hits the particle before it is fully cooled down), which are important close to the star. Note that it is needed to iterate the procedure to get a self-consistent solution since the emissivity of the PAHs and VSGs depend on the local radiation field that is determined by the radiative transfer. However, convergence is very fast since most of the heating of these grains is done by direct UV radiation from the star.

\begin{figure}[htbp]
  \centering
  \includegraphics[width=0.5\textwidth]{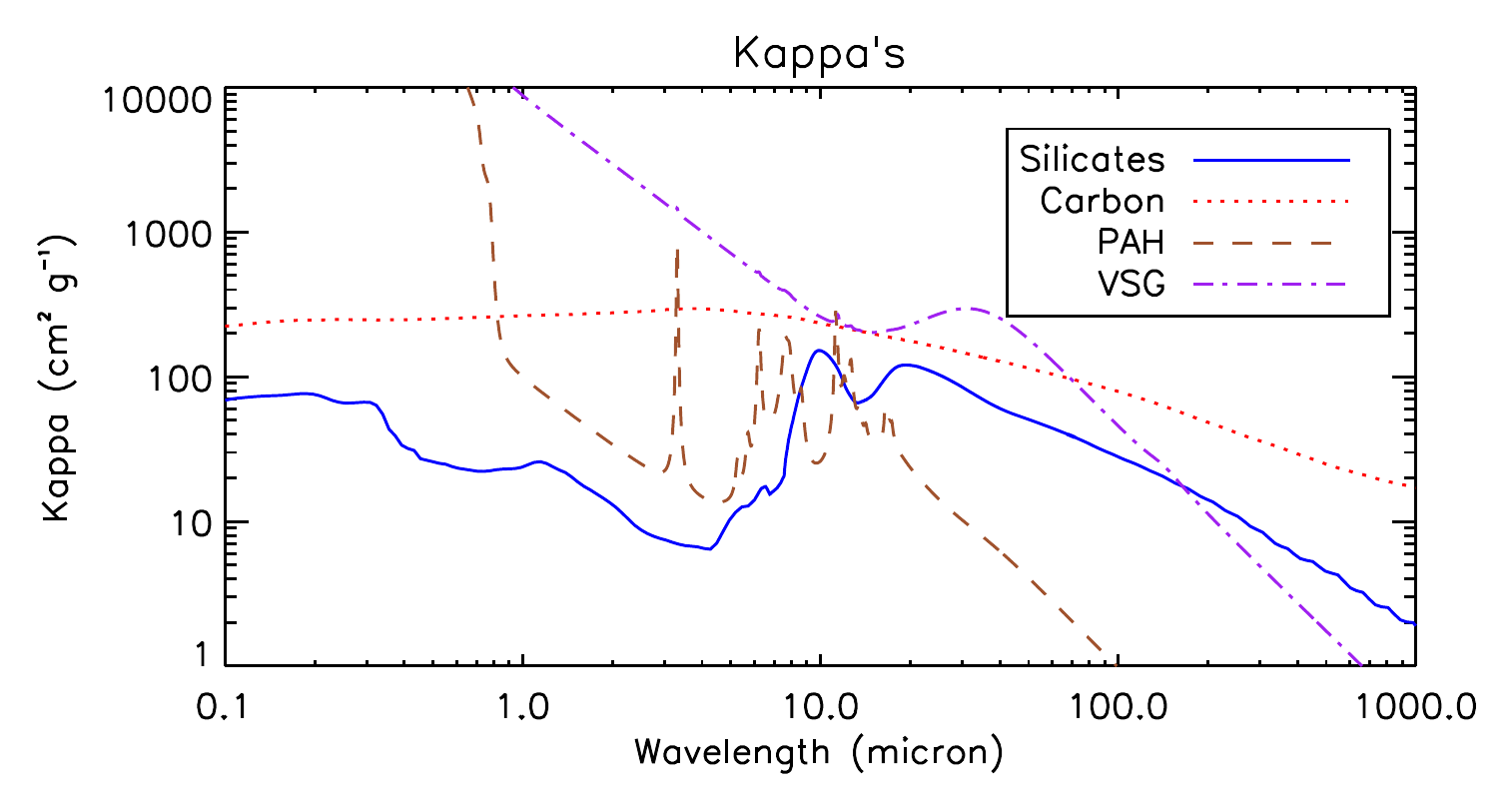} 
  \caption{Examples of absorption crosssection per unit mass  used in this paper for amorphous olivine (0.1$\upmu$m -- 1mm, $p = -3.5$) and amorphous carbon (0.1$\upmu$m -- 1mm, $p = -3.5$) and PAHs and VSGs with sizes of 0.001 and 0.01 $\upmu$m respectively. The halo consists of 100\% carbon. For the disk components, we used a composition of 80\% silicates and 20\% amorphous carbon in the entire disk.}
  \label{fig:kappas}
\end{figure}

Examples of the computed optical properties used in our model are shown in Figure \ref{fig:kappas}. The halo component in the models consists of 100\% carbon. For the disk components, we used a composition of 80\% silicates and 20\% amorphous carbon in the entire disk. Thus the opacity is dominated by silicate grains in the UV and mid- to far-IR wavelength domain while at optical, NIR and mm wavelengths, the carbon grains dominate the opacity.

\section{The disk size in the Q-band }
\label{sec:modelQ}

\begin{table*}[htdp]
\tiny
\caption{   \label{tab:modelfit}Best-fit model parameters}
\begin{center}
\begin{tabular}{llllllll}

\hline
\hline
Object		&	M$_{dust}$			&M$_{halo}$ &R$_{inner disk/halo}$ 	&	R$_{wall}$ 		&R$_{out}$			&	a				& p 				\\
			&		[M$_{\odot}$]		&		[M$_{\odot}$]		&	 [AU]				&	 [AU]				&	 [AU]				&[a$_{min}$, a$_{max}$]	&  					\\
\hline
HD\,97048	&	6.0$\times$10$^{-4}$	&	\dots	&	0.3 - 2.5			&	34$^{+4}_{-4}$ 	& 500				&\{0.5$\upmu$m, 1mm\}	& -3.5	\vspace{0.1cm}	\\	
HD\,169142	&	0.8$\times$10$^{-4}$	&	0.31$\times$10$^{-12}$	&	0.1- 0.2			&	23$^{+4}_{-4}$		& 235 				&\{0.5$\upmu$m, 1mm\}	& -3.5	\vspace{0.1cm}			\\	
HD\,135344\,B 	&	1.0$\times$10$^{-4}$	&	0.47$\times$10$^{-12}$	&	0.1- 0.3 			&	30$^{+4}_{-3} $ 	& 200				&\{1.0$\upmu$m, 1mm\}	& -4.0	\vspace{0.1cm}			\\	
Oph IRS 48	&	3.0$\times$10$^{-5}$	&	0.50$\times$10$^{-12}$	&	0.1 -0.3 			&	63$^{+4}_{-4} $	& 235				&\{0.1$\upmu$m, 1mm\}	& -4.0			\\	

\vspace{0.4cm}\\

\end{tabular}
\end{center}

\end{table*}%

\subsection{Modeling goal}

Our goal is to reproduce the SED and the azimuthally averaged Q-band brightness profile shown in Figure \ref{fig:SEDQ}. In this section, we will first show that a disk with a continuous density distribution fails to fit the data. Instead, a multiple-component disk model, which includes an inner disk or an inner halo, a gap, a wall, and an outer disk fits the observations best. We find that the inner edges of the outer disks are located in a temperature range of $\sim$100--150 K and dominate the emission in the Q-band. Therefore, we robustly constrain the location of this wall by fitting our model to the observed Q-band images. We will discuss the results for each object individually. In the end of this section, we briefly discuss the free parameters, assumptions and other possible degeneracies.

\subsection{Models with a continuous density distribution fail} 
It is well known that the model fits to SEDs are highly degenerate in the derived geometry (e.g. \citealt{2011Andrews}). However, resolved images in the Q-band can break this degeneracy. Fitting the observed Q-band size provides a solid measurement of the location of the inner edge of the outer disk dominated by micronsized grains. We have explored a wide range of parameters but fail to reproduce the observed Q-band sizes and SEDs using models with a continuous density distribution (i.e., no gap or depleted inner region). Test-modeling shows that the emission at Q-band wavelengths originates in dust located in the inner $\lesssim$10 AU for disks with a continuous density profile from the inner rim at the dust-sublimation radius up to the outer disk. This implies the convolved Q-band sizes of such disk models have a similar size as the PSF and are thus not resolved that for an observation taken with an 8 meter class telescope of a disk located at $\sim$150 pc. For a diffraction limit observation, these sizes are 0.50'' at 18.8 $\upmu$m  and 0.63'' at 24.5 $\upmu$m. Since all the sources in our sample are significantly resolved in the Q-band images, we consistently fail to reproduce our imaging data with continuous disk models.

Another possible solution to broaden the Q-band size is to include stochastically heated VSG to the disk. For HD\,97048,  only a very high abundance ($\gtrsim$ 3 $\times$ 10$^{-6}$ M$_{\odot}$) of VSGs broadens the Q-band image to the observed size. However, this model gives a very bad fit to the shape of the SED: the flux is over-predicted in the MIR range, and the shape of the SED is too flat (i.e., especially since the characteristic emission bump between $\sim$15-30 $\upmu$m is poorly fitted). For HD\,169142 and HD\,135344\,B, we fail to reproduce the observed Q-band size with any abundance of VSG. We conclude that continuous density models fail to simultaneously fit the SED and Q-band size and consider the gapless disk solution unlikely. 

\subsection{Fitting strategy for a multiple-component disk model}

We find that an inner hole/gap (i.e., a depletion of $\gtrsim10^2$ of the surface density in the inner region) is required to fit the Q-band size and SED. The effect of a gap is that the inner edge of the outer disk has a high surface brightness and a higher temperature at larger distances from the star.  The MIR bumps at $\sim$20 $\upmu$m seen in all SEDs hint that the inner and outer disks are decoupled. These bumps follow the shape of a modified black-body with a temperature of $\sim$150 K, suggesting that a disk component with a similar temperature is dominating the SED. Our fitting procedure can be  summarized as follows:

\begin{enumerate}
\item Start with a disk with a continuous density profile and fit the Far-Infrared (FIR) to mm photometry to a grain size power-law index of p between 3.0 and 4.0. 
\item Make a gap in the disk by decreasing the surface density by 15 orders of magnitude. This will result in a ``wall'' structure at the inner edge of the outer disk. Choose the radius of the inner edge of the outer disk, so that the convolved model image fits the observed Q-band image size. 
\item Put in an optically thick inner disk. This will increase the flux in the NIR but casts a shadow on the outer disk and therefore reduces the flux at MIR and FIR wavelengths.   
\item If an optically thick inner disk fails to produce enough NIR flux or casts too much shadow on the outer disk, add or replace with an optically thin inner halo to fit the NIR flux.  
\item Choose the minimum size (between 0.1 $\upmu$m and 1 $\upmu$m) of the grains in the disk to fit the flux in the MIR and FIR to the SED.
\end{enumerate}

We have tested various different approaches and find that we are able to fit the data by using this procedure using a largely similar set of free parameters and assumptions for the grains and in the physical structure of the disk (see also Section \ref{sec:modeldiscription}). This disk fitting procedure allows us to fit all the Q-band images and SEDs (Figure \ref{fig:SEDQ}). The N-band is not yet considered, since there are multiple candidates to be responsible for the emission. These candidates will be discussed in Section \ref{sec:modelN}, and it will be shown that we follow a valid approach, since it does not influence the Q-band size.

The spatial resolution and sensitivity of our Q-band data is not sufficient to study the effects of asymmetries in the brightness profile for which could be caused by the inclination or complex structures of the disk wall. Therefore we \emph{azimuthally average} the radial brightness profiles in order to compare the observed and model images (see Figure \ref{fig:SEDQ}). With the assumption that the wall of the outer disk has a vertical shape, we fit the radius of the inner edge of the outer disk, so that the convolved model image fits the observed brightness profile. The model image is convolved with the PSF (i.e., the observed calibration star). The uncertainty in the distance of the objects translates into the dominant uncertainty in constraining the radius of the wall.

The inner disk structure is not well known in Herbig Ae/Be stars. It has been found that the NIR flux is often much higher than expected for an inner dusty disk in hydrostatic equilibrium \citep{2009Acke}. As a solution, we use an optically thin but geometrically thick shell of dust around the star to fit the NIR flux. Replacing the optically thick inner disk with an optically thin halo does not affect the Q-band size of the models. We assume a 100\% amorphous carbon distribution in the halo because it has a steep opacity profile and thus gives the best fit to the SED. Alternatively, large ($\gtrsim 10 \upmu$m) silicate grains or metallic iron can also be used.  Although the physical nature of a halo is not yet understood, it has some useful properties. While a puffed up inner disk could also be used as a solution to fit the NIR flux of the disk, test modeling shows that it casts a significant shadow and therefore requires the entire outer disk to have a pressure scale-height, which is $\sim$2-3 times higher than for a hydrostatic disk. A halo does not cast a significant shadow on the outer disk so that the temperature is kept high by direct irradiation and thereby the scale height remains high in the outer disk. We prefer to maintain hydrostatic equilibrium in the outer disks of our models. We note that the use of a halo instead of an inner disk does not have an influence on the derived gap size. See \citet{2012Honda} for more discussion of the use of a halo in our disk modeling. An extended review about the inner regions of protoplanetary disks can be found in \citet{2010DullemondMonnier}. 

Table \ref{tab:modelfit} shows an overview of the basic parameters used in our best-fit models. The wavelength range between $\sim$5--14 $\upmu$m (including the N-band continuum and PAH emission) is studied in Section \ref{sec:modelN}.

\begin{figure*}[htbp]
   \centering
   \includegraphics[width=0.24\textwidth]{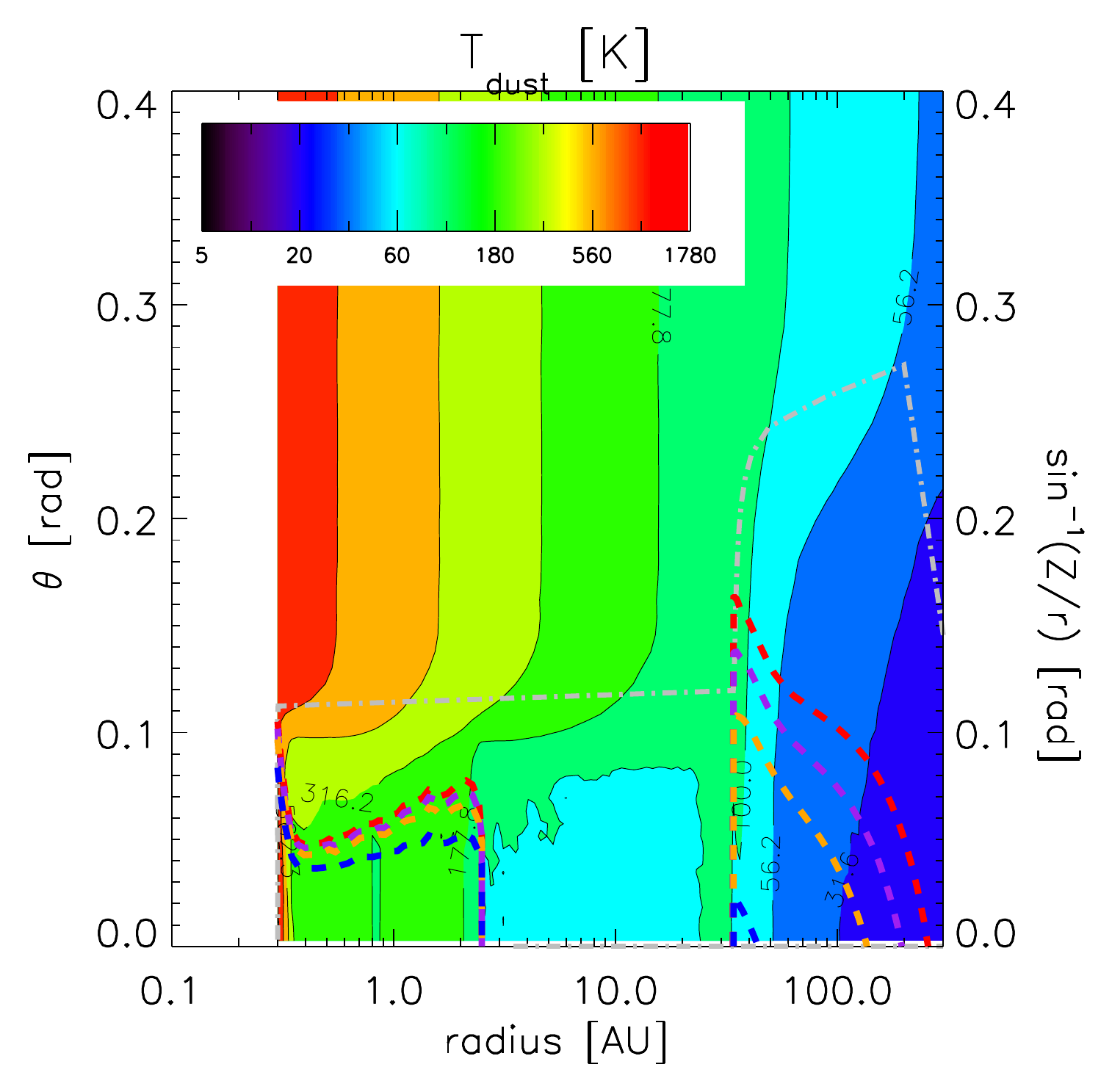}
   \includegraphics[width=0.24\textwidth]{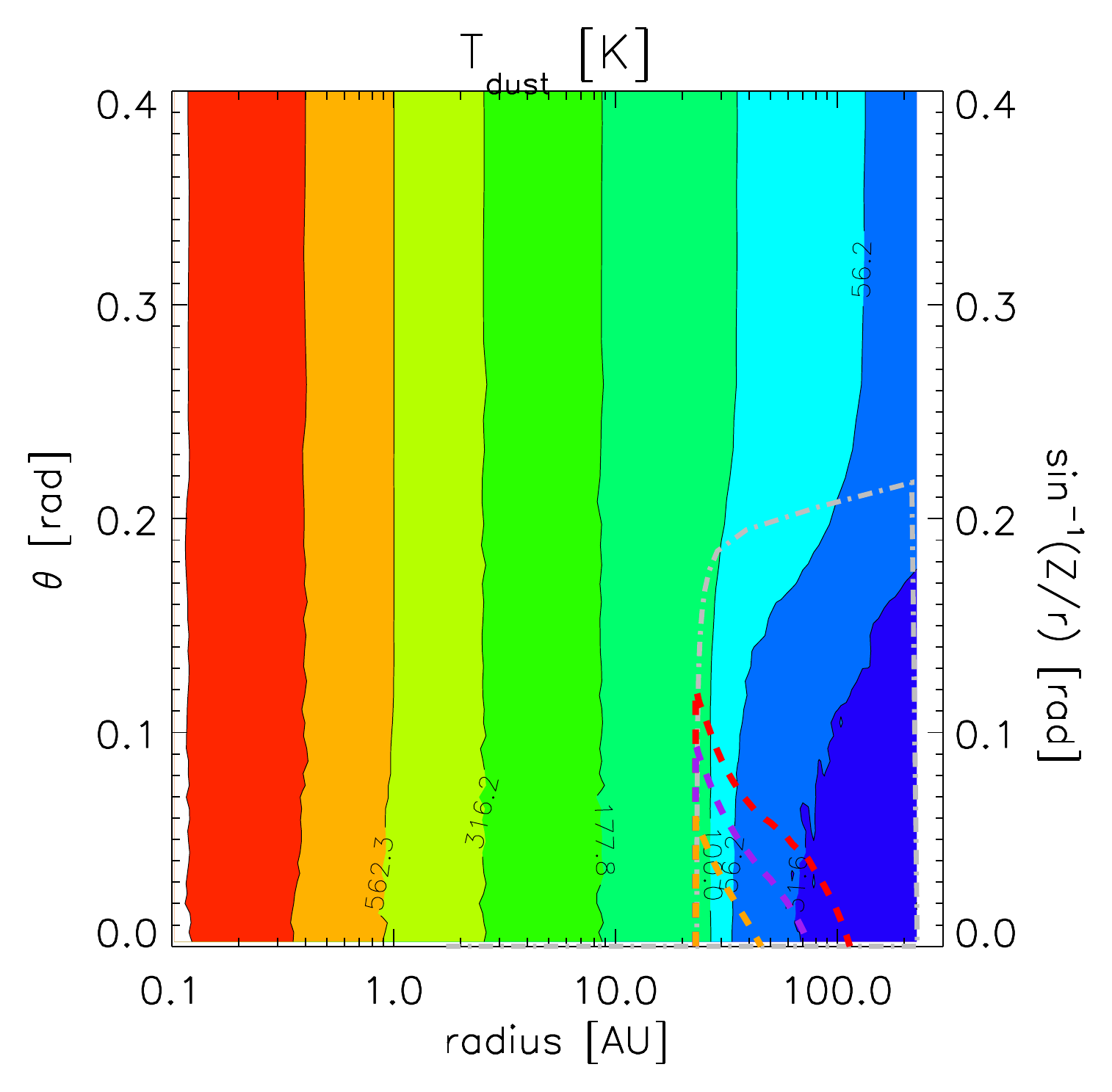}
   \includegraphics[width=0.24\textwidth]{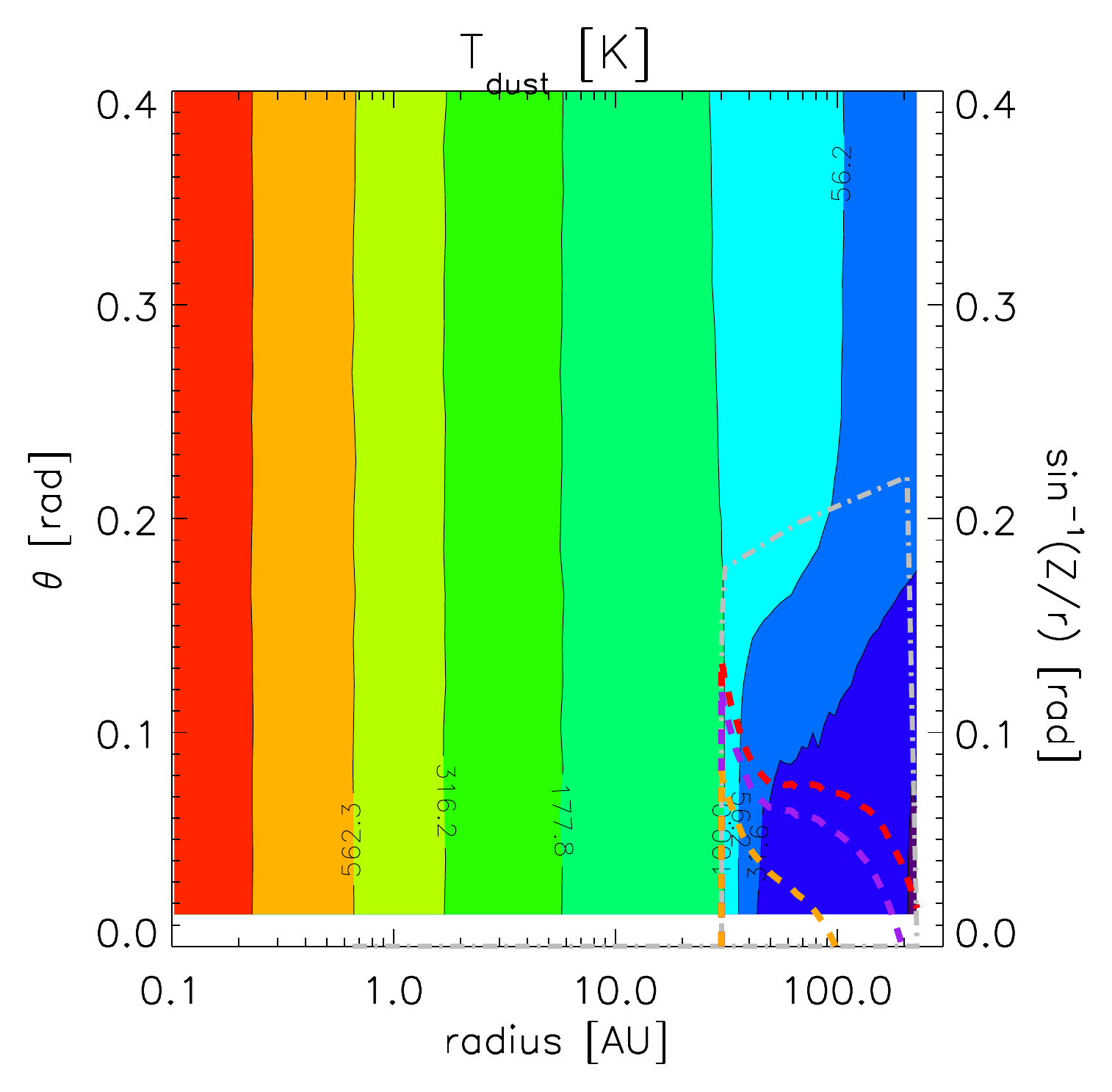}
   \includegraphics[width=0.24\textwidth]{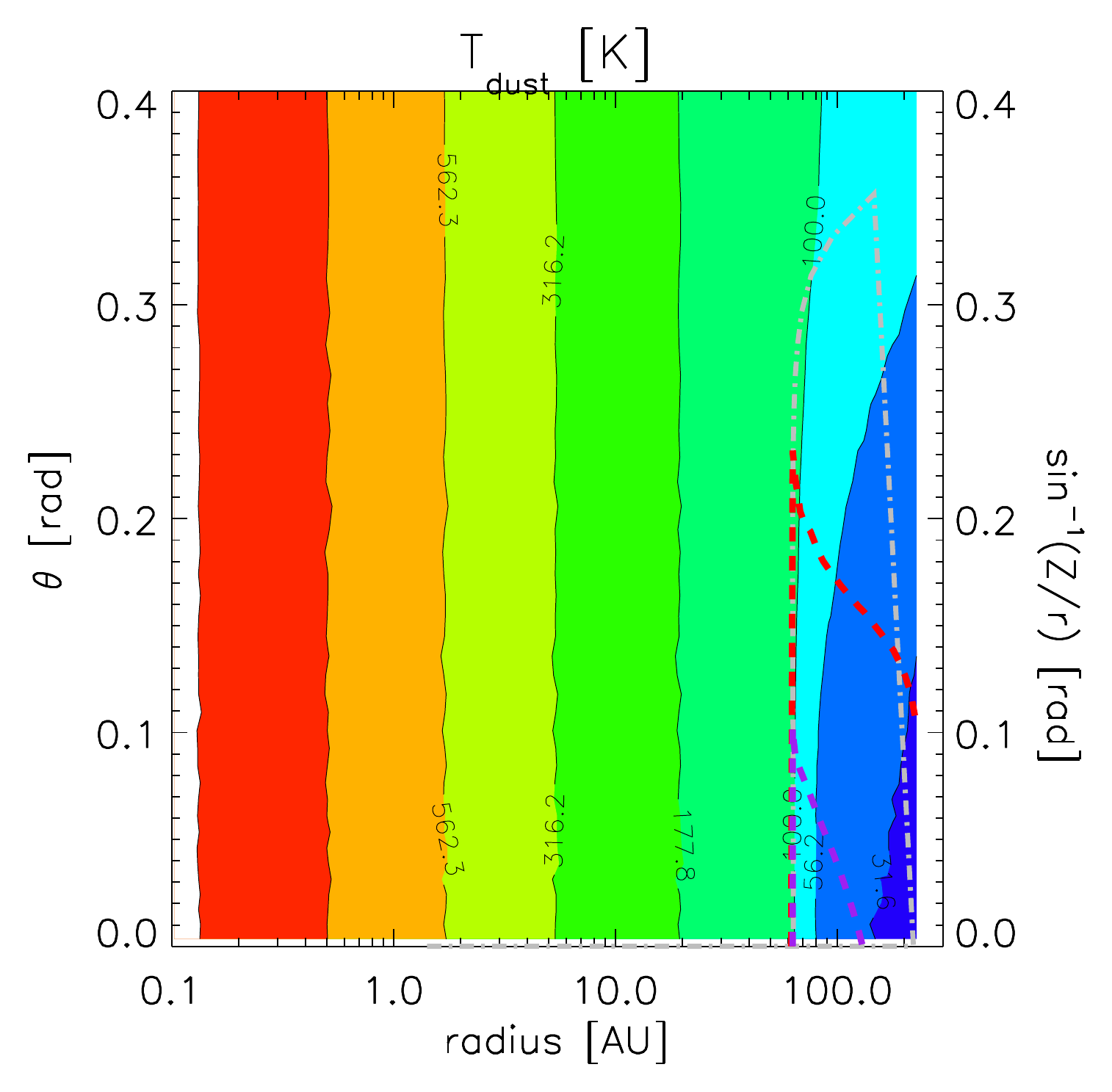}
   \includegraphics[width=0.24\textwidth]{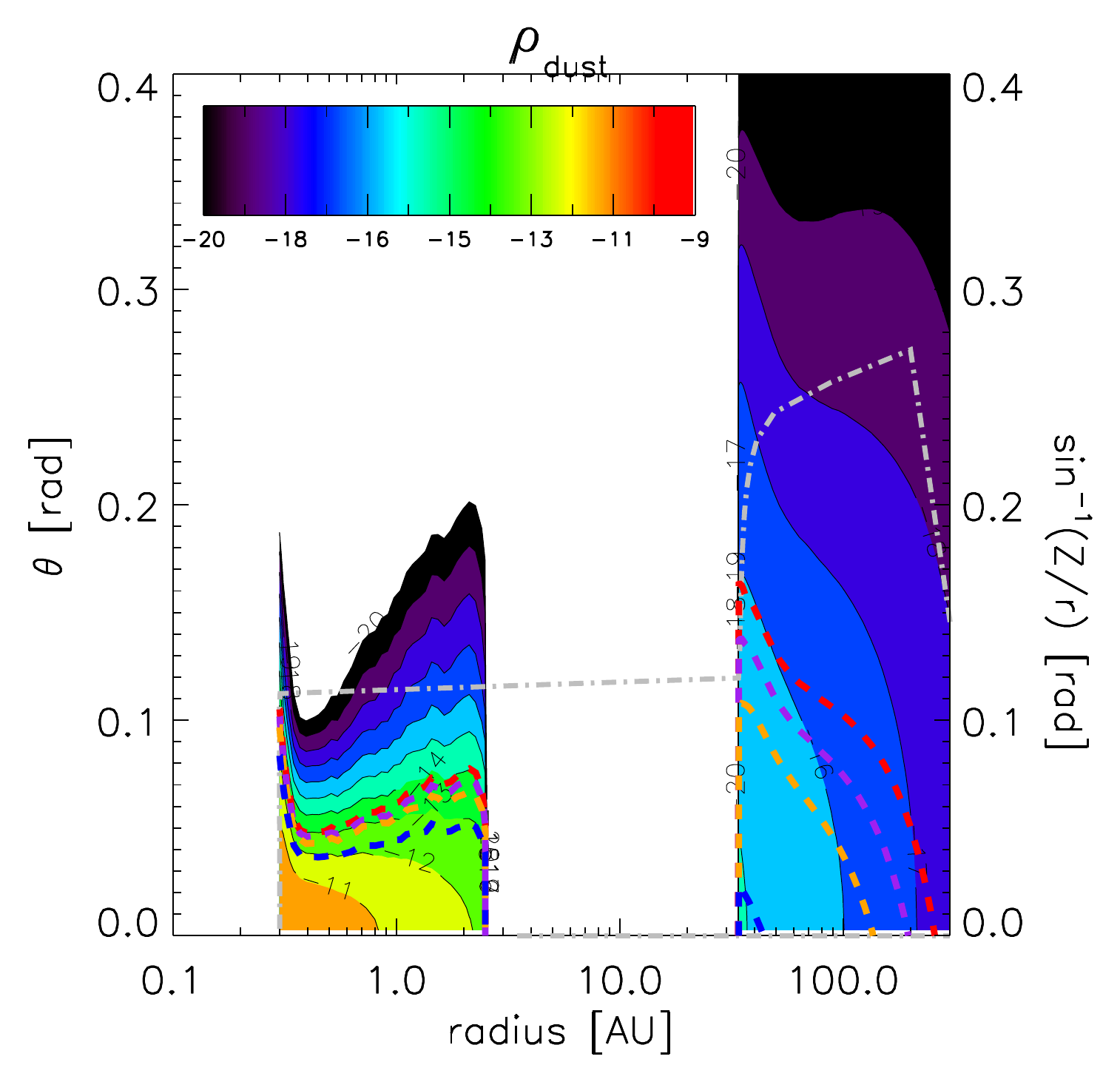}
   \includegraphics[width=0.24\textwidth]{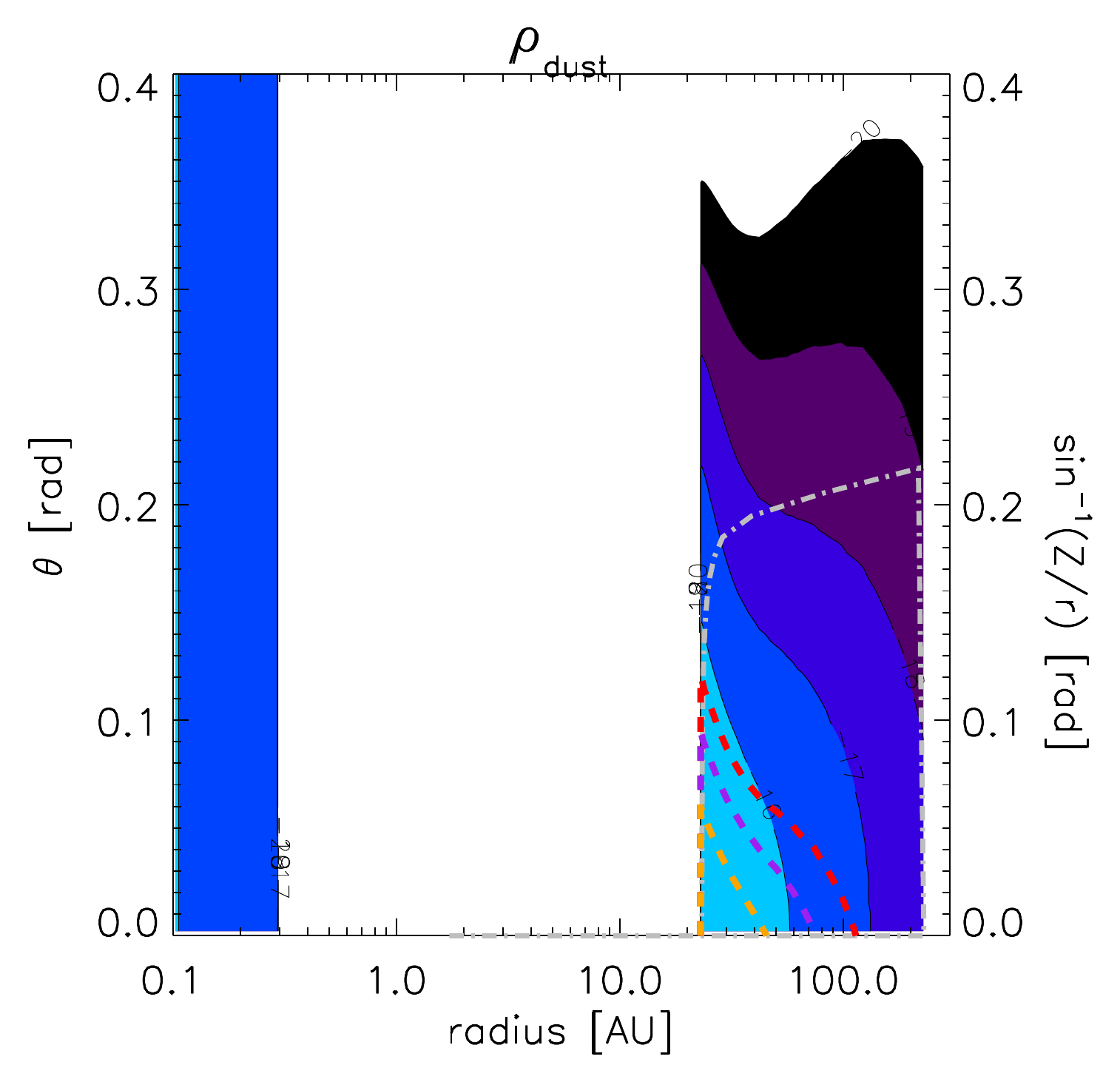}
   \includegraphics[width=0.24\textwidth]{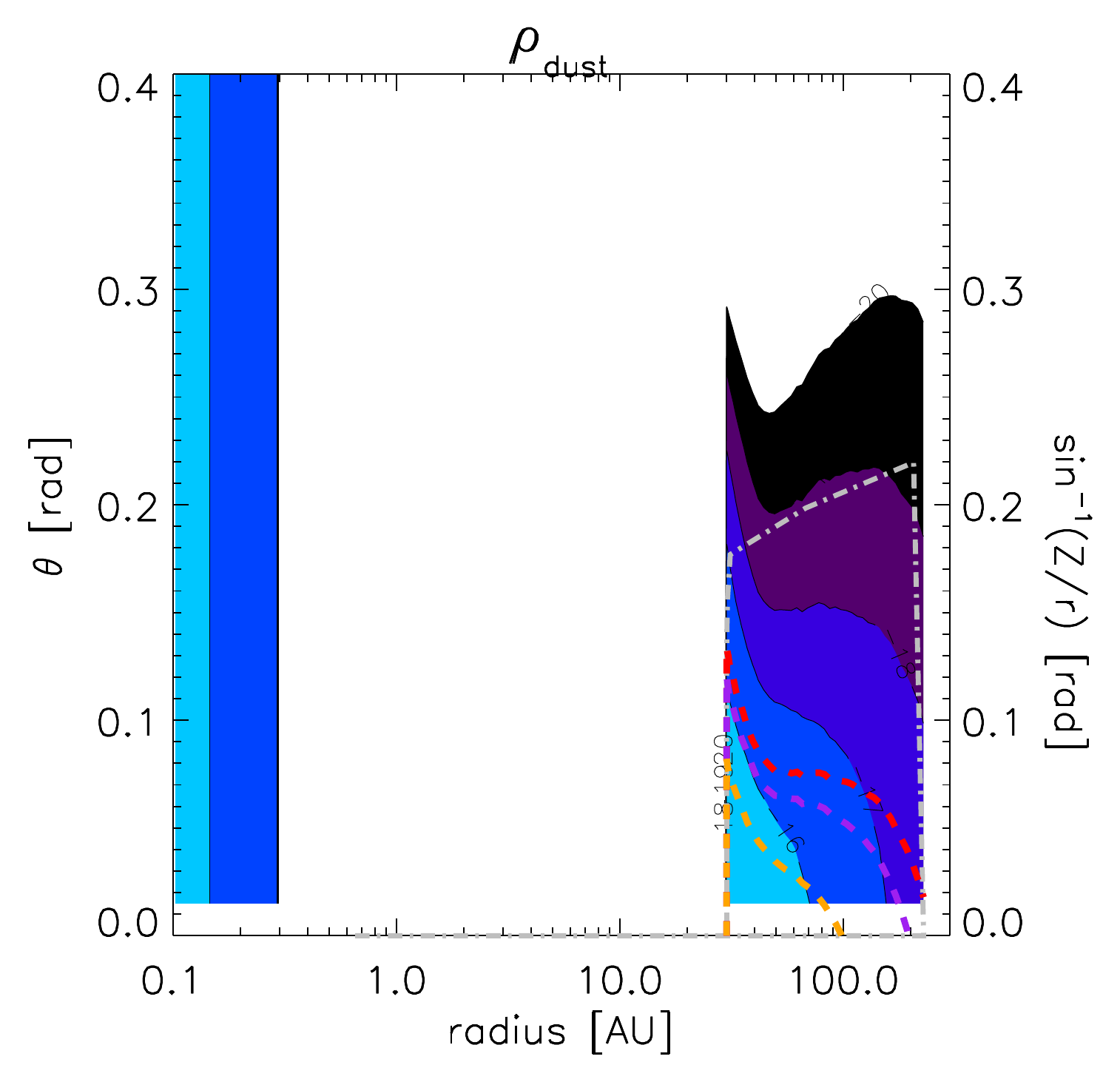}
   \includegraphics[width=0.24\textwidth]{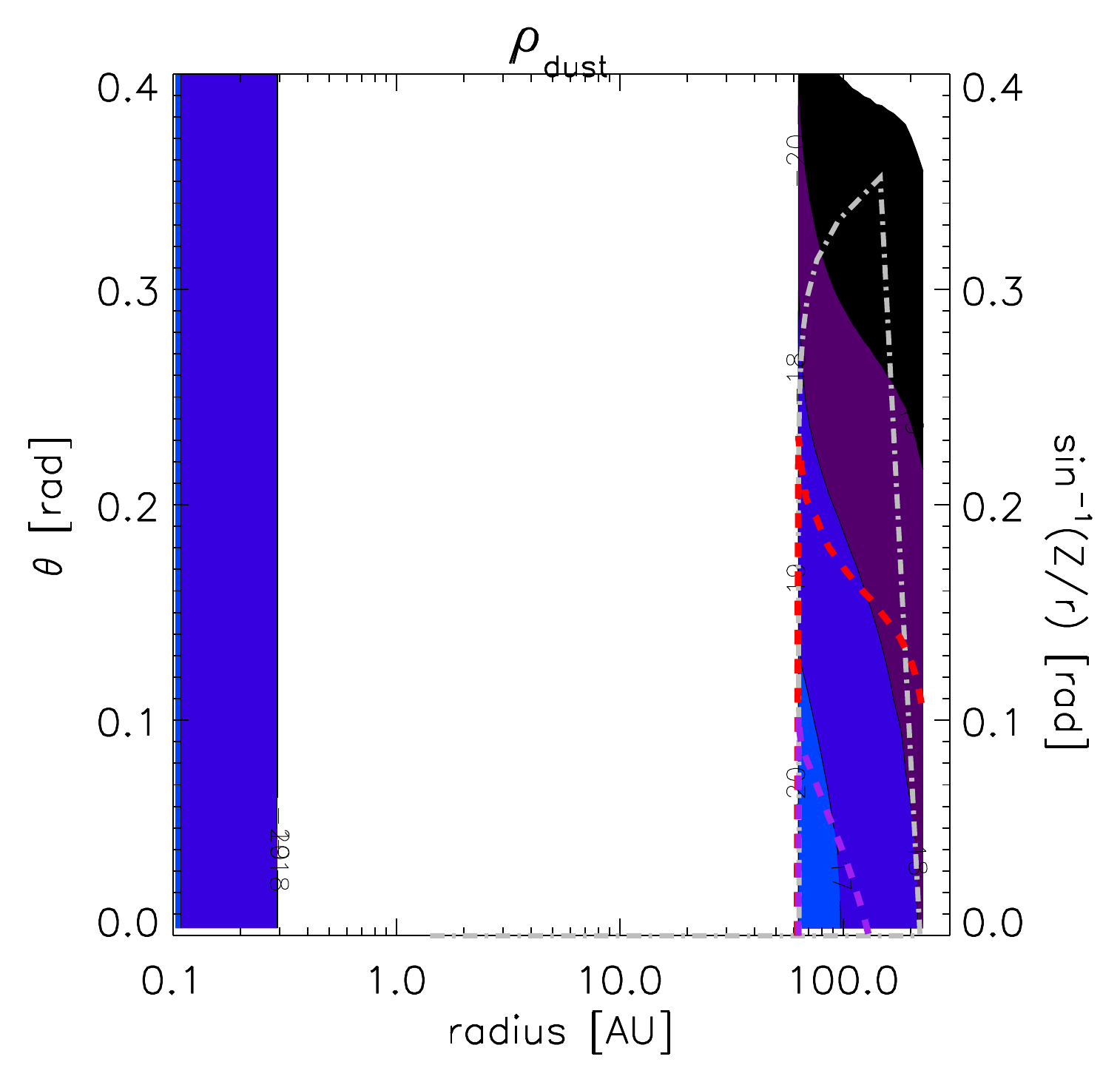}

   \caption{Top: radial temperature (K). Bottom: density (g cm$^3$) structures of the disks. From left to right: HD\,97048, HD\,169142, HD\,135344\,B, and Oph IRS 48.  The dashed lines represent the $\tau$ = 1 surfaces as seen perpendicular from the disk. From top to bottom, the colors are red, purple, orange, and blue and represent the wavelengths 1, 20, 70, and 1000 $\upmu$m. The gray dot-dashed line shows the $\tau$ = 1 surface in radial direction in scattered light ($\lambda$ = 0.5 $\upmu$m). Note that HD\,97048 has a thick inner disk, and therefore the radial $\tau$ = 1 surface starts at the rim of the inner disk.}
   \label{fig:temp}
      \label{fig:temp}
\end{figure*}

\begin{figure*}[htbp]
  \centering
  \includegraphics[width=0.5\textwidth]{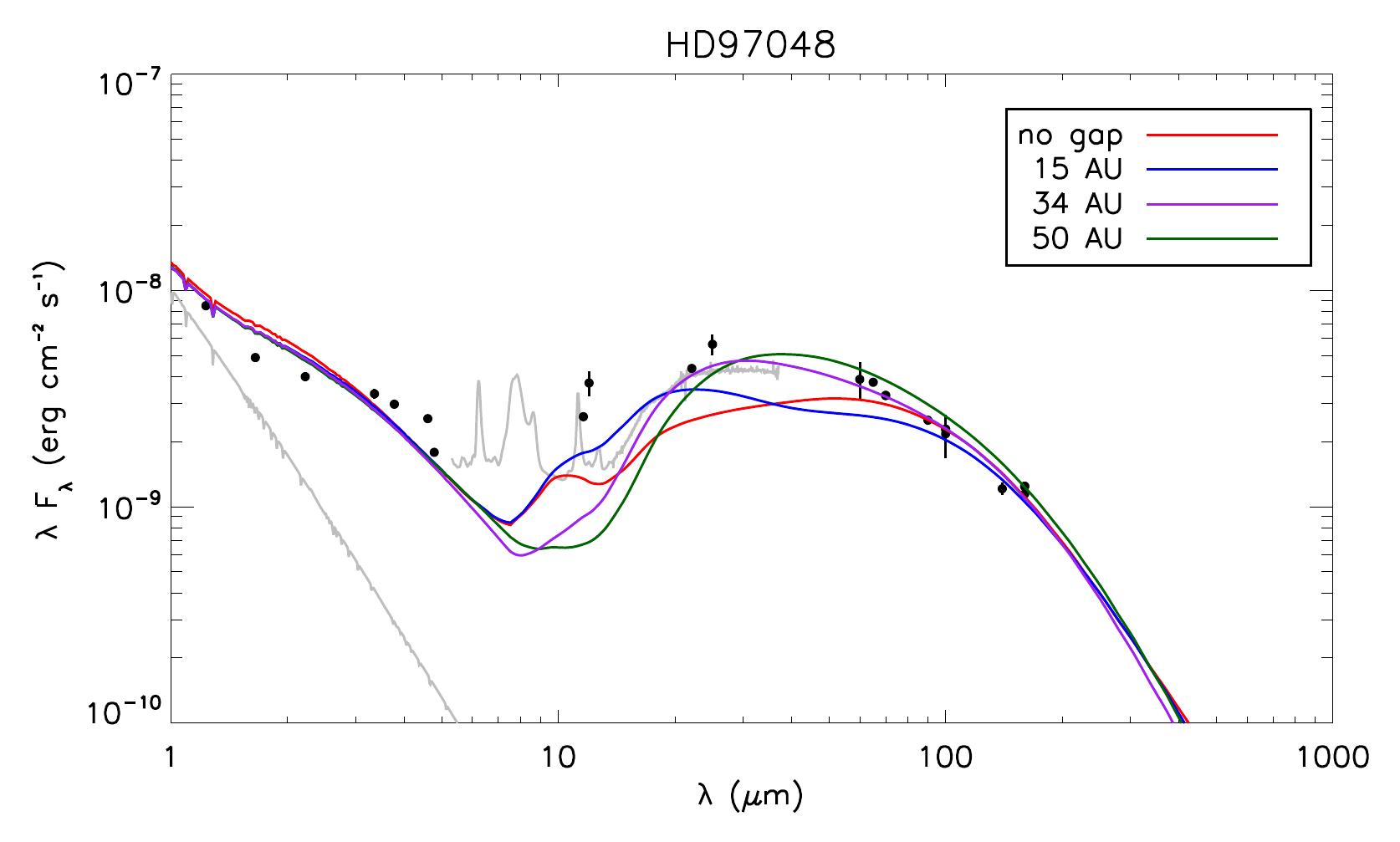} 
  \includegraphics[width=0.33\textwidth]{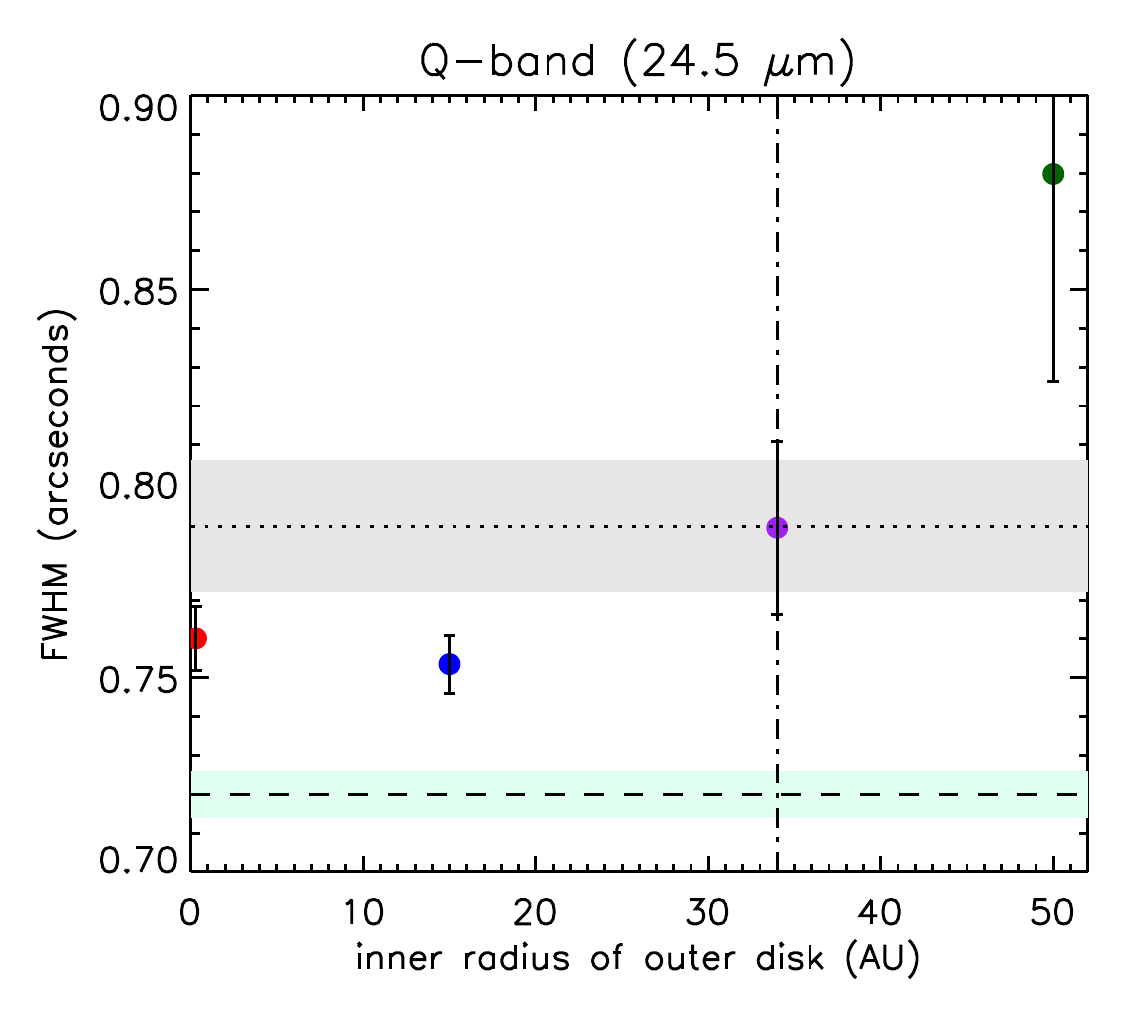} 
  \includegraphics[width=0.5\textwidth]{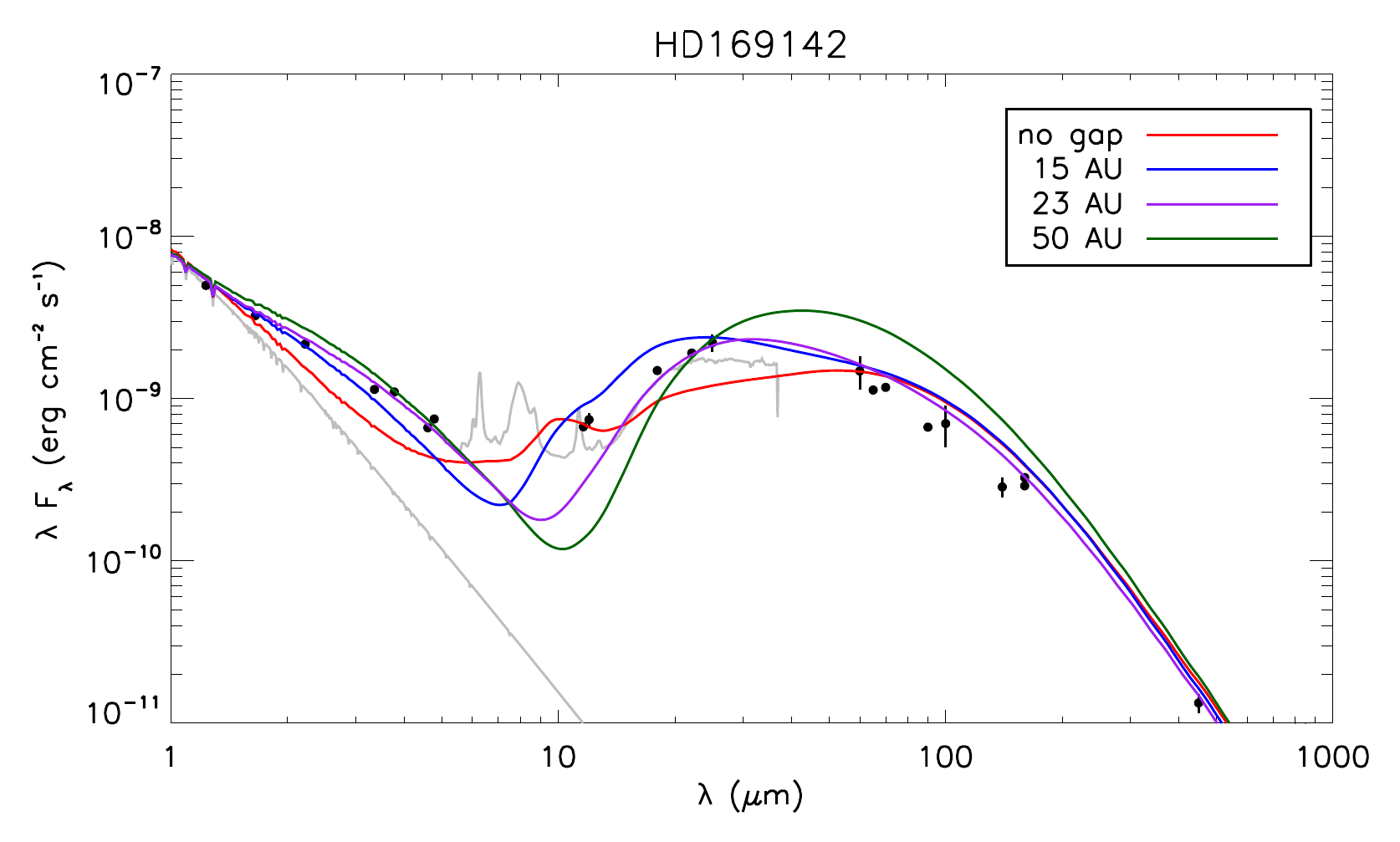} 
  \includegraphics[width=0.33\textwidth]{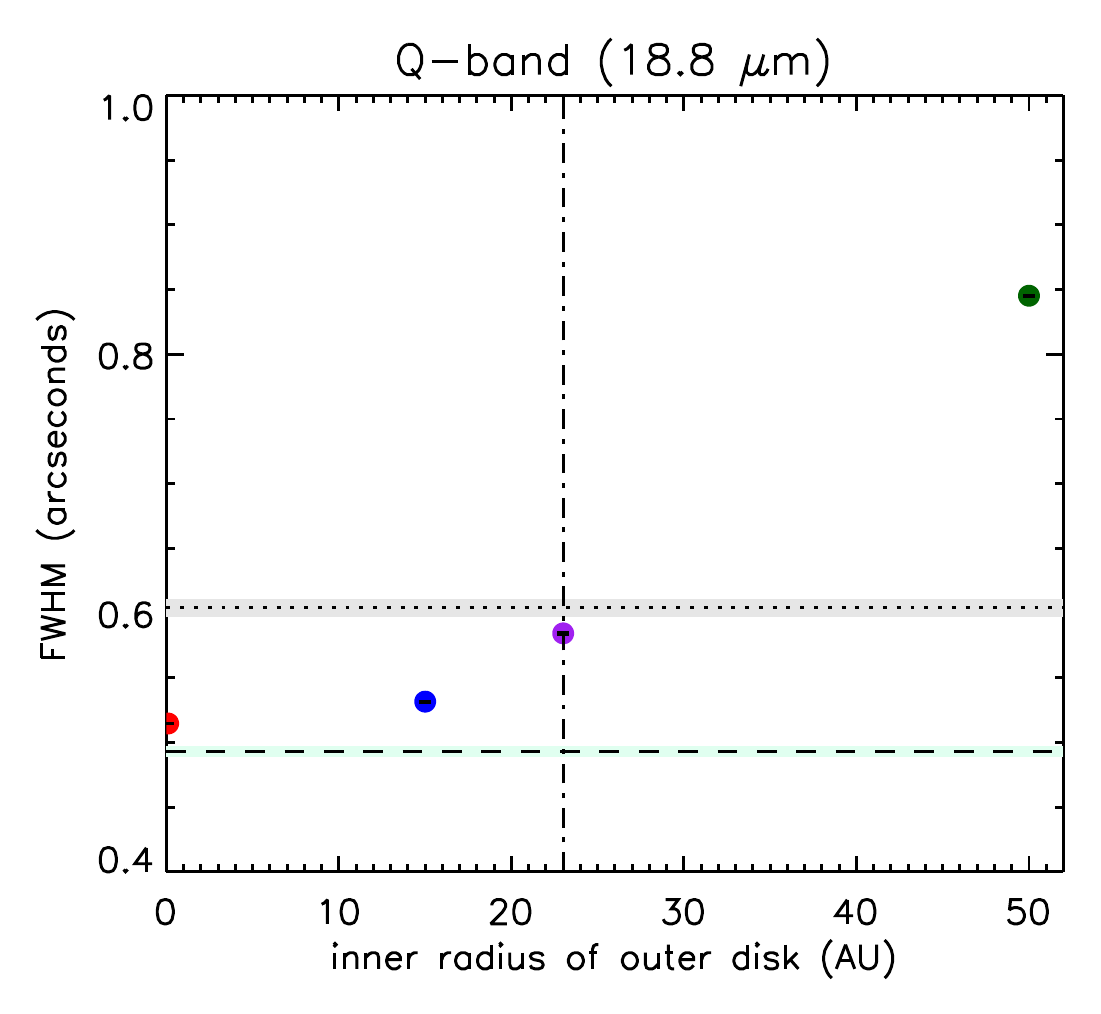} 
  \includegraphics[width=0.5\textwidth]{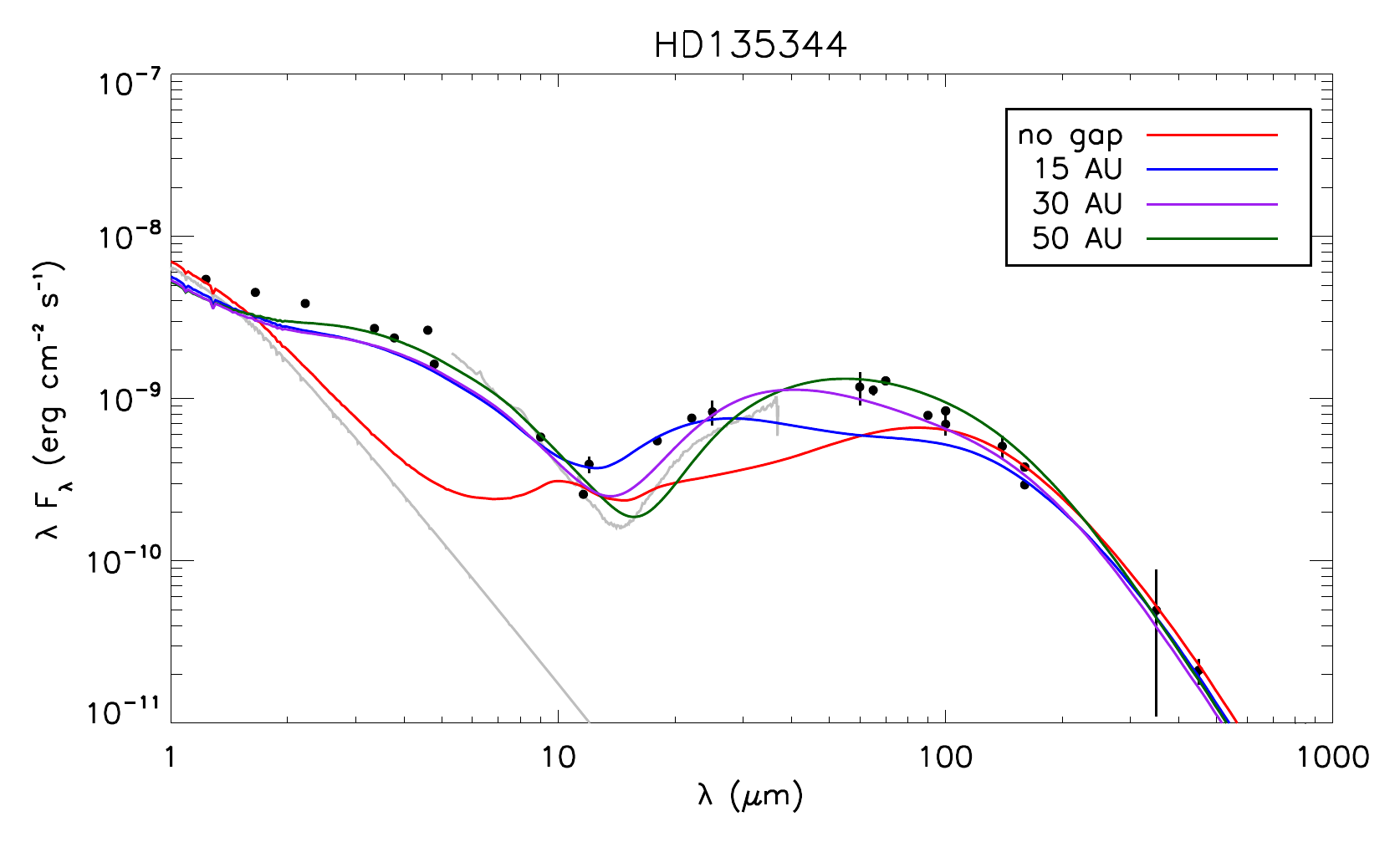} 
  \includegraphics[width=0.33\textwidth]{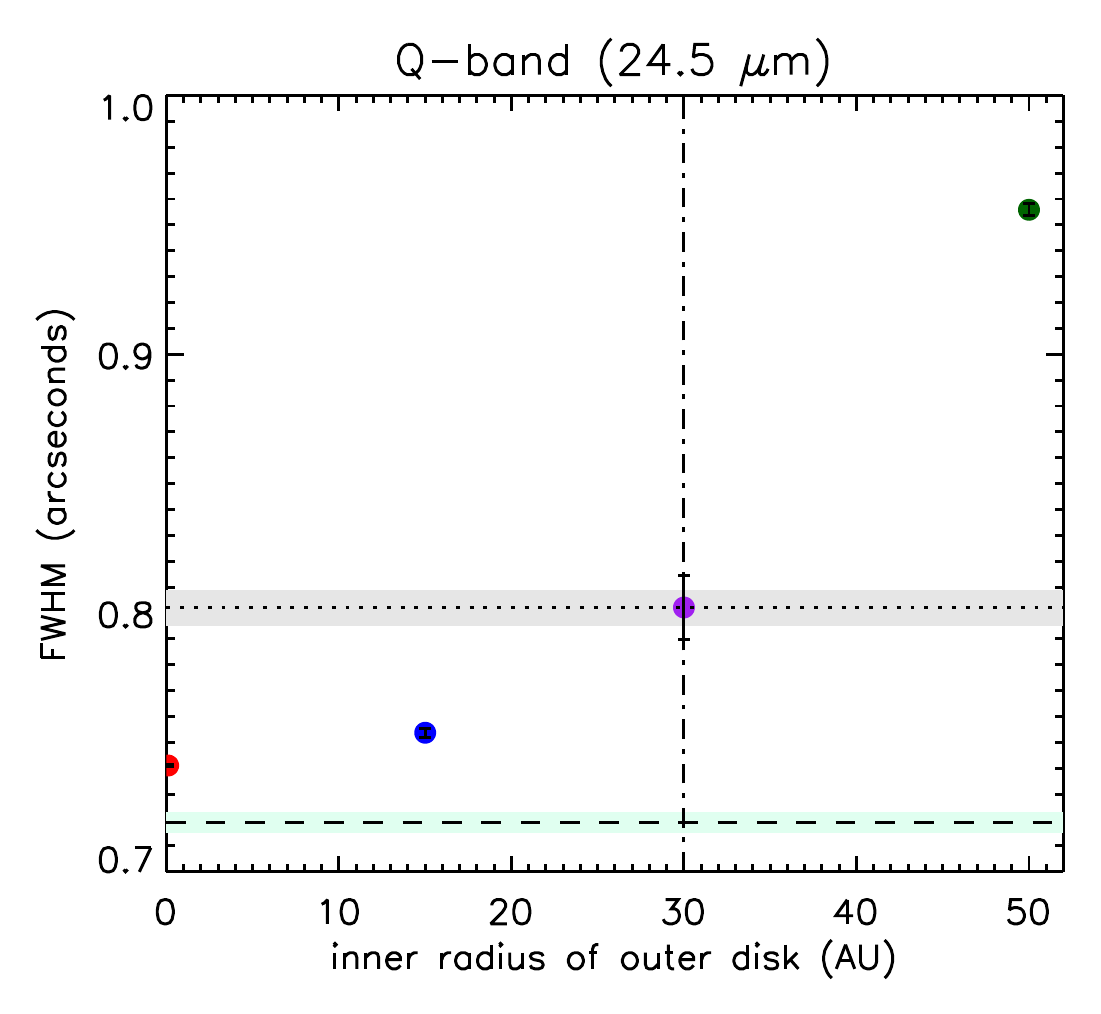} 
  \includegraphics[width=0.5\textwidth]{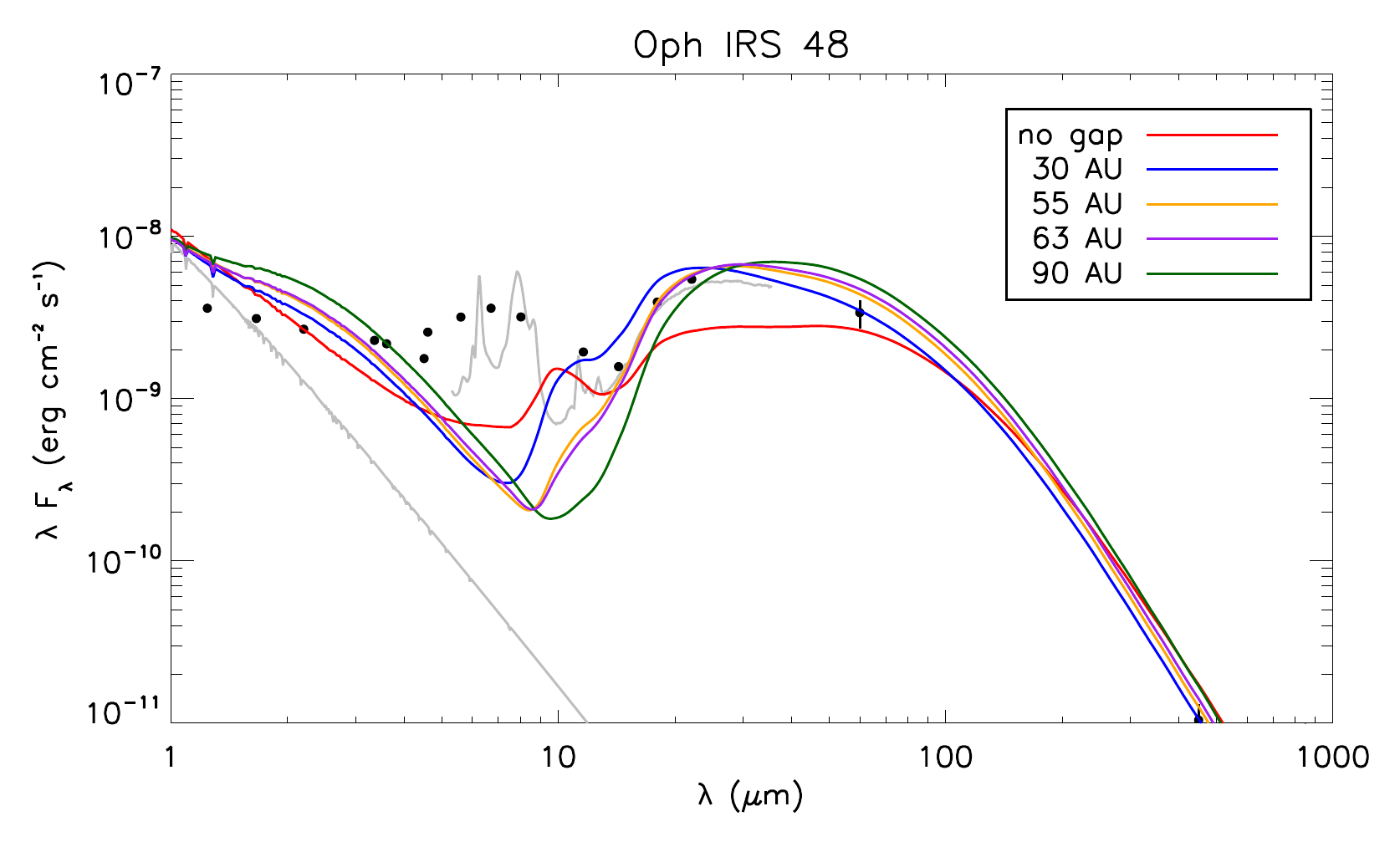} 
  \includegraphics[width=0.33\textwidth]{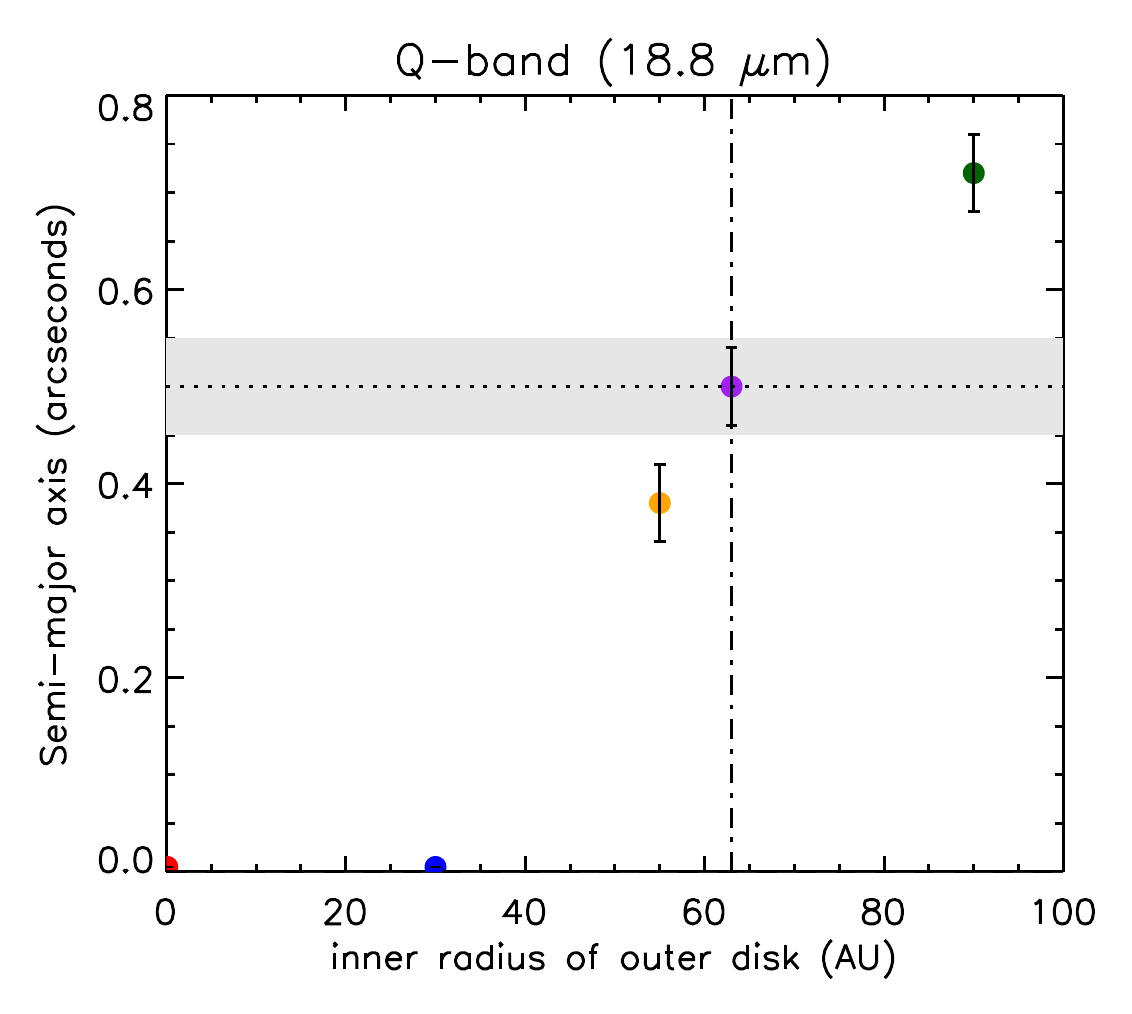} 
  
  \caption{Parameter study showing the SED and Q-band FWHM as a function of the inner radius of the outer disk. From top to bottom: HD\,97048, HD\,169142, HD\,135344\,B and Oph IRS 48. Left: the SEDs of models with a different location of the inner edge of the outer disk. Right: the FWHM sizes of these four models as a function of the location of the outer disk wall, given for the Q-band at 18.8 $\upmu$m (HD\,169142 and Oph IRS 48) and 24.5 $\upmu$m (HD\,97048 and HD\,135344\,B). The horizontal dotted and dashed lines give the FWHM sizes of the observations and calibration stars (PSFs) respectively. The $\pm 1  \upsigma$ variation in the FWHM as a function of the position angle is given by the grey and light cyan boxes (observation and PSF) and the error bars of the models. The colors on the left and right figures correspond to similar models. The vertical dashed-dotted line shows the chosen parameter of our best-fit model. We note that for Oph IRS 48 the semi-major axis of the MIR image is fitted, but not the FWHM. }
  \label{fig:endgap}
\end{figure*}

\subsection{Results of the Model Fitting}

Below, we review the fitting procedure for each source individually and discuss their uncertainties. The decomposed SEDs and Q-band fits are shown in Figure \ref{fig:SEDQ}, and the radial temperature and density structure are shown in Figure \ref{fig:temp}. The star and disk parameters for the best-fitting models are given in Table \ref{tab:modelfit}. 

\subsubsection{Best-fit model HD\,97048}
We find the best-fitting model with the radius of the inner edge of the outer disk set at 34$^{+4}_{-4}$ AU. This is the first time that a disk gap has been reported for HD\,97048.  
An optically thick inner disk with inner and outer radii of 0.3-2.5 AU fits the NIR flux in a natural way so no halo is required. 

We can compare our best-fit model to observations at other wavelengths. Polarimetric differential images in the H and K band by \citet{2012Quanz} show emission from $\sim$0.1''--1.0'' ($\sim$16 up to 160 AU). We have computed theoretical images of the disk of HD\,97048 in polarized scattered light using the disk parameters derived from the fitting of the SED and the Q-band image. We find that the convolved polarized scattered light image of our best-fit model does not match the observed brightness profile. Up to $\sim$ 0.175'', our models predict a reduced brightness due to the gap. At 34 AU ($\sim$0.22''), our model shows a bump due to the higher surface area of the inner edge of the outer disk. From 55 AU ($\sim$0.35''), the surface brightness of the outer disk starts to dominate and decreases further as a constant power-law. 
A possible solution to this discrepancy between the observed scattered light images and our predictions may be that the gap is not entirely empty but filled with small ($<$1$\upmu$m) particles that scatter efficiently. This solution has been suggested to explain the missing cavities in the polarized scattered light images of transitional disks, while SMA observations detected large gaps \citep{2013Grady, 2012Dong}. 



\subsubsection{Best-fit model  HD\,169142}
The averaged size of the Subaru/COMICS 18.8 and 24.5 $\upmu$m images are best fitted with the wall of the outer disk set at 23$^{+3}_{-5}$ AU \citep{2012Honda}. We note that this radius is a compromise to the fit to both wavelengths since it predicts a slightly smaller size than observed at 18.8$\upmu$m and slightly overpredicts the size at 24.5$\upmu$m. The discrepancy between the 18.8 and 24.5 $\upmu$m observations has been considered in the derived error.

An optically thick inner disk does not produce enough flux to fit the NIR flux in the SED; therefore, we added a halo close to the star. There are currently no other observations in the literature with sufficient spatial resolution to test our results. 

\subsubsection{Best-fit model  HD135344B}
The radial brightness profile is best fitted with the inner edge of the outer disk set at 30$^{+5}_{-3} $ AU. The Spitzer spectrum clearly shows two different temperature components. This shape is consistent with the NIR excess originating from a remarkably strong component of  T$\sim$1500 K. Since the SED between $\sim$ 5-10 $\upmu$m follows a Rayleight-Jeans slope, there is no strong contribution in the NIR from lower temperature ($<$1000 K) dust. 

Subaru/HiCIAO images by \citet{2012Muto} shows scattered light emission as close as 0.2'' ($\sim$28 AU) from the star while SMA sub-mm imaging by \citet{2009Brown} showed a gap size of $\sim$39 AU. Our Q-band size is consistent with the HiCIAO disk size. The origin and significance of the differences in derived radii are still unclear. However, a model solution is easily found, since the Q-band is dominated by micron-sized grains and the sub-mm by large ($>$100 $\upmu$m) particles. Removing the millimeter-sized grains in the disk below 39 AU does not affect the fit in the Q-band or the SED.

\subsubsection{Best-fit model Oph IRS 48}
Due to a larger gap size relative to the other stars in the sample, the observed Q-band image of Oph IRS 48 shows a clear extended ring of emission peaking at $\sim$0.55" \citep{2007aGeers} from the center. The radius from the center to the peak of this ring represents the location of the inner edge of the outer disk from the star. We have fitted the inner edge of the outer disk of our model to the radius of the peak position of the observed ring (the semi-major axis). The images are best fitted with the inner edge of the outer disk at 63$^{+4}_{-4} $ AU. We do not attempt to fit the asymmetry in brightness which could be caused by e.g. Rossby wave instabilities \citep{2001Li} causing axial asymmetries and differences in grain sizes in the disk. The calibration star taken during the observation (the PSF) is of low quality. Therefore we have fitted a Gaussian to the calibration star and used this to convolve the theoretical model image. 
 
The star has a high extinction (Av$\approx$11-13) and no photometry is available in the optical, therefore de-reddening introduces a large uncertainty in the luminosity of the star \citep{2012aBrown, 2010McClure}. Since the luminosity directly affects the temperature distribution of the outer disk, this translates to an uncertainty in fitting the SED. Modeling shows, however, that the relative brightness distribution at 18.8 $\upmu$m is not influenced by a factor of 2-3 higher or lower in stellar luminosity. So the derived wall radius does not change. 

We note that the observed radial brightness profile (Figure \ref{fig:SEDQ}) is significantly brighter than our model at a radius of $\sim$0.6-0.8'' from the star. This could be explained by a larger PSF during the observations, a higher stellar luminosity, differences in grain properties as a function of radius or by a complex disk asymmetric structure of the outer disk. 

The outer radius of the disk is unknown, so we choose an outer radius of 235 AU similar to HD\,169142. The modeling is not sensitive to this choice. SMA observations at 880$\upmu$m of \citet{2012bBrown} indicate a hole up to $\sim$13 AU. SMA and new ALMA 690 GHz observations (van der Marel et al. in prep) of this source at sub mm wavelengths are in conflict with each other. While SMA images indicate a hole at 13 AU, there is no indication of this in recent ALMA observations. ALMA reveals a highly asymmetric dust structure, indicating that the SMA image is actually misaligned due to its vertically elongated beam and thus interpreted differently. 

\subsection{Short summary of Q-band analysis}

We have applied a similar modeling approach to all four sources and derived the outer radius of the disk gap from the Q-band imaging. The decomposed SEDs (Figure \ref{fig:SEDQ}) show that the wall component (the first 3 AU of the outer disk) dominates the emission at Q-band wavelengths. Therefore, the size of the resolved Q-band images are a sensitive diagnostics for the radius where the emission is emitted. For transition disks with a typical distance of $\sim$150 pc and observed with an 8 meter class telescope, resolved Q-band images can be a good tracer of the gap size if the dust in the wall reaches temperatures of $\sim$100--150 K. 

To study the robustness of our results we have performed a parameter study. We have taken the best-fit models and studied the effect of variations in the following parameters: the location of the inner edge of the outer disk, the size of the inner disk, the surface density power-law, the stellar luminosity, the distance, the minimum grain size and the silicate-to-carbon ratio. We discuss the influence of VSG in Section \ref{sec:modelN}. We find that the angular size of the Q-band image is only sensitive to the angular size of the wall radius, where, of course, the absolute radius of the wall depends directly on the adopted distance. Figure \ref{fig:endgap} quantifies the results of the parameter study on the location where the outer disk starts. The models show that the FWHM of the Q-band generally increases as the inner wall of the outer disk moves outward. The distance uncertainty results in the dominant error on the fit to the radius of the wall, which is typically in the order of a few AU.

\section{The disk size in the N-band}

\label{sec:modelN}
\subsection{The origin of emission in the N-band}
The spatial resolution of a diffraction-limited observation in the N-band is a factor of two higher than in the Q-band. However, while the flux in the Q-band comes predominantly from micron-sized grains in the wall of the outer disk, the origin of flux in the N-band is more difficult to constrain, since it is a composite of a) thermal dust from the inner and outer disk,  b) stochastically heated very small carbonaceous grains (VSGs), and c) polycyclic aromatic hydrocarbons (PAHs). In this section we demonstrate that the N-band cannot be used as a reliable tracer for the gap size.

We use VLT/VISIR data (see section \ref{sec:VISIR}) to study the N-band sizes. Figure \ref{fig:FWHMprofiles} shows the FWHM profiles of the VISIR observations in our sample. From this profile, the radial extent of the 8.6 and 11.2 $\upmu$m PAH features compared to the continuum can be studied. We study the general trends of the N-band sizes caused by dust, VSGs, and PAHs and study the relation to the disk geometry.

It is important to emphasize that the plateau of continuum emission between the strong PAH features at 8.6 and 11.2 $\upmu$m is not from PAH emission. This is illustrated by the VLT/VISIR observations of the FWHM profile of HD\,97048 in the N-band (see Figure \ref{fig:FWHMprofiles}). The FWHM increases by $\sim$11 $\upmu$m, peaks at 11.2 $\upmu$m and decreases back to the continuum size at 11.6 $\upmu$m. If the plateau of continuum emission at 10.5 $\upmu$m would also originate in PAH molecules, then the FWHM size would not be different compared to the size at 11.2 $\upmu$m. 

Below, we first discuss two possible origins of continuum emission between the PAH features; thereafter, we compare their sizes with the disk size in the PAH 8.6 and 11.2 $\upmu$m filters.


\subsection{Thermal dust continuum}   
\label{sec:Ndustcontinuum}
\begin{figure*}[htbp]
  \centering
  \includegraphics[width=0.5\textwidth]{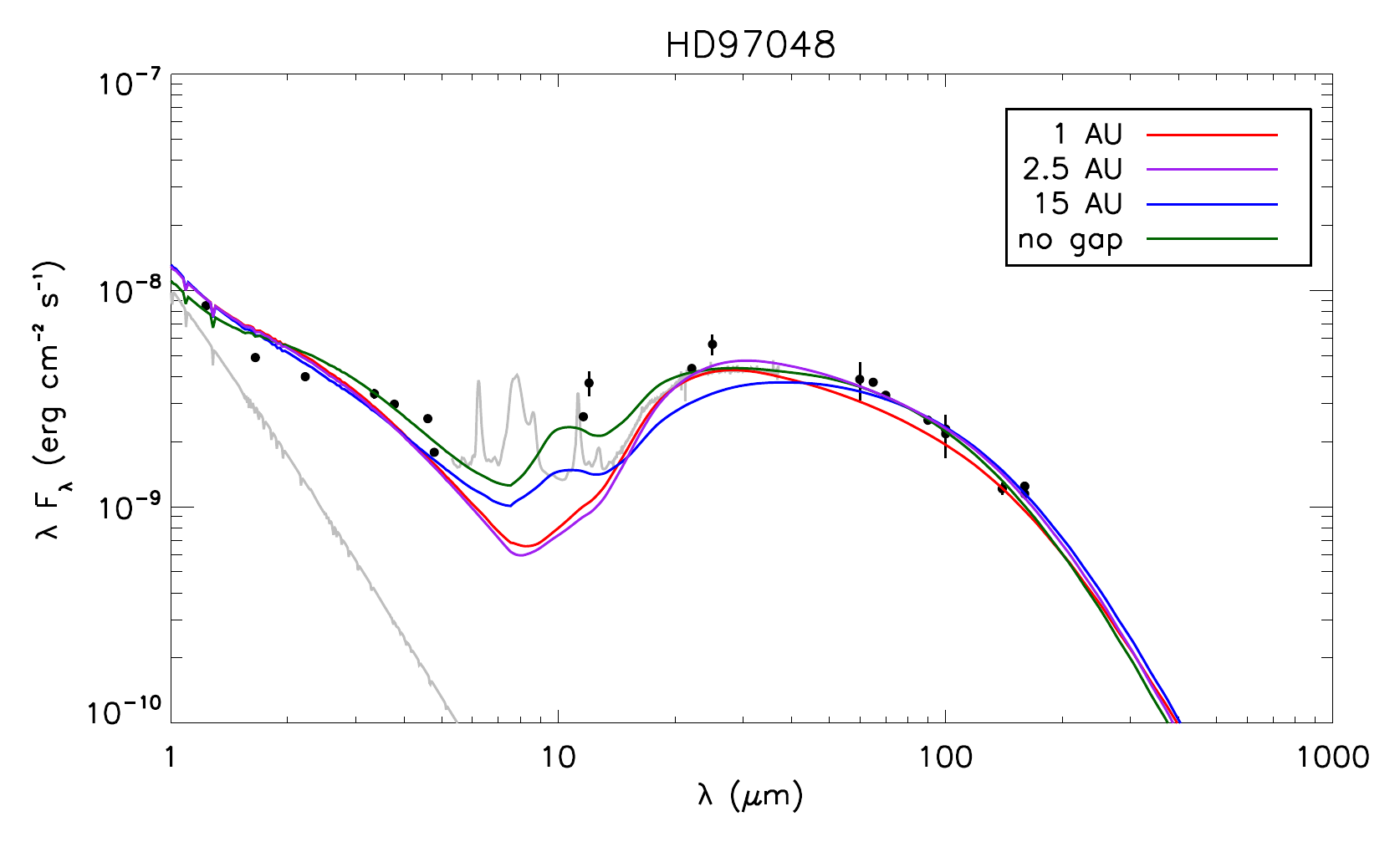} 
  \includegraphics[width=0.33\textwidth]{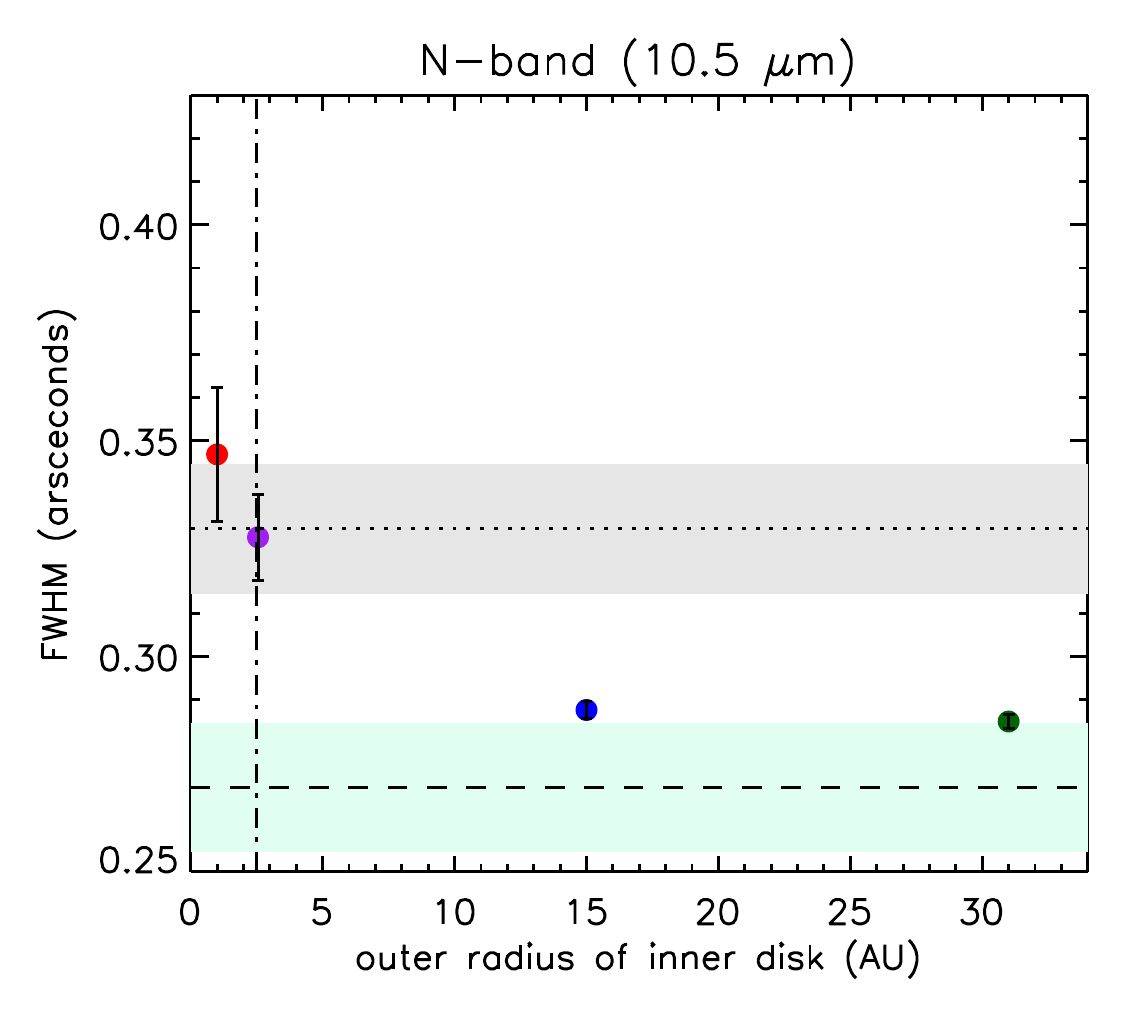} 

  \caption{Models of HD\,97048 with increasing size for the outer radius of the inner disk. The larger the inner disk, the smaller the N-band size. The grey and light cyan boxes and the error bars indicate the $\pm 1  \upsigma$ error on the FWHM of the azimuthally averaged radial brightness profiles for the target and PSF, respectively. Observed data of HD\,97048 is from \citet{2006Doucet}}
  \label{fig:startgapHD97048}
\end{figure*}

Thermal emission from dust grains with a temperature of $\sim$200--400 peaks in the N-band. Depending on the stellar temperature, luminosity, the disk structure and grain size, dust of this temperature is typically located in the inner $\lesssim$15 AU in the disk. In the best-fit models of the transitional disks of our sample, we have assumed that the gaps are depleted in dust. Continuum emission in the N-band, therefore, originates in the Rayleight-Jeans and Wiens tails of the inner and outer disk respectively. This is illustrated by the decomposed SEDs in Figure \ref{fig:SEDQ}, in which the separated disk components are shown. In this scenario, the FWHM size in the N-band is a measure for the relative contribution of the inner disk versus the outer disk. 

HD\,135344\,B is not resolved with respect to the PSF. In the decomposed SEDs in Figure \ref{fig:SEDQ} the flux at 10.5$\upmu$m for HD\,135344\,B is dominated by the halo. Since the halo is very close to the star (0.1--0.3 AU), the N-band is also not resolved in our best-fit model and is in agreement with the observations. 


The size of the continuum emission in the N-band is influenced by the radial size of the inner disk. In Figure \ref{fig:startgapHD97048}, we show a small parameter study for HD\,97048 in which the outer radius of the inner disk is increased. We compare the FWHM of our models with the \citep{2006Doucet} observations of HD\,97048 since they are of better quality and have a low seeing (their observation of the PSF is diffraction limited). If the contribution from the inner disk becomes larger in the N-band, the FWHM of the model image decreases. For small inner disks of $\lesssim$5 AU, this has no effect on the Q-band size. However, the N-band is nearly as small as the PSF and the Q-band also decreases (see Figure \ref{fig:endgap}) for models with a larger inner disk or no gap. We can fit the observed N-band size by choosing an inner disk from 0.3 to 2.5 AU, though an inner disk of this size does not fit the SED (see Figure \ref{fig:startgapHD97048}).

 \subsection{Continuum emission from Very Small Grains}
\label{sec:VSG}

\begin{figure*}[htbp]
  \centering
  \includegraphics[width=0.5\textwidth]{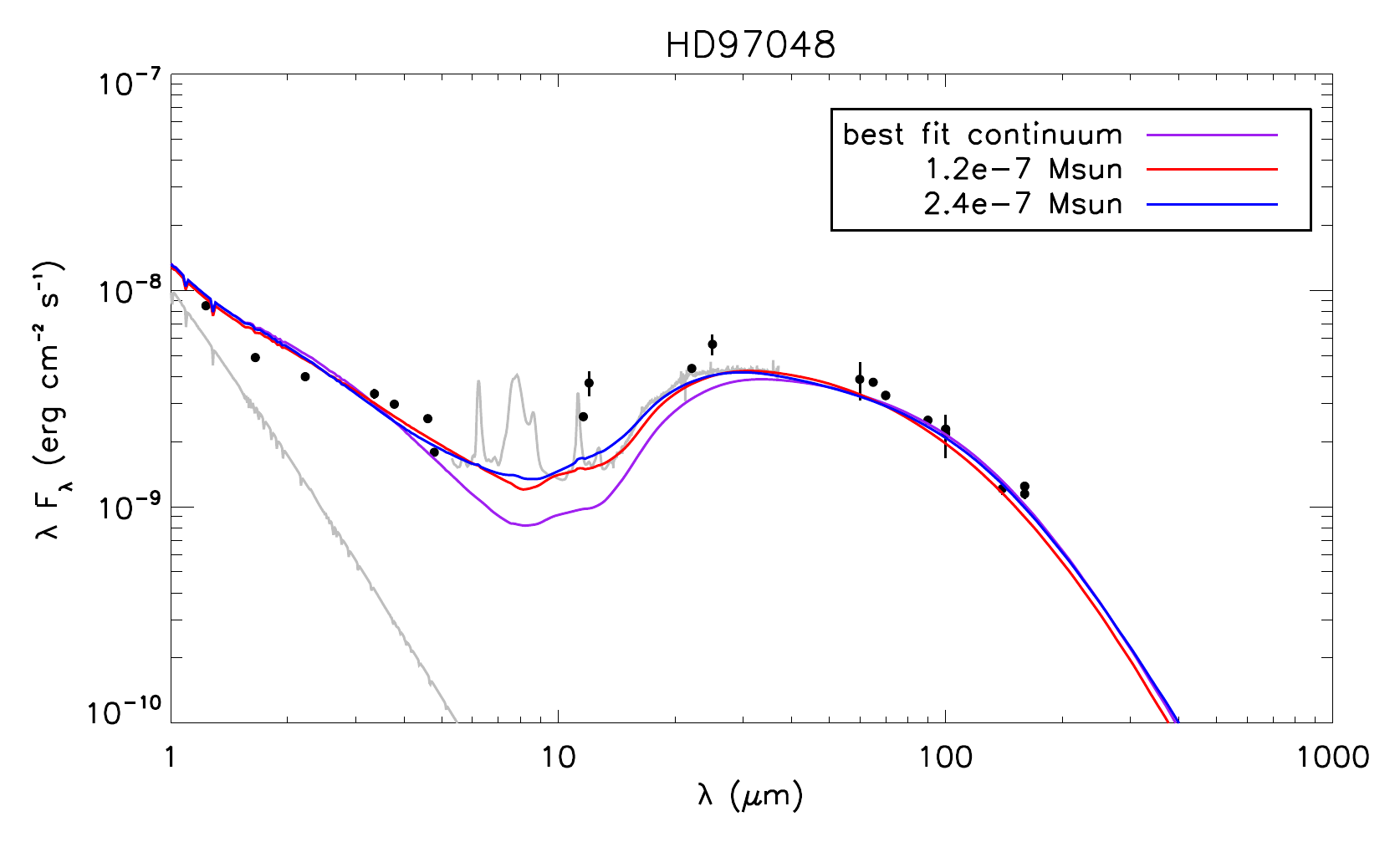} 
  \includegraphics[width=0.33\textwidth]{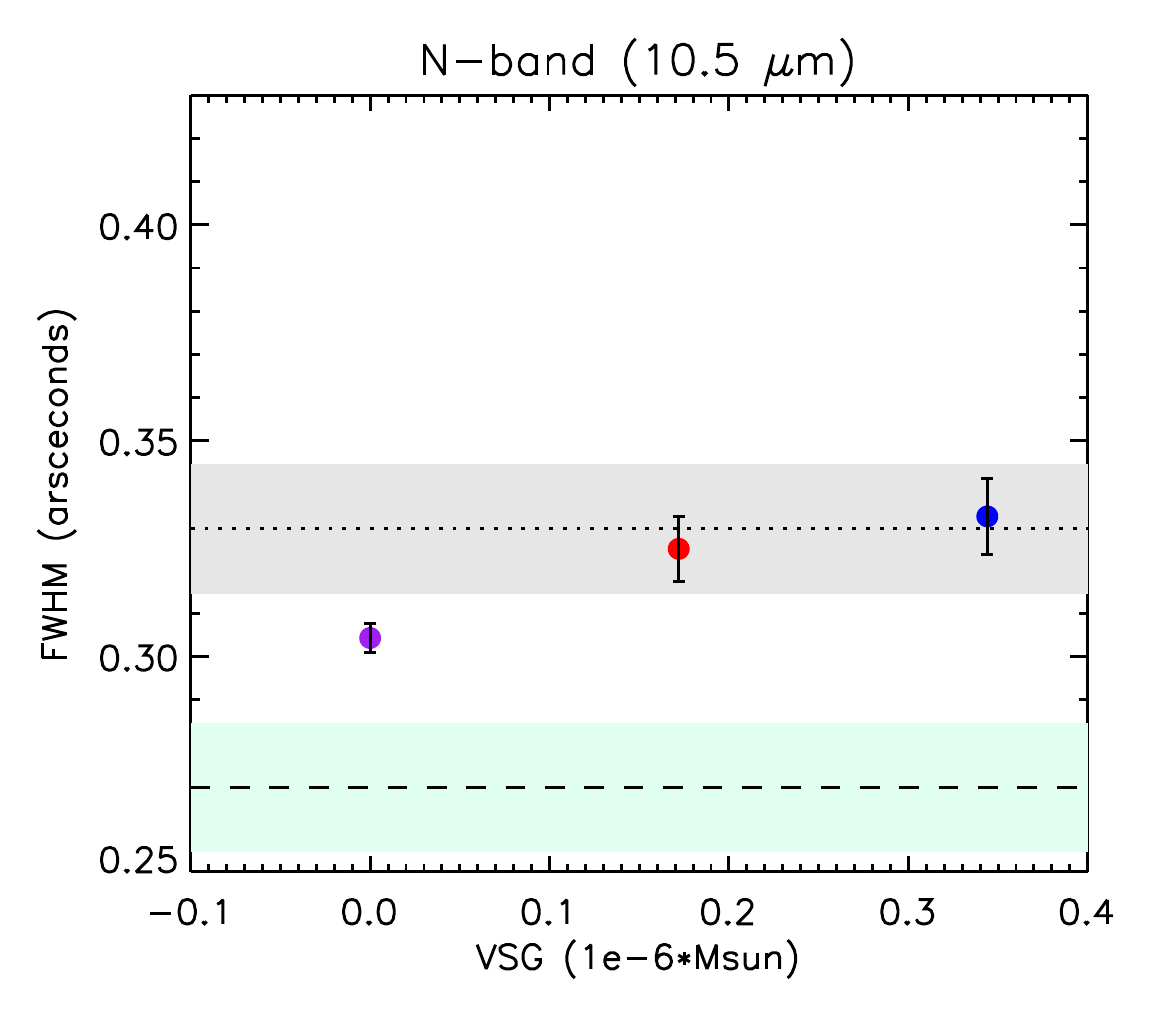} 

  \caption{Models of HD\,97048 with very small grains added to the disk. We note that a larger inner disk of 0.3-5 AU is chosen in this modeling grid to fit the observations. The higher the abundance of VSG, the larger the N-band size. The grey and light cyan boxes and the error bars indicate the $\pm 1  \upsigma$ error on the FWHM of the azimuthally averaged radial brightness profiles for the target and PSF, respectively. Observed data of HD\,97048 is from \citet{2006Doucet}.} 
  \label{fig:VSGHD169142}
    \label{fig:VSGHD97048}
\end{figure*}

Besides the inner disk, emission in the N-band can also originate in stochastically heated, very small carbonaceous grains that can be transiently warm far from the star. Similar to PAH molecules, VSG are excited by UV photons and re-emit at longer wavelengths. These very small grains are expected in the size range of $\sim$20--30 $\AA$. For comparison, recall that a typical 50 C-atom interstellar PAH molecule is $\sim$6 $\AA$, and the interstellar dust grains have sizes of $\sim$50--3000 $\AA$ \citep{2005Tielens}. For a disk with a high abundance of VSGs, the image size is influenced by the radial density distribution of VSGs in the disk and the number of UV photons that interact with the VSGs. Thus, the N-band size becomes larger for an increasing abundance of VSG in the outer regions of a protoplanetary disk. 

To demonstrate the effect of VSG on the disk size, we have performed a small parameter study using the best-fit model for HD\,97048 with a larger (0.3-5 AU) contribution from the inner disk and an abundance of VSGs in the entire disk. Figure \ref{fig:VSGHD97048} shows the SEDs and FWHM sizes in the N-band of these models.
We can compensate for an increased contribution from a larger inner disk (which makes the disk size in the N-band smaller) by including VSG, which emits at 10 $\upmu$m in the outer disk and hence fit the observed N-band size \citep{2006Doucet} as well as the SED. 
The VSG increase the Q-band size but only negligibly ($\sim$0.005''). Hence, the derived radii of the outer disks through the analysis of the Q-band (See Section \ref{sec:modelQ}) is not affected by including or neglecting VSG emission in the Q band. 

The spectrum of HD\,135344\,B serves as an ideal example of an SED with very weak PAH and VSG emission. It has a stellar temperature of 6590 K, and as a result, its UV field is low compared to the other sources in our sample, which may explain the weakness of the PAH features. The contribution of VSG seems low, since all the N-band excess observed in the spectrum is originating from the hot inner disk.

\subsection{N-band continuum summary}
We conclude that it is difficult to constrain the origin of emission in the N-band. An improved fit to the SED and the N-band image was found in Section \ref{sec:VSG} for HD\,97048 by assuming a larger inner disk size and by including VSG in the modeling. Since we do not have information on the actual size, structure and dust composition of the inner disk however, we are degenerate in determining to what degree thermal dust or VSG are responsible for emission in the SED and the disk sizes in the N-band. Detailed analysis of multiwavelength data with sufficient spatial resolution may provide more insight on this issue. The most important insight of our modeling effort in this section is that the Q-band is a more reliable tracer for disk gaps than the N-band for the sources in our sample (i.e. group Ib Herbig stars at a distance of $\sim$150 pc).


\subsection{The disk size in the PAH features}

\label{sec:modelPAH}
PAHs are the carriers of the characteristic features observed in the 5--15$\upmu$m range. Studies of the relative emission profiles have been carried out to find a relationship between the chemical and physical characteristics of these large molecules (e.g., size, charge state and excitation) and the physical conditions of a region (e.g., density, temperature, radiation field and metallicity). After direct irradiation from the photospheric UV radiation field, PAH molecules presumably located on the surfaces and/or inner rims of protoplanetary disks cool through emission in their mid-IR vibrational modes. PAH molecules are thought to be co-spatial with the gas. PAH emission is often found spatially extended in protoplanetary disks of Herbig stars \citep{2004Boekel, 2006Habart}. If PAH molecules in the outer disk are illuminated by  UV photons, they can be excellent tracers of the outer disk structure. As \citet{2006Lagage} and \citet{2007Doucet} show for HD\,97048, imaging in the PAH filter can be used to derive the (flaring) structure of the disk surface.  

Using spatially resolved VLT/VISIR observations, we compare the sizes of the PAH emitting regions in the N-band for the objects in our sample. Figure \ref{fig:FWHMprofiles} shows the FWHM plots as a function of wavelength in the N-band. For HD\,97048, the FWHM size in the PAH feature is larger than in the continuum as expected. However, this behavior is not observed in the other sources: the sizes of the PAH features in HD\,169142 are as large as the continuum; they are unresolved in HD\,135344\,B; and most surprisingly, the FWHM of the 11.3 $\upmu$m PAH feature in Oph IRS 48 is smaller than the continuum \citep{2007bGeers, 2007aGeers}. We recall that the 3.3$\upmu$m PAH feature is extended in HD\,97048, HD\,169142 and Oph IRS 48 \citep{2006Habart, 2007bGeers}. Can we understand the very different behavior of our sample in the spatial extent of the emission in the PAH bands and the 10 micron continuum?
\begin{figure}[htbp]
  \centering
  \includegraphics[width=0.5\textwidth]{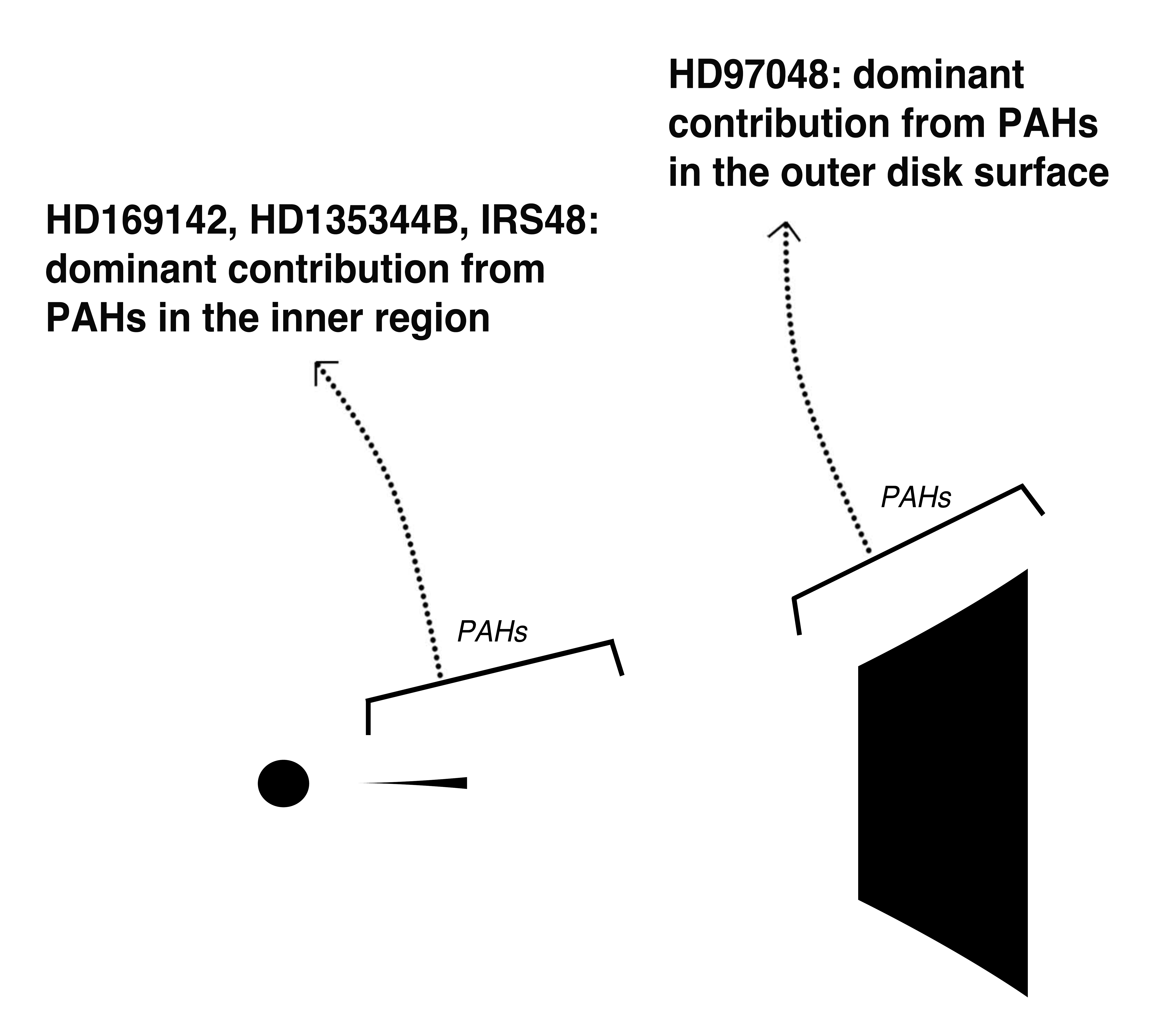} 

  \caption{Sketch of a Herbig group Ib star. In the VISIR data, we find that PAH emission in the spectra of HD\,169142, HD\,135344\,B,  and Oph IRS 48 is dominated by PAHs located in the inner region (i.e. $\lesssim$50 AU). While in HD\,97048, the PAH emission is dominated by PAHs in the outer disk surface (as shown by \citealt{2006Lagage, 2007Doucet}).  }
  \label{fig:cartoon_PAH}
\end{figure}

We have examined the scenario that PAH emission is emitted from the disk surface (i.e., not from the halo). We have fitted the peak fluxes of the observed PAH features by adding PAH molecules in the disk with a fixed abundance with respect to the other dust components. See Section \ref{sec:modeldiscription} for details of the PAH implementation in the radiative transfer code. For HD\,97048, HD\,169142, HD\,135344\,B, and Oph IRS 48, the abundance ratio (fraction of total dust mass) fitting the observations are respectively \{3.5, 5, 10, 100\} $\times$10$^{-4}$, and the is result in a total mass in PAHs are respectively \{1.7, 0.4, 1.0, 3.0\} $\times$10$^{-7}$ M$_{\odot}$.  If no VSG are included, we find that the FWHM sizes at the PAH wavelengths are resolved and significantly larger than the continuum. Except for HD\,97048, these models are not consistent with the FWHM observations from VISIR. If VSG are included and have a large contribution to the continuum emission in the N-band as explained in Section \ref{sec:VSG}, the observed angular size in the PAH band is still larger compared to the continuum. This is because the contrast between the flux from PAHs and the continuum from thermal dust is higher than the contrast between emission from VSG and thermal dust. Therefore, the angular size of the disk is larger in the PAH band than in the N-band continuum for any of these models include PAHs and fit the SED.

For HD\,97048, PAH emission has a dominant contribution from PAHs in the outer disk and thus have a larger observed size in the PAH bands of VISIR. A closer look at the brightness distribution of the 11.2$\upmu$m VISIR image of HD\,97048 \citep{2007Doucet} shows no clear sign of a gap. Instead, it shows a relatively strong inner component between 0--0.2'' (0--30 AU). The PAH model in \citet{2007Doucet}, which assumes a continuous density in the disk and no gaps in the inner disk structure, overpredicts the PAH surface brightness of the outer disk by a factor of 2 between 0.2--1.5'' (30--240 AU). This suggests that there is a significant contribution from the inner disk to the PAH flux, so the relative contribution from inner and outer disk to the PAH flux seems to be different from what \citet{2007Doucet} modeled. However, the observed FWHM of the disk in the PAH band still has to come out larger than the continuum.

The VISIR observations of HD\,169142 show that the observed FWHM of the continuum and the PAH features are similar in size. The contrast between the size in the PAH filter and the continuum can be decreased, if VSG are added to the disk and dominate the continuum emission in the N-band. In that case, the continuum size also increases and the difference between the size in the PAH filter versus the continuum becomes smaller. However, the addition of VSG does not result in a equal disk size between the continuum and the PAH band; therefore, an extra component of PAHs in the inner disk is needed.

For HD\,135344\,B, the PAH features are fitted with a higher abundance ratio than for HD\,97048. Because the UV field of HD\,135344\,B is lower, the line fluxes of the PAH features are also weaker. Our disk model with PAH emission dominated by PAHs in the outer disk predicts that the PAH features would be spatially resolved by VISIR. This is, however, not seen in our VISIR data in Figure \ref{fig:FWHMprofiles}. No PAH features were also observed within the correlated flux spectrum of HD\,135344\,B measured with MIDI \citep{2008Fedele}, leading these authors to draw the conclusion that the PAH emission must be more extended than the continuum (i.e. $\gtrsim$5--10 AU).  These measurements can only be reconciled with our VISIR data in a situation in which the bulk of the N-band continuum emission comes from a narrow ring very close to the star (0.05-1.8 AU), and that the emitting PAHs are located in another ring/structure small enough not be be resolved by VISIR outside this region.

In Oph IRS 48, the FWHM of the 11.3 $\upmu$m PAH feature is even smaller than the continuum \citep{2007bGeers}. Direct imaging observations by VISIR in the PAH band at 11.3 $\upmu$m \citet{2007aGeers} show that the PAHs are centrally located and do no show a similar ring structure as seen at 18.8 $\upmu$m continuum, which traces the outer disk. This can only be explained if PAH emission has a higher contribution from PAHs located in the inner regions (possibly from within the gap) compared to PAHs in the surface of the outer disk. 

\subsection{PAH-band summary}
Modeling the N-band FWHM sizes of the 11.3 $\upmu$m PAH feature of our sample has demonstrated that PAH emission is not always dominated by emission from PAHs in the outer disk surface. A strong contribution from PAHs in the inner regions (probably including the gap itself) can explain the small FWHM of the VISIR 11.3 $\upmu$m PAH feature of HD\,169142, HD\, 135344\,B and Oph IRS 48. Measuring the size of a disk in the PAH filter does contain information about the disk size. However, the relative contributions to the total PAH flux from the inner and outer disks can vary strongly from source to source, making PAHs a more complex tracer of disk structure than previously assumed. See Figure \ref{fig:cartoon_PAH} for a sketch of this geometry.


\section{Discussion}
\label{sec:discussion}

\subsection{The weakness of silicate emission features in some sources is caused
\label{sec:weaksilicates}
by large disk gaps}

Spectroscopic ISO/SWS \citep{2001Meeus} and Spitzer/IRS observations \citep{2010Juhasz} revealed that the 10 $\upmu$m silicate feature is detected in the majority of the Herbig Ae/Be stars (8 out of 53 lack the feature in the Spitzer sample). Flaring and flat objects with silicate features are respectively called group Ia and group IIa; those without silicate features are called group Ib and group IIb. Although this may be a consequence of the limited number of Herbig stars in the sample, objects without evidence of silicate emission are so far \emph{only} found among flaring group I sources.

\begin{figure}[htbp]
  \centering
  \includegraphics[width=0.5\textwidth]{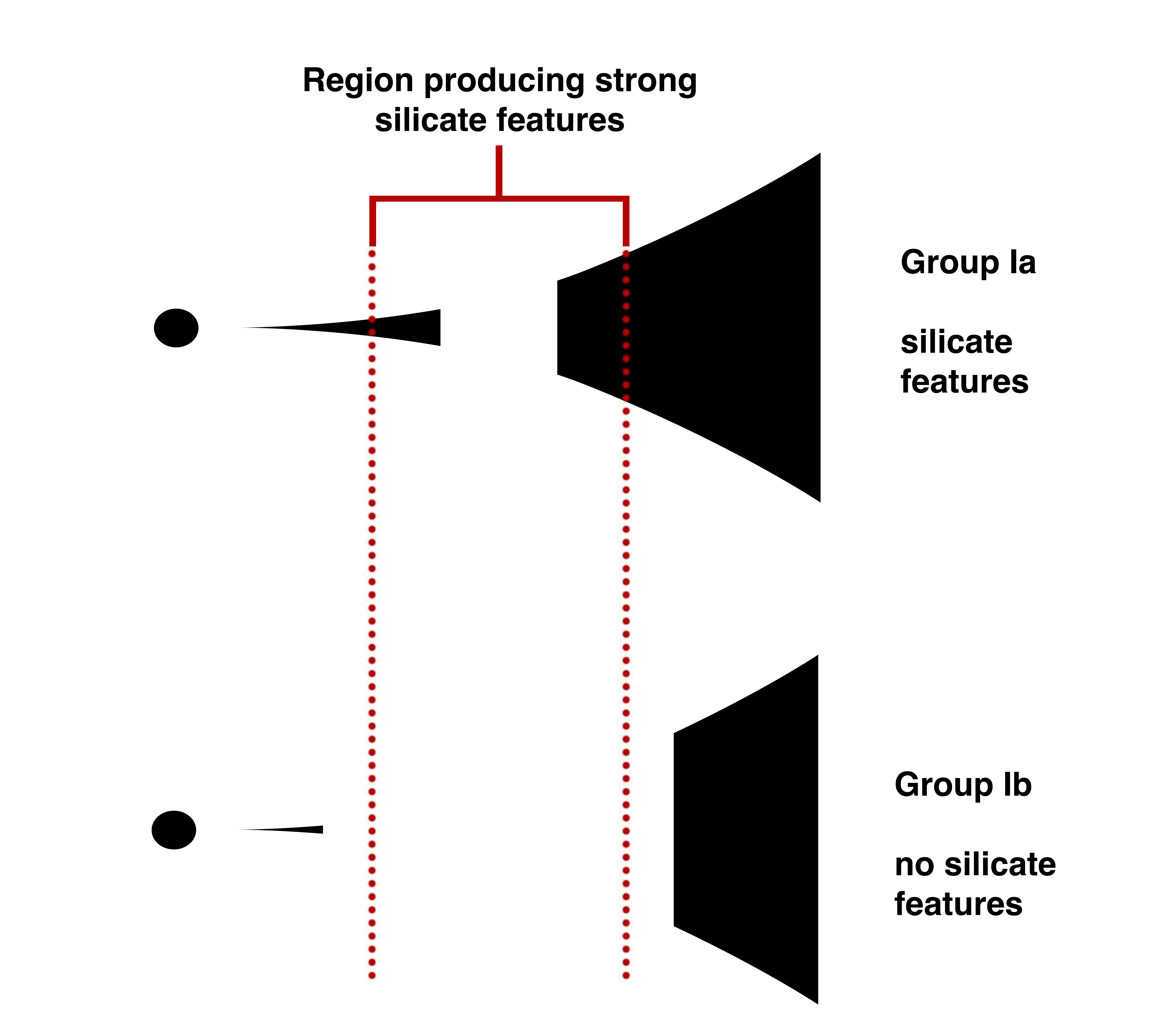} 

  \caption{Sketch showing the region where (sub-) micron-sized silicate grains have a temperature (T$\sim$200--400 K) where they produce strong silicate features. Transitional group Ia sources (top) show silicate features, because the inner disk and/or the outer disk wall have a sufficient amount of silicate dust in this region. In the group Ib objects of our sample (bottom), the silicate dust is severely depleted in this region, and thus, no silicate features are produced. }
  \label{fig:cartoon_silicate}
\end{figure}

Until now, it was not well understood how the absence of these features is connected to the physical disk structure. 
An explanation supported by our study is because of the geometry: the (sub-) micron-sized silicate dust suitable for prominent feature emission (T$\sim$200--400 K) are depleted in the disk due to the large size of inner holes or gaps. Objects \emph{without} silicate features are only found in group I Herbig Ae/Be stars, which suggests that gaps are connected to the flaring geometry. Since we found that Herbig Ae/Be stars in our sample are best fitted with large gaps separating the inner and outer disk, we suggest that a decreasing strength of the silicate feature is related to the severe depletion of (sub-) micron-sized silicate grains of $\sim$200--400 K. Its weakness can serve as a tracer of large dust-depleted inner regions of protoplanetary disks. Figure \ref{fig:cartoon_silicate} shows a sketch of this scenario for transitional group I objects. Group Ia sources that have a gap show silicate features, because the inner disk (e.g. HD\,142527, \citealt{2011Verhoeff}) and/or the outer disk wall (e.g. HD\,100546, \citealt{2011Mulders}) still have a sufficient amount of silicate dust in the critical temperature regime ($\sim$200--400 K region). In group Ib objects (the objects in our sample: HD\,97048, HD\,169142, HD\,135344B, and IRS 48), our study shows that the silicate dust is severely depleted in this region. 

HD\,97048 is a prototypical group I source, and so far no gap has been reported. Now that our study finds an inner gap (or a region of lower dust density), it strengthens the case for disk gaps as the reason for the absence of the silicate features in the spectra of group Ib, and more generally, the classification of group I sources as transitional disks. 

\subsection{Gaps in Herbig Ae/Be objects}

It was already suggested by \citet{1994Waelkens} based on SED analysis that disks around Herbig Ae/Be stars may have large gaps separating an inner and outer disk. It was proposed that larger circumstellar bodies could form in these gaps, typically, in the temperature range of 200 to 500 K. With the improved spatial resolution of current telescopes, the prediction of large gaps as a common feature in Herbig Ae/Be stars seems to be confirmed.

Other group I Herbig stars \emph{without} silicate features that are not included in our sample are HD\,100453, HD\,34282, and HD\,141569 (with distances of 112, 164, and 99 parsec, respectively). The first two sources are also resolved in the Q-band by Gemini/T-ReCS observations \citep{2011Marinas}. HD\,141569 has generated much debate in the literature, but it is now believed that this disk is in transition to a debris disk (\citealt{2005Wyatt, 2009Reche}). Our analysis outlined in Section \ref{sec:modelQ} suggests that significantly resolved Q-band images of Herbig Ib stars indicate disk gaps. Since all Herbig Ib stars are resolved in the Q-band, we suggest that all Herbig Ib objects have similar disk structures with large gaps between the inner and outer disk. 

Flaring Herbig Ae/Be stars \emph{with} silicate features and detected gaps are AB Aur \citep{2010Honda}, HD\,142527 \citep{2006Fukagawa, 2006Fujiwara}, HD\,36112 \citep{2010Isella}, and HD\,100546 \citep{2003Bouwman, 2010Benisty}. These stars still do have silicate emission, despite a gap. 

In recent years there has been much speculation about Herbig group I objects being transitional disks. However, there are also many flaring Herbig Ae/Be stars for which no gaps are yet reported. Most of these objects are located at larger distances and therefore more difficult to spatially resolve. Other targets are just not yet observed at full resolution and/or have not been subject to detailed radiative transfer analysis. 

There is only limited information available on the spatial structure of flat (group II) disks. \citet{2011Marinas} reported faint extended emission at 12$\upmu$m from HD\,144668 and HD\,163296, but no resolved group II objects in the Q-band\footnote{We follow the classification in \citet{2010Juhasz} of HD\,34282 as a group Ib source. Thus, no group II source in the sample of \citet{2011Marinas} is resolved in the Q-band.}. The disks of HD\,163296 and HD\,31648 (MWC 480) are spatially resolved in the sub-mm \citep{2011Sandell} and observed in scattered light \citep{2008Wisniewski, 2012Kusakabe}; however, no gaps are reported.  Since they show the highest sub-mm excess among flat objects and show CO gas lines in the sub-mm \citep{2005Dent, 1997Mannings, 2011Qi}, they are considered as transitional between flaring and flat. UV observations of HD\,104237 point also indirectly, towards a small disk; for example, obscuration of a background microjet in the UV constrained the circumstellar disk to r $<$ 0.6'' (70 AU). There are no other flat disks identified in the literature with observational evidence of the disk structure beyond $\gtrsim$50 AU. 

It is still an open question whether the inner disk structures of flaring group I and flat group II disks are different. However, there are observations that may indicate that flat group II sources could have optically thicker inner disks. First, there are no group II objects \emph{without} silicate features. Thus, group II sources do not seem to have large gaps in the temperature range of $\sim$200--400 K (see Section \ref{sec:weaksilicates}). Second, a photometric study of \citet{2009Acke} demonstrated a correlation between higher optically thick inner rims which resulted in lower outer disk brightness due to shadowing effects. This correlation is strongest in flat group II sources suggesting that their inner disks are optically thicker than in flaring group I objects. Third PAH and other gas lines in flat group II sources, such as far-IR CO \citep{2012Meeus}, milimeter CO \citep{2005Dent} and near-IR H2 \citep{2011Carmona} emission are much weaker or not present. While a lower far-IR continuum flux can be ascribed to dust settling, it is not yet understood why gas lines are much weaker in group II sources. Opposite to what is observed, modeling by \citet{2007bDullemond} predicts that PAH features become \emph{stronger} as the dust settles down to the mid-plane. A possible solution may be that group II sources have optically thicker inner disks casting a greater shadow on the outer disk, thereby reducing the strength of the gas lines and the FIR continuum flux even more.

\begin{figure}[htbp]
  \centering
  \includegraphics[width=0.5\textwidth]{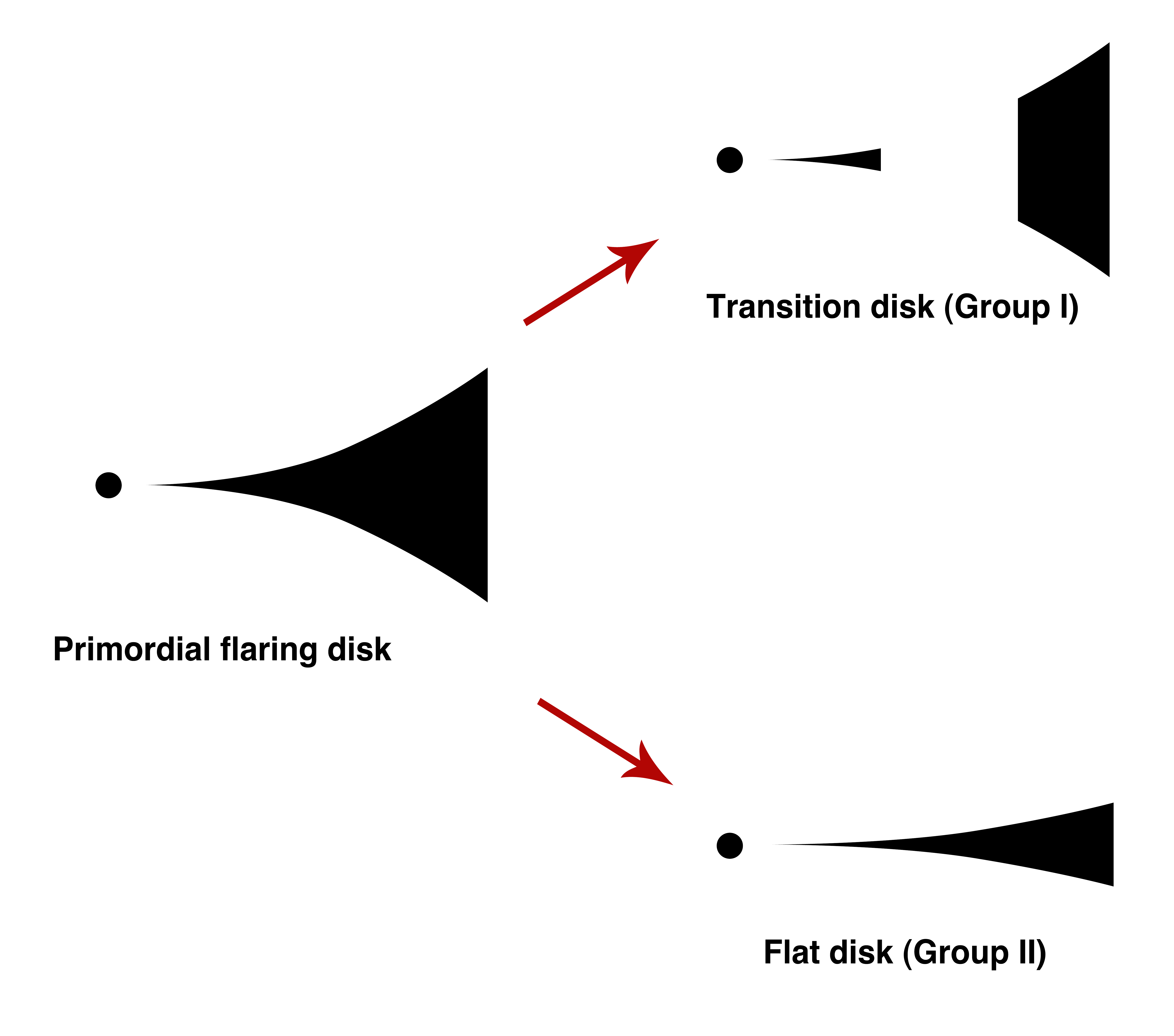} 

  \caption{Sketch of an evolutionary scenario for disk evolution in Herbig Ae/Be objects (after \citealt{2010Currie}). Both groups have evolved from a common ancestor (i.e. primordial, possibly embedded, flaring disk structure). In transitional group I objects, gap formation proceeds the collapse of the outer disk. In group II objects, grain growth and dust settling have flattened the outer disk.}
  \label{fig:cartoon_evolution}
\end{figure}

\subsection{Consequences for the evolutionary link between group I and group II }

While gaps are found in an increasing number of group I Herbig stars, there is no evidence of gaps in group II sources. Since it is likely that the gap is formed by a low mass companion, which is probably a planet \citep{1996Pollack}, this planet will maintain the gap throughout the further evolution of the disk (e.g. \citealt{2007Armitage}). If group II objects do not have gaps then the evolutionary link from group I to group II is no longer evident. As a solution to this problem, we suggest that a typical `primordial flaring disk' with a continuous density distribution (i.e. no gaps) in the disk may be a `common ancestor' for transitional group I and flat group II objects (See the sketch in Figure \ref{fig:cartoon_evolution}). From that starting point, a flaring disk can evolve into transitional group I objects by gap-formation or a self-shadowed group II source in which the outer disk has collapsed. We stress that this scenario is very speculative but may however help to redefine the evolutionary link between Herbig group I and II sources. A similar scenario for T-Tauri stars has been proposed by \citet{2010Currie}. 

Not included in Figure \ref{fig:cartoon_evolution} is the hypothetical possibility of a disk in which both effects are present (planet formation and the collapse of the outer disk). New key questions need to be addressed to understand any putative evolutionary link between group I and group II objects. Can transitional group I sources with large gaps still evolve to flat group II? Or reversely, can flat group II objects become group I sources when/if gaps form? Better constraints on the structure of group II disks by direct imaging are essential to address these issues. 

The SED and spectral characteristics of group I sources seem to indicate the dominant presence of a high wall irradiated by the star. As more and more Q-band observations have become available, it has become clear that these walls are indicative of gaps in the disk structure. As we do not observe similar characteristics of walls in group II sources, we infer that these sources do not have disks with gaps.

\begin{figure}[htbp]
  \centering
  \includegraphics[width=0.5\textwidth]{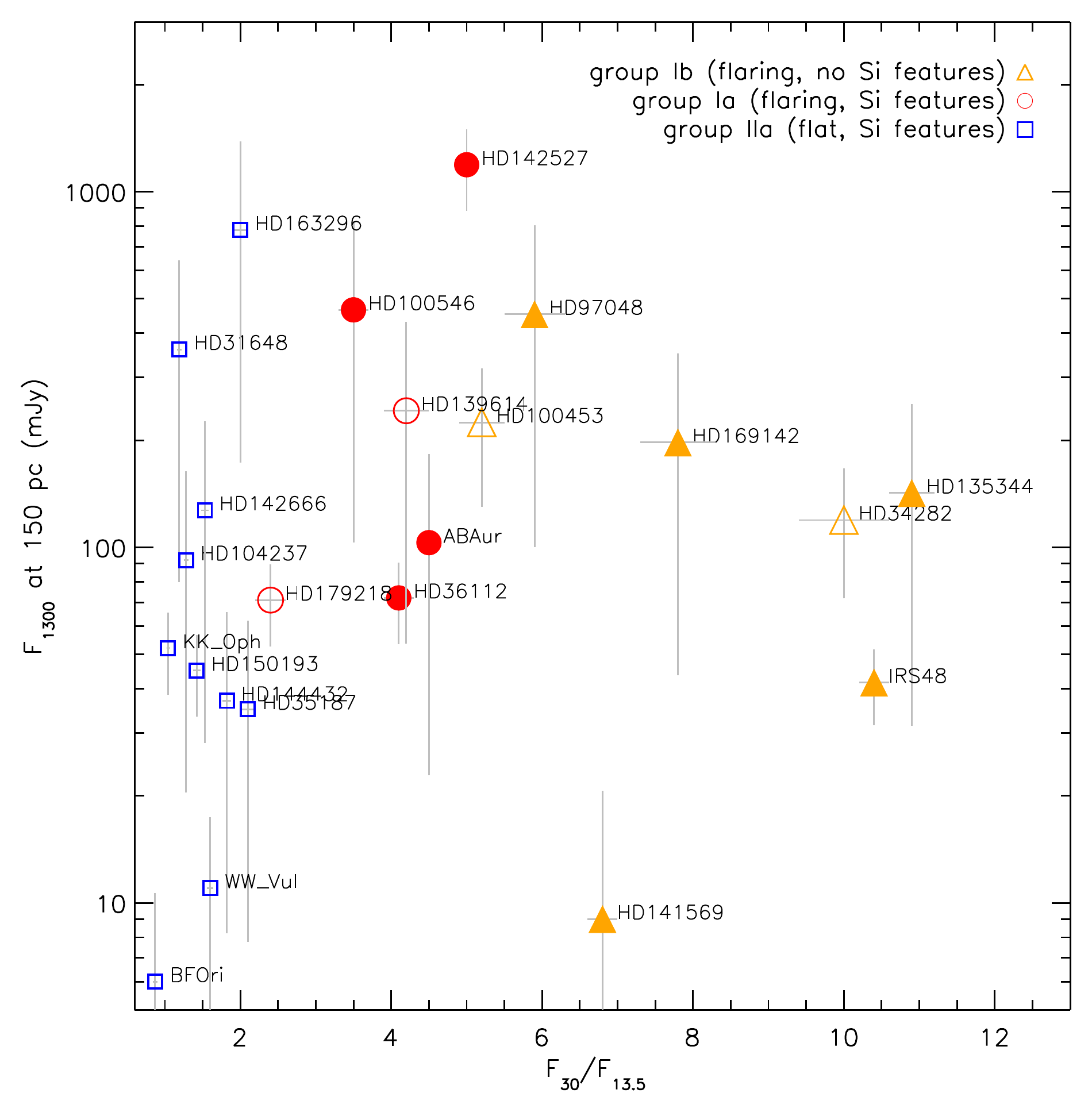} 

  \caption{For a sample of Herbig stars, the sub-mm flux at 1.3 mm (normalized at a distance of 150 pc) is compared to the MIR spectral index given by the flux ratio at 30.0 and 13.5 $\upmu$m. Filled symbols indicate sources for which gaps have been reported in the literature. The sub-mm flux can be used as a proxy for the disk mass while the MIR spectral slope  may trace gaps in the temperature range T$\sim$200--400 K. Because gaps are found in most group I objects, the evolutionary link from group I to group II is no longer obvious. }
  \label{fig:SEDcorrelation}
\end{figure}

More insight into the different evolutionary pathways of Herbig stars can be obtained by examining the color-magnitude diagram on Figure \ref{fig:SEDcorrelation}. The objects are taken from the sample of \citet{2010Acke} for which sub-mm photometric data was available. The filled symbols indicate the Herbig stars for which gaps have been reported in the literature. This diagram can be used to compare the inner and outer disk evolution among Herbig stars. The y-axis shows the flux at 1.3 mm that is widely used in estimates for the total disk mass in milimeter grains (e.g., \citealt{2005Andrews, 2007Andrews}).  As disks lose their mass during their evolution, they will move down in this diagram. There is no significant difference between group I (red dots and orange triangles, F$_{30}$/F$_{13.5} \gtrsim 2.2$) and group II (blue squares) in disk mass. On the x-axis however, the absence of silicate features (orange triangles) is strongly correlated with the MIR spectral index (F$_{30}$/F$_{13.5} \gtrsim 5.1$), as expected for a disk with a large gap and a vertical wall of T$\sim$100--150 K at the inner edge of the outer disk. Since no gaps have yet been observed in group II objects, it seems unlikely that group Ib objects (with large inner cavities and no silicate features) continue their evolutionary path on this diagram as group IIa objects (with silicate features). Instead, the end-stage of the evolution of a group Ib object is likely to be represented by a group IIb object (though group IIb objects have so far not been identified in infrared spectral surveys of Herbig Ae/Be stars) or a disk/debris system, such as HD\,141569.

Candidates for the `common ancestor' group may be Herbig group Ia objects without gaps. Since many group I objects are too far away to be spatially resolved, improved spatial resolution is needed to identify whether these objects are transitional or have continuous disks. Other candidates are heavily embedded YSOs (i.e. Lada Class I sources). Examples may be present in an ISO study of 46 Herbig Ae/Be stars \citep{2004AckeAncker} where a small sample of six objects has been identified (labeled as group III) and display the amorphous 10 micron feature in absorption. It was believed that these objects possess disks whose luminosity is dominated by viscous dissipation of energy due to accretion. These objects are deeply embedded systems and hence, are fundamentally different from the other sample stars.  

T-Tauri stars are more numerous than Herbig objects, and therefore, better statistics are available in studies of the SED shape in the context of star and disk properties. In recent Spitzer studies, large samples of T-Tauri stars have been studied (e.g. \citealt{2010Cieza,2012Cieza, 2011Currie, 2013Oliveira}). SEDs classifications of T-Tauri stars reveal a similar variety as for Herbig Ae/Be stars. Using radiative transfer modeling, these SEDs are identified as objects with large inner cavities, transitional disks with obvious gaps and objects with homologously depleted (flattened) outer disks. However, the dust mineralogy (grain sizes, temperatures and crystallinity fractions) does not show any correlation to either stellar and disk characteristics or mean cluster age in the 1--10 Myr range as shown by \citep{2013Oliveira}.

A variety of effects are thought to be responsible for disk evolution. Such as planet formation (\citealt{1993Lissauer,1997Boss}), grain growth and settling \citep{2004bDullemond, 2005Dullemond}, photo-evaporation (e.g. \citealt{2006Alexander, 2009Gorti}), viscous accretion \citep{1998Hartmann}, the magneto-rotational instability \citep{2007Chiang}, and dynamical interactions between the disk and stellar or substellar companions \citep{1994Artymowicz}. The timescales and strengths of these disk clearing mechanisms highly depend on intrinsic disk properties such as initial disk mass, composition and angular momentum. Properties are poorly constrained and likely have a large dispersion. The duration and stability of each evolutionary phase, with either a cleared inner gap or a flattened outer disk are therefore currently one of the main challenges of the disk-evolution field.


\section{Conclusions}
\label{sec:conclusions}
We have modeled the spectral energy distribution and N- and Q-band imaging of type Ib Herbig AeBe stars using the radiative transfer code MCMax. The best-fitting models of the protoplanetary disks indicate that the inner regions of these disks contain large gaps. As a consequence, directly irradiated vertically extended outer disk walls emerge on the inner edge of the outer disk. We conclude:

\begin{itemize}
\item We observed HD\,97048 and HD\,135344\,B with the T-ReCS instrument on the Gemini South telescope. Both sources are extended at 24.5$\upmu$m with respect to the calibration stars. The derived source sizes are 0.789$\pm$0.017'' and 0.802$\pm$0.007'', respectively.
\item For all four sources in our sample, the observed Q-band sizes require disk models with an inner-hole/gap, making them transitional disks. We fail to model the Q-band images using models with a continuous density distribution (i.e. no gaps). Although SED modeling is degenerate in terms of the source structure, our study suggests that the typical SEDs of Herbig group Ib sources is an indicator of gaps. We find that Q-band images are required to unambiguously infer the presence and sizes of gaps.
\item Radiative transfer modeling constrains the radius of the inner edge of the outer disk at 34$^{+4}_{-5}$  AU for HD\,97048, 23$^{+3}_{-5}$ AU for HD\,169142, 30$^{+2}_{-1} $ AU for HD\,135344\,B and 63$^{+4}_{-4} $ AU for Oph IRS 48 . 
\item We have selected Herbig group Ib sources that show a pronounced increase in flux at about 20$\upmu$m and are resolved in the Q-band. Hence, the emitting dust temperature of the wall is typically $\sim$100--150 K.
\item The absence of silicate features in Herbig Ib objects is a natural consequence of the presence of large gaps in the inner disks of our models. The depletion of dust in the inner regions results in weaker silicate features.  
\item The observations show that the continuum emission in the N-band is not due to emission in the wings of PAHs. In general, this continuum emission can be due to VSGs or to thermal emission from the inner or outer disk. For most of our sources, we cannot decide which of these is the dominant source of emission. Therefore, Q-band imaging is a more reliable tracer of gaps.
\item Disks with PAH emission do not automatically imply a disk structure, where PAH emission is dominated by PAHs on the surface of the outer disk.
\item If the conclusion holds up that all group I sources are transitional and group II sources have no gaps, then the evolutionary link from group I to group II is no longer obvious. A possible solution is that both groups have evolved from a common ancestor (i.e., gapless flaring-disk structure). In transitional group I objects, gap formation has proceeded the collapse of the outer disk while grain growth and dust settling have flattened the outer disk in flat group II objects.      

\end{itemize}

\thanks{The authors thank Vincent Geers for providing the reduced data from his 2007 papers. K.M. is supported by a grant from the Netherlands Research School for Astronomy (NOVA). Based on observations collected at the European Southern Observatory, Chile under program IDs, 075.C-0540A, 075.C-0540C and 079.C-0602A. 
M.M. acknowledges funding from the EU FP7-2011 under Grant Agreement nr. 284405.
We are grateful to all of the staff members of the Subaru Telescope.
We also thank Ms. Hitomi Kobayashi and Dr. Yuji Ikeda at Kyoto-Nijikoubou Co., Ltd.
This research was partially supported by KAKENHI
(Grant-in-Aid for Young Scientists B: 21740141) by the Ministry of
Education, Culture, Sports, Science and Technology (MEXT) of Japan. 
This publication makes use of data products from the Wide-field Infrared Survey Explorer, which is a joint project of the University of California, Los Angeles, and the Jet Propulsion Laboratory/California Institute of Technology, funded by the National Aeronautics and Space Administration. This publication makes use of data products from the Two Micron All Sky Survey, which is a joint project of the University of Massachusetts and the Infrared Processing and Analysis Center/California Institute of Technology, funded by the National Aeronautics and Space Administration and the National Science Foundation.
}



\bibliography{bib.bib} 

\clearpage


\begin{appendix} 
\section{Photometry}
\label{sec:photometry}
\emph{Herschel/PACS}: For HD\,97048 and HD\,135344\,B far-IR photometry was derived from Herschel/PACS scan maps at 70, 100 and 160 micron. The PACS photometric data were reduced with the mini 
scanmap pipeline in HIPE version 8.1.0 (calTree version 32). The 
photometry was extracted using apertures of 12", 15" en 20" for 70, 100
and 160 micron, respectively. The fluxes were aperture and color-corrected,
and have a maximum error of 5\%, according to the PACS manual. 

\emph{Other photometry} is collected from the literature. The photometric data is presented in Tables \ref{tab:photometry970} for HD\,97048, \ref{tab:photometry169} for HD\,169142, \ref{tab:photometry135} for HD\,135344\,B and \ref{tab:photometryIRS} for Oph IRS 48. Upper limits and erroneous data are not included. The photometry is sorted by the central wavelengths of the photometric filters. Magnitudes have been converted to fluxes (in Jansky) for comparison in the SED. The conversion has been done using zero magnitudes provided in the instrument documentations and, where possible the spectral index were considered. The photometry listed in the tables is not corrected for extinction, while the photometry presented in the SEDs in Figure \ref{fig:SEDQ} are corrected, according to the stellar extinction given in Table \ref{tab:sample}, the parameter $R=3.1$ and the extinction law presented in \citet{1999Fitzpatrick}.

\begin{table}[htdp]
\tiny
\caption{  \label{tab:photometry970} Photometric data HD\,97048 (not corrected for extinction) used in this study }
\begin{center}
\begin{tabular}{l r  r @{ $\pm$ } l  c  }
\hline
\hline
band ID  &  \multicolumn{1}{c}{$\uplambda$ [$\upmu$m]}& \multicolumn{2}{c}{F$_{\nu}$ [Jansky] }& reference \\
\hline
IUE 15 &   0.15 & 0.053 &  0.001 & a\\
IUE 18 &    0.18 &  0.066 &   0.001 & a  \\
IUE 22 &    0.22 &  0.054 &   0.002 & a\\
IUE 25 &    0.25 &  0.091 &   0.002 &a \\
IUE 30 &    0.30 &  0.192 &   0.004 &a \\
Walraven W &    0.32 &  0.163 &   0.003 &b  \\
Johnson U &    0.36 &  0.458 &   0.010 &b  \\
Walraven U &    0.36 &  0.350 &   0.005 & b  \\
Walraven L &    0.38 &  0.724 &   0.017 & b  \\
Walraven B &    0.43 &  1.216 &   0.014 & b  \\
Johnson B &    0.44 &  1.374 &   0.018 & b  \\
Walraven V &    0.54 &  1.531 &   0.014 & b  \\
Johnson V &    0.55 &  1.629 &   0.015 & b  \\
Cousins R &    0.64 &  1.574 &   0.021 & b  \\
Cousins I &    0.79 &  1.667 &   0.035 & b  \\
Near-IR J &    1.23 &  2.280 &   0.042 & b  \\
Near-IR H &    1.65 &  1.967 &   0.037 & b  \\
Near-IR K &    2.22 &  2.301 &   0.043 & b  \\
WISE 1		& 	3.35& 3.039 & 0.195 & c\\
Near-IR L &    3.77 &  3.089 &   0.146 & b  \\
WISE 2		&	4.60& 3.273 & 0.127 & c\\
Near-IR M &    4.78 &  2.382 &   0.112 & b  \\
WISE 3		&	11.60& 8.690& 0.088 & c\\
IRAS 12 &   12.00 & 12.88 &   1.91 & d  \\
WISE 4		&	22.10& 27.923  & 0.332 & c\\
IRAS 25 &   25.00 & 40.89 &   4.99 & d  \\
IRAS 60 &   60.00 & 67.93 &  15.64 & d  \\
AKARI S65	& 65.00	&	71.36	& 2.40 & e \\	
 PACS 70 		& 70.00	&66.91& 	3.35 & f \\
 AKARI S90	& 90.00	&	66.04	&3.80& e	\\
 PACS 100 	&100.00	&66.75& 	3.34 & f \\ 
IRAS 100 &  100.00 & 63.54 &  16.45 & d  \\
AKARI S140	&140.00&	49.58	&3.72& e\\
AKARI S160	&160.00&	53.67	&3.56& e\\
 PACS 160 	& 160.00	&58.21& 	2.91 & f \\
1300 micron & 1300.00 &  0.452 &   0.352 & g  \\

\hline
\end{tabular}
\end{center}

\textbf{References:} 
\textbf{a)} IUE archival data
\textbf{b)} \citealt{2001deWinter}
\textbf{c)} WISE All-Sky Data Release
\textbf{d)} IRAS Point-source catalogue
\textbf{e)} AKARI/IRC mid-IR all-sky Survey 
\textbf{f)} This paper
\textbf{g)} \citealt{1994Henning} 

\end{table}%

\begin{table}[htdp]
\tiny
\caption{  \label{tab:photometry169} Photometric data HD\,169142 (not corrected for extinction)  used in this study }
\begin{center}
\begin{tabular}{l r r @{ $\pm$ } l  c  }
\hline
\hline
band ID  &  \multicolumn{1}{c}{$\uplambda$ [$\upmu$m]}& \multicolumn{2}{c}{F$_{\nu}$ [Jansky] }& reference \\
\hline

ANS 15 &    0.15 &  0.002 &   0.0002 & a  \\
ANS 18 &    0.18 &  0.077 &   0.001 & a  \\
ANS 22 &    0.22 &  0.163 &   0.003 & a  \\
ANS 25 &    0.25 &  0.199 &   0.004 & a  \\
Walraven W &    0.32 &  0.566 &   0.012 & b  \\ 
ANS 33 &    0.33 &  0.519 &   0.010 & a  \\
Johnson U &    0.36 &  0.841 &   0.019 &c \\
Walraven U &    0.36 &  0.758 &   0.011 & b  \\
Walraven L &    0.38 &  1.183 &   0.028 & b  \\
Walraven B &    0.43 &  1.794 &   0.021 & b  \\
Johnson B &    0.44 &  1.986 &   0.026 & c  \\
Walraven V &    0.54 &  2.084 &   0.019 & b  \\
Johnson V &    0.55 &  2.207 &   0.020 & c  \\
Cousins R &    0.64 &  1.963 &   0.026 & c  \\
Cousins I &    0.79 &  2.004 &   0.041 & c  \\
Near-IR J &    1.23 &  1.828 &   0.034 & c  \\
Near-IR H &    1.65 &  1.667 &   0.031 & c  \\
Near-IR K &    2.22 &  1.534 &   0.029 & c  \\
WISE 1		& 	3.35		& 1.238		& 0.047	& d\\
Near-IR L &    3.77 &  1.349 &   0.064 & c  \\
WISE 2		&	4.60		& 0.995 		& 0.021 		& d\\
Near-IR M &    4.78 &  1.172 &   0.055 & c  \\
WISE 3		&	11.60	& 2.575		& 0.028 		& d\\
IRAS 12 &   12.00 &  2.95 &   0.29 & e  \\
AKARI S18 	&	18.00	&	8.90	&	0.23	&	f	\\
WISE 4		&	22.10	& 14.021		& 0.103 		& d\\
IRAS 25 &   25.00 & 18.39 &   2.24 & e  \\
IRAS 60 &   60.00 & 29.57 &   6.81 & e  \\
AKARI S65	&	65.00	&	24.45	&	0.10	&	f	\\
 PACS 70 		& 70.00	&27.35 	&  0.03 & g \\
 AKARI S90	&	90.00	&	19.99	&	1.23		&	f	\\
IRAS 100 &  100.00 & 23.37 &   6.74 & e  \\
AKARI S140	&	140.00	&	13.32	&	1.89		&	f 	\\
AKARI S160	&	160.00	&	15.47	&	0.46	&	f	\\
 PACS 160 	& 160.00	&17.39 &  0.05 & g \\
 450 micron &  450.00 &  2.00 &   0.275 & h  \\
800 micron &  800.00 &  0.553 &   0.431 & c  \\
850 micron &  850.00 &  0.40 &   0.02 & h  \\
1100 micron & 1100.00 &  0.286 &   0.223 & c  \\
1300 micron & 1300.00 &  0.197 &   0.153 & c  \\
2000 micron & 2000.00 &  0.070 &   0.079 & c  \\

 \hline

\end{tabular}
\end{center}

\textbf{References: a)} 
\textbf{a)} IUE archival data
\textbf{b)} \citealt{1989vanderVeen} 
\textbf{c)} \citealt{1996Sylvester}
\textbf{d)} WISE All-Sky Data Release
\textbf{e)} IRAS Point-source catalogue
\textbf{f)} AKARI/IRC mid-IR all-sky Survey 
\textbf{g)} \citealt{2010Meeus}
\textbf{h)} \citealt{2008DiFrancesco}
\end{table}%

\begin{table}[htdp]
\tiny
\caption{  \label{tab:photometry135} Photometric data HD\,135344\,B  (not corrected for extinction) used in this study }
\begin{center}
\begin{tabular}{l r r @{ $\pm$ } l  c  }
\hline
\hline
band ID  &  \multicolumn{1}{c}{$\uplambda$ [$\upmu$m]}& \multicolumn{2}{c}{F$_{\nu}$ [Jansky] }& reference \\
\hline

Johnson U 		&    0.35 &   0.414&   0.009 & a \\
Johnson B 		&    0.44 &  1.023 &   0.014 & a \\
Johnson V 		&    0.55 &  1.393 &   0.013 & a  \\
Cousins R 		&    0.64 &  1.633 &   0.022 & a  \\
Cousins I 			&    0.79 &  1.862 &   0.039 & a  \\
Near-IR J 			&    1.23 &  2.060 &   0.038 & a \\
Near-IR H 		&    1.65 &  2.365 &   0.044 &a  \\
Near-IR K 		&    2.22 &  2.766 &   0.051 &a \\
WISE 1		& 	3.35		& 2.967		& 0.183 		& b\\
Near-IR L 			&    3.77 &  2.923 &   0.138 & a \\
WISE 2		&	4.60		& 3.994 		& 0.169 		& b\\
Near-IR M 		&    4.78 &  2.564 &   0.121 & a \\
AKARI S09 		&	9.00		&	1.727	&	0.063	&	c	\\
WISE 3		&	11.60	& 0.990		& 0.013 		& b\\
IRAS 12 			&   12.00 &  1.57 &   0.19 & d  \\
AKARI S18 		&	18.00	&	3.29	&	0.014	&	c	\\
WISE 4		&	22.10	& 5.571		& 0.056 		& b\\
IRAS 25 			&   25.00 &  6.89 &   1.21 & d  \\
IRAS 60 			&   60.00 & 23.59 &   5.43 & d  \\
AKARI S65		&	65.00	&	24.41	&	1.68	&	c	\\
 PACS 70 			& 70.00	&30.00	& 1.50 & e\\
 AKARI S90		&	90.00	&	23.65	&	1.23		&	c	\\
 PACS 100 		&100.00	&28.01	& 1.40 & e \\
IRAS 100 			&  100.00 & 23.15 &   3.43 & d  \\
AKARI S140	&	140.00	&	23.73	&	3.81		&	c	\\
AKARI S160	&	160.00	&	15.69	&	0.95	&	c	\\
 PACS 160 		& 160.00	&20.21	&  1.01 & e \\
350 micron 		&  350.00 &  5.794 &   4.509 & f \\
450 micron 		&  450.00 &  3.175 &   0.575 & a  \\
800 micron 		&  800.00 &  0.569 &   0.443 & a  \\
850 micron 		&  850.00 &  0.489 &   0.024 &a \\
1100 micron 		& 1100.00 &  0.208 &   0.162 & a  \\
1300 micron 		& 1300.00 &  0.142 &   0.110 & a \\

\hline

\end{tabular}
\end{center}

\textbf{References: }
\textbf{a)}   \citealt{1996Sylvester}
\textbf{b)} WISE All-Sky Data Release
\textbf{c)} AKARI/IRC mid-IR all-sky Survey 
\textbf{d)} IRAS Point-source catalogue
\textbf{e)} This paper
\textbf{f)} \citealt{1999Gezari}

\end{table}%

\begin{table}[htdp]
\tiny
\caption{  \label{tab:photometryIRS} Photometric data Oph IRS 48  (not corrected for extinction) used in this study }
\begin{center}
\begin{tabular}{l r  r @{ $\pm$ } l  c  }
\hline
\hline
band ID  &  \multicolumn{1}{c}{$\uplambda$ [$\upmu$m]}& \multicolumn{2}{c}{F$_{\nu}$ [Jansky] }& reference \\
\hline

2MASS J  		&    1.24 &  0.0946 &   0.0025 &a \\
2MASS H  	&    1.66 &  0.3050 &   0.0182 & a \\
2MASS K  	&    2.20 &  0.6182 &   0.0146 & a \\
WISE 3.3		& 	3.35& 1.294& 0.062& b\\
IRAC 1  		&    3.60 &  1.400 &   0.075 & c \\
IRAC 2  		&    4.50 &  1.630 &   0.082 & c \\
WISE 4.6		&	4.60& 2.454 & 0.102 & b\\
IRAC 3  		&    5.60 &  4.050 &   0.022 & c \\
ISOCAM  		&    6.70 &  5.87 &   0.110 & d \\
IRAC 4  		&    8.00 &  6.530 &   0.035 & c \\
WISE 11		&	11.60& 6.262& 0.069 & b\\
ISOCAM  		&   14.30 &  6.49 &   0.099 & d \\
AKARI  S18	&   18.00 & 21.04 &   0.049 & e \\
WISE 4		&	22.10& 36.472& 0.467 & b\\
IRAS 60  		&   60.00 & 65.46 &   13.01 & f \\
450 micron	&  450.00 &  1.56 &   0.39 & g \\
850	micron 	&  850.00 &  0.18 &   0.01 & g \\
1300 micron 	& 1300.00 &  0.06 &   0.01 & g \\

\hline

\end{tabular}
\end{center}

\textbf{References: }
\textbf{a)} 2MASS All-Sky Catalog of Point Sources 
\textbf{b)} WISE All-Sky Data Release
\textbf{c)} \citealt{2009Cieza}
\textbf{d)}  \citealt{2001Bontemps} 
\textbf{e)} AKARI/IRC mid-IR all-sky Survey 
\textbf{f)}  IRAS Point-source catalogue
\textbf{g)}  \citealt{2007Andrews}


\end{table}%

\end{appendix} 
\clearpage
\end{document}